\documentclass[twocolumn]{aastex62}

\DeclareGraphicsExtensions{.ps,.png}

\usepackage{amsmath}
\usepackage{amstext}
\usepackage{apjfonts}
\usepackage{graphicx}
\usepackage{savesym}
\usepackage{amssymb}
\usepackage{latexsym}
\usepackage{longtable}
\usepackage{xcolor}
\usepackage{times}
\usepackage{hyperref}
\usepackage{xspace}
\usepackage{multirow}

\renewcommand{\deg}{\mbox{$^{\circ}$}}

\newcommand{\omc}{\mbox{$\omega$ Cen~}}

\shorttitle{New Photometric Calibration of WFC3-UVIS and WFC3-IR} 
\shortauthors{Calamida et al.}

\begin{document}

\title{New Photometric Calibration of the Wide Field Camera 3 Detectors}

\correspondingauthor{Annalisa Calamida}
\email{calamida@stsci.edu}

\author[0000-0002-0882-7702]{Annalisa Calamida}
\affiliation{Space Telescope Science Institute - AURA, 3700 San Martin Drive, Baltimore, MD 21218, USA}

\author{Varun Bajaj}
\affiliation{Space Telescope Science Institute - AURA, 3700 San Martin Drive, Baltimore, MD 21218, USA}

\author[0000-0001-9736-8335]{Jennifer Mack}
\affiliation{Space Telescope Science Institute - AURA, 3700 San Martin Drive, Baltimore, MD 21218, USA}

\author[0000-0002-4775-7292]{Mariarosa Marinelli}
\affiliation{Space Telescope Science Institute - AURA, 3700 San Martin Drive, Baltimore, MD 21218, USA}

\author{Jennifer Medina}
\affiliation{Space Telescope Science Institute - AURA, 3700 San Martin Drive, Baltimore, MD 21218, USA}

\author{Aidan Pidgeon}
\affiliation{Space Telescope Science Institute - AURA, 3700 San Martin Drive, Baltimore, MD 21218, USA}

\author{Vera Kozhurina-Platais}
\affiliation{Space Telescope Science Institute - AURA, 3700 San Martin Drive, Baltimore, MD 21218, USA}

\author{Clare Shanahan}
\affiliation{Space Telescope Science Institute - AURA, 3700 San Martin Drive, Baltimore, MD 21218, USA}

\author[0000-0002-4814-2492]{Debopam Som}
\affiliation{Space Telescope Science Institute - AURA, 3700 San Martin Drive, Baltimore, MD 21218, USA}

\begin{abstract}
We present a new photometric calibration of the WFC3-UVIS and WFC3-IR detectors based on observations collected from 2009 to 2020 for four white dwarfs, namely GRW+70~5824, GD~153, GD~71, G191B2B, and a G-type star, P330E. These calibrations include recent updates to the {\it Hubble Space Telescope} primary standard white dwarf models and a new reference flux for Vega.
Time-dependent inverse sensitivities for the two WFC3-UVIS chips, UVIS1 and UVIS2, were calculated for all 42 full-frame filters, after accounting for temporal changes in the observed count rates with respect to a reference epoch in 2009. We also derived new encircled energy values for a few filters and improved sensitivity ratios for 
the two WFC3-UVIS chips by correcting for sensitivity changes with time. Updated inverse sensitivity values for the 20 WFC3-UVIS quad filters and for the 15 WF3-IR filters were derived by using the new models for the primary standards and the new Vega reference flux and, in the case of the IR detector, new flat fields. However, these values do not 
account for any sensitivity changes with time. The new calibration provides a photometric internal precision better than 0.5\% for the wide-, medium-, and narrow-band WFC3-UVIS filters, 5\% for the quad filters, and 1\% for the WFC3-IR
filters. As of October 15, 2020, an updated set of photometric keywords are populated in the WFC3 image headers.
\end{abstract}


\section{Introduction}\label{sec:intro}
The Wide Field Camera 3 (WFC3) instrument was installed on the \emph{Hubble Space Telescope} (\emph{HST}) during
the last servicing mission, Servicing Mission 4, on June 24, 2009. 

WFC3 has greatly advanced the imaging capabilities of \emph{HST} thanks to a combination of broad wavelength coverage, 
wide field of view, and high sensitivity. Composed of two optical/ultraviolet CCD detectors, or chips, UVIS1 and UVIS2, and a near-infrared (NIR) HgCdTe array, WFC3 can deliver high-resolution imaging over the wavelength range 2000 to 17000 \AA~ with a variety of wide-, intermediate-, and narrow-band filters. For more details on the
detectors we refer the reader to the WFC3 Instrument Handbook\footnote{https://hst-docs.stsci.edu/wfc3ihb}.

The WFC3-UVIS detectors were built from two different CCD wafers and so  
have different quantum efficiencies, more significantly in the ultra-violet (UV) regime 
($\lambda \lesssim$ 4,000 \AA), where UVIS2 is more sensitive.
In addition, the sensitivity of both detectors changes with time and 
the rate of change is different for each of them (\citealt{gosmeyer2016, shanahan2017a, khandrika2018}, \citealt[hereinafter CA21]{calamida2021}).
A change of sensitivity with time also seems to be present in the IR detector, the characterization of which is still ongoing (\citealt{bohlin2019, kozhurina2020}, \citealt[hereinafter BA20]{bajaj2020}).

In 2016, the WFC3 team implemented a chip-dependent photometric calibration and new values 
of the inverse sensitivities for UVIS1 and UVIS2 were provided \citep[hereinafter DE16]{deustua2016}, 
and later improved by using updated CALSPEC\footnote{https://www.stsci.edu/hst/instrumentation/reference-data-for-calibration-and-tools/astronomical-catalogs/calspec} models for the \emph{HST} primary spectrophotometric standard white dwarfs (WDs, \citealt[hereinafter DE17]{deustua2017}).
However, these values did not take into account the sensitivity change of the WFC3-UVIS detectors. 
As documented in \citet{khandrika2018} and CA21,
sensitivity changes are up to 0.2\% per year, according to the filter and the chip, resulting in differences of 
more than 2\%  in flux between 2009, when WFC3 was installed, and the current epoch. 
Due to the sensitivity changes being different for UVIS1 and UVIS2, 
as well as small errors in the flat field between the four readout amplifiers, the 2017 ratios of the observed count rates across the UVIS1 and UVIS2 detectors were off by up to 2\% for some filters \citep[hereafter CA18]{calamida2018}. 

The CALSPEC models for the \emph{HST} primary spectrophotometric standard WDs, 
GD153, GD71 and G191B2B, were updated in March 2020 \citep[hereinafter BO20]{bohlin2020}, and the Vega reference grey flux increased by $\approx$ 0.9\%. 
Overall, the standard WD fluxes increased by $\approx$ 2\% for wavelengths in the range 0.15 - 0.4 $\mu$m, and $\approx$ 1.5\% in the range 0.4 - 1 $\mu$m.
Therefore, WFC3-UVIS inverse sensitivities were updated in October 2020 to take into account these new CALSPEC reference fluxes and the different time sensitivity changes of the detectors.
%(CALSPECv11).
%\footnote{https://www.stsci.edu/hst/instrumentation/reference-data-for-calibration-and-tools/astronomical-catalogs/calspec}). 
The 20 quad filter inverse sensitivities were also updated to incorporate the new models, but 
did not include any time-dependent correction since 
no observations in these filters are available beyond 2010.

The WFC3-IR inverse sensitivities were last presented in \citet{kalirai2011}, and were only based on the first 1.5 years of photometric measurements of the three \emph{HST} primary standard WDs and the G-type standard P330E. In October 2020, the IR inverse sensitivities were updated by using all the observations collected through August 2020,
%for the same stars and for the \emph{HST} secondary standard WD, GRW70+70~5824 (hereinafter GRW70), 
the new CALSPEC models for the \emph{HST} primary WDs, and the new Vega reference flux; however, no time-dependent correction was applied  (BA20). 

The new WFC3-UVIS and WFC3-IR photometric calibrations also include observations of the standard WD GRW70+70~5824 (hereinafter GRW70). An improved spectral energy distributions (SED) of GRW70, based on new STIS observations with the $G430L$ grating, was added to the CALSPEC database.  The observed SED of GRW70 was upgraded to one of the best secondary \emph{HST} standards: routine monitoring of this star with STIS, ACS and WFC3 showed similar time-dependent changes as seen for the three \emph{HST} primary WDs (GD153, GD71, G191B2B), with no suggestion of systematic variability of GRW70 to within a limit of $\approx$1\% (BO20, Bohlin 2022, private communication). 

In this manuscript, we describe how the new inverse sensitivities for the 42 full-frame and the 20 quad WFC3-UVIS filters, and the 15 WFC3-IR filters were derived.
%illustrating the work of the WFC3 Instrument Science Reports (WFC3-ISRs) of CA21 and BA20. 
In both cases, approximately 10 years of photometry collected for three primary and one secondary \emph{HST} standard WDs, and the standard G-type star P330E were used. Their updated SEDs based on the new CALSPEC models were also used. More accurate encircled energy (EE) corrections were derived for a few WFC3-UVIS filters by normalizing the images using the newly derived time-dependent corrections and ratios. 
%Also, new color term transformations for the UVIS1 and UVIS2 WFC3 detectors 
%in the UV filters were derived, following the recipe of \citet{calamida2018}, and by using new photometric data for the standard stars. I
In this article, we also test the improved WFC3-UVIS photometric calibration by comparing multi-band and multi-epoch photometry for the Large Magellanic Cloud (LMC) open cluster NGC~1978 to a set of theoretical models.

The structure of the manuscript is as follows. In \S 2 we illustrate 
the observations used in this work and the data reduction process, in \S 3 we describe the data analysis process.
\S 4 presents the new EE corrections for WFC3-UVIS while \S 5 describes the process to derive
new in-flight corrections for UVIS1 and UVIS2, and \S 6 the process to derive the new
inverse sensitivities for the WFC3-UVIS and IR detectors. \S 7 compares the new to the old inverse sensitivities and in \S 8 we validate the new WFC3-UVIS time-dependent photometric calibration.
The next section provides an example on
how to perform WFC3-UVIS flux calibration, 
%and how to apply the color term transformations, 
and we summarize the results in \S 10.

\section{Observations and data reduction}\label{sec:obs}
\subsection{WFC3-UVIS}\label{redu_uvis}
Observations for four CALSPEC standard WDs, GRW70, GD153, GD71, G191B2B, 
and the G-type standard star, P330E, were collected with WFC3-UVIS between June 2009 and November 2019 during calibration and a few General Observer (GO) programs. 

The WFC3-UVIS detectors, UVIS1 and UVIS2, are $4k\times2k$ CCDs with a pixel scale 
of 0.04\arcsec/pixels, for a total combined field of view of 162\arcsec$\times$ 162\arcsec.
Each detector is divided in two amplifiers, A and B on UVIS1, and C and D on UVIS2. A scheme of the WFC3-UVIS detector and its amplifiers can be found in Fig.~1.2 of Section 1.2 of the Data Handbook\footnote{https://hst-docs.stsci.edu/wfc3dhb/chapter-1-wfc3-instruments/1-2-the-uvis-channel}.

Using a range of available sub-arrays, targets may be positioned on specific regions of the detector, and the total observing overhead is reduced by reading out only a fraction of the array. 
In order to mitigate the effects of charge transfer inefficiency, sub-arrays are defined at each of the UVIS  detector corners closest to the readout amplifiers 
(see Section 6.4\footnote{https://hst-docs.stsci.edu/wfc3ihb/chapter-6-uvis-imaging-with-wfc3/6-4-uvis-field-geometry\#id-6.4UVISFieldGeometry-6.4.4} of the Instrument handbook for details). 
For the WFC3 flux calibration, the five CALSPEC standards are typically observed in the smallest 512$\times$512 pixel corner sub-arrays, namely UVIS1-C512A-SUB (on amplifier A), UVIS1-C512B-SUB (B), UVIS2-C512C-SUB (C), and UVIS2-C512D-SUB (D). Two sub-array positions are observed for each detector in order to check the accuracy of the flat field calibration.

Exposure times for each filter were optimized to obtain a minimum Signal-to-Noise ratio 
($S/N$) of $\approx$ 100, and on average $S/N \approx$ 500 per exposure. 
Table~\ref{table:1} lists the proposal program numbers, 
the standard star names and filters for the observations included in this work.

Images were processed through the WFC3 pipeline, {\it calwf3}, version {\it 3.5.0}, which used the image 
photometry table (IMPHTTAB) available in November 2019, {\it 1681905hi\_imp.fits}, which corresponds to the latest WFC3-UVIS
photometric calibration of DE17. {\it calwf3} processes the images through the bias correction, dark subtraction, flat-fielding, gain conversion and charge transfer efficiency (CTE) correction. 
{\it calwf3} version {\it 3.5.0} used the original version of the CTE correction \citep{anderson2014} and the PCTETAB {\it zcv2057mi\_cte.fits}; a new CTE correction was implemented in April 2021 by the WFC3 team and is currently available with {\it calwf3} version {\it 3.6.0} and later \citep{anderson2021}, and uses the PCTETAB {\it 54l1347ei\_cte.fits}.

Standard stars were observed in the four UVIS 512$\times$512 corner sub-arrays, positioned close to the readout amplifiers, where the charge transfer inefficiency effects are smaller. Also, all the observed standards are bright ($V \lesssim$ 13.5 mag), and so less affected by the charge transfer inefficiency. 
However, we decided to test the effect of the new CTE correction on the standard star observations. We processed several images of GRW70 in a few filters with both the old and the new CTE correction. Aperture photometry was performed by using the same parameters and results were compared: count rates for GRW70 differed by no more than $\approx$ 0.01\% in all the filters examined. 

The {\it \_flc} images processed through {\it calwf3} were also multiplied by the pixel area map (PAM, \citealt{kalirai2010}), to correct for differences in the area of each pixel on 
the sky due to the geometric distortion of the UVIS1 and UVIS2 detectors.

\LTcapwidth=\textwidth
{ \footnotesize
\begin{longtable*}{ll p{0.75cm}p{0.75cm}p{0.75cm}p{0.75cm}p{0.75cm}p{0.75cm}p{0.75cm}p{0.75cm}p{0.75cm}p{0.75cm}p{0.75cm}p{0.75cm}}
%\centering
%\begin{scriptsize}
\caption{\footnotesize Program number of the regular calibration or GO proposal, standard star name and filters for the WFC3-UVIS observations included in this analysis. These data can be found on the MAST archive by using the following DOI: 10.17909/gvre-t314. \label{table:1}}\\
\hline
Program & Star  & \multicolumn{12}{c}{Filters} \\
\hline
\hline
\endfirsthead
\hline
11426 & GRW70 & F218W & F225W & F275W & F280N & F300X & F336W & F343N & F373N & F390M & F390W & F395N & F410M \\
& & & & & & & & & & F438W & F467M & F606W & F814W  \\
\hline
\multirow{2}{*}{11450} & & F218W& F225W& F275W& F280N& F300X& F336W& F343N & F350LP & F373N & F390M&F390W&F395N \\
& {GD153} & F410M&F438W&F467M&F469N & F475W& F475X& F487N& F502N& F547M& F555W& F600LP & F606W\\
& & F621M&F625W& F656N& F658N& F665N& F673N& F689M& F763M & F775W& F814W&F845M& F953N \\
\hline
11557 & GRW70 & & & & & & & & & & & & F475W \\
\hline
\multirow{2}{*} & & F200LP & F218W& F225W& F275W& F280N& F300X& F336W& F343N& F350LP & F373N& F390M&F390W\\
& & F395N& F410M& F438W& F467M& F469N& F475W& F475X& F487N& F502N& F547M& F555W& F600LP \\
& {G191B2B} & F606W& F621M& F625W& F631N& F645N& F656N& F657N& F658N& F665N& F673N&F680N& F689M \\ 
& & & & & & & & F763M& F775W& F814W& F845M& F850LP & F953N  \\
{11903} & {GD153} & F225W& F275W& F336W& F350LP &  F390W& F438W& F467M& F469N& F475W& F502N& F547M& F555W \\
& & & & & & & & & & F606W& F775W& F814W& F850LP  \\
&{GD71} & F350LP & F390W& F438W& F467M&  F469N& F475W& F502N& F547M& F555W& F606W& F775W& F814W \\
& &  & & & & & & & & & & &  F850LP \\
& {P330E} & F200LP & F218W& F225W& F275W& F300X& F336W& F350LP & F390W& F410M& F438W& F467M&F475W  \\
& &  & F475X & F547M& F555W& F600LP & F606W& F621M& F625W& F689M& F775W& F814W& F850LP  \\
\hline
11907 & GRW70 & &  F218W& F225W& F275W& F336W& F390M& F390W& F438W& F475W& F547M& F606W& F814W  \\
\hline
\multirow{2}{*}{12333}& {GRW70} & F218W& F225W& F275W& F300X& F336W& F390M& F390W& F438W& F467M& F469N& F475W& F502N\\
& & & & & & & & & F547M& F555W& F606W& F814W& F850LP \\
\hline
\multirow{2}{*}{12698} & {GRW70} & F218W& F225W& F275W& F300X& F336W& F390M& F390W& F438W& F467M& F475W& F502N & F547M\\
& & & & & & & & & &  F555W& F606W& F814W& F850LP \\
\hline
13088 & GRW70  & & & & & & F218W& F225W& F275W& F336W& F438W& F606W& F814W  \\
\hline
\multirow{2}{*}& & F200LP & F218W& F225W& F275W& F280N& F300X& F336W& F343N& F350LP & F373N& F390M& F390W\\
& {GD153} & F395N& F410M& F438W& F467M& F469N& F475W& F475X& F487N& F502N& F547M& F555W& F600LP\\
&  & F606W& F621M& F625W& F631N& F645N& F656N& F657N& F658N& F665N& F673N& F680N& F689M\\
{13089}  & & & & & & & & F763M& F775W& F814W& F845M& F850LP &  F953N \\
&  & F200LP & F218W& F225W& F275W& F280N& F300X& F336W& F343N& F350LP & F373N& F390M& F390W\\
& {P330E} & F395N& F410M& F438W& F467M& F469N& F475W& F475X& F487N& F502N& F547M& F555W& F600LP\\
& & F606W& F621M&F625W& F631N& F645N& F656N& F657N& F658N& F665N& F673N& F680N& F689M\\
& & & & & & & & F763M& F775W& F814W& F845M& F850LP &  F953N \\
\hline
13574 & GRW70  & & & & & & F218W& F225W& F275W& F336W& F438W& F606W& F814W \\
\hline
\multirow{2}{*} &  & F200LP & F218W& F225W& F275W& F280N& F300X& F336W& F343N& F350LP& F373N& F390M& F390W\\
& {GD153} & F395N& F410M& F438W& F467M& F469N& F475W& F475X& F487N& F502N& F547M& F555W& F600LP\\
& &  F606W& F621M&F625W& F631N& F645N& F656N& F657N& F658N& F665N& F673N& F680N& F689M\\
{13575}  & & & & & & & & F763M& F775W& F814W& F845M& F850LP&  F953N \\
& & F200LP& F218W& F225W& F275W& F280N& F300X& F336W& F343N& F350LP& F373N& F390M& F390W\\
& {P330E}  &  F395N& F410M& F438W& F467M& F469N& F475W& F475X& F487N& F502N& F547M& F555W& F600LP\\
& &  F606W& F621M&F625W& F631N& F645N& F656N& F657N& F658N& F665N& F673N& F680N& F689M\\
& & & & & & & & F763M& F775W& F814W& F845M& F850LP&  F953N \\
\hline
\multirow{2}{*}{13711} & {G191B2B} & & & & & & & & F275W& F336W& F475W& F625W& F775W \\
&   {GD153} & & & & & & & & F275W& F336W& F475W& F625W& F775W \\
&   {GD71}  & & & & & & & & F275W& F336W& F475W& F625W& F775W \\
\hline
\multirow{2}{*}& & F200LP & F218W& F225W& F275W& F300X& F336W& F350LP& F390M&F390W& F410M& F438W& F467M\\
& {G191B2B} &  F475W& F475X& F547M& F555W& F600LP& F606W& F621M& F625W& F689M& F763M& F775W& F814W\\
{14018}  & & & & & & & & & & & & F845M& F850LP \\
& {GRW70}  & F218W& F225W& F275W& F300X& F336W& F390M&F390W& F410M& F438W& F467M& F475W& F547M\\
& & & & & & & & & & F555W& F606W& F814W& F850LP \\
\hline
\multirow{2}{*}{14021} & {GD153} & F218W& F225W& F275W& F336W& F350LP& F438W& F475W& F547M& F555W& F600LP & F606W& F621M\\
& & & & & & & & & & F625W&  F775W& F814W& F845M \\
& {P330E} & F275W& F336W& F350LP& F438W& F475W& F547M& F555W& F600LP& F606W& F621M& F625W&  F775W\\
& & & & & & & & & & & F814W& F845M& F850LP \\
\hline
\multirow{2}{*} & {G191B2B} & F218W& F225W& F275W& F336W& F438W& F475W& F547M& F555W& F600LP& F606W& F621M& F625W\\
& & & & & & & & & & & F775W& F814W& F845M \\
&  {GD153} & F218W& F225W& F275W& F336W& F350LP& F438W& F475W& F547M& F555W& F600LP& F606W& F621M\\
{14384} & & & & & & & & & & F625W&  F775W& F814W& F845M \\
&  {GD71} & F218W& F225W& F275W& F336W& F350LP& F438W& F475W& F547M& F555W& F600LP& F606W& F621M\\
& & & & & & & & & & F625W& F775W& F814W& F845M \\
&  {P330E} & F275W& F336W& F350LP& F438W& F475W& F547M& F555W& F600LP& F606W& F621M& F625W&  F775W\\
& & & & & & & & & & & F814W& F845M& F850LP \\
\hline
\multirow{2}{*}{14815} & {GD153} & & & & & & F218W& F225W& F275W& F336W& F438W& F606W& F814W \\
&  {GRW70} & & & & & &  F218W& F225W& F275W& F336W& F438W& F606W& F814W \\
\hline
\multirow{2}{*} & {G191B2B} & F218W& F225W& F275W& F336W& F438W& F475W& F547M& F555W& F600LP & F606W& F621M& F625W\\
& & & & & & & & & & & F775W& F814W& F845M \\
&  {GD153} & F218W& F225W& F275W& F336W& F350LP& F438W& F475W& F547M& F555W& F600LP & F606W& F621M\\
{14883}  & & & & & & & & & & F625W&  F775W& F814W& F845M \\
& {GD71} & F218W& F225W& F275W& F336W& F350LP& F438W& F475W& F547M& F555W& F600LP & F606W& F621M\\
& & & & & & & & & & F625W& F775W& F814W& F845M \\
& {P330E} & F275W& F336W& F350LP& F438W& F475W& F547M& F555W& F600LP& F606W& F621M& F625W&  F775W\\
& & & & & & & & & & & F814W& F845M& F850LP \\
\hline
\multirow{2}{*}  & {G191B2B} & F218W& F225W& F275W& F336W& F438W& F475W& F547M& F555W& F606W& F621M& F625W& F657N\\
& & & & & & & & & & & F775W& F814W& F953N \\
& {GD153} & F218W& F225W& F275W& F336W& F350LP& F438W& F475W& F547M& F555W& F600LP& F606W& F621M\\
{14992} & & & & & & & & & F625W& F657N& F775W& F814W& F845M \\
& {GD71} & F218W& F225W& F275W& F336W& F438W& F475W& F547M& F555W& F606W& F621M& F625W&  F657N\\
& & & & & & & & & & & F775W& F814W& F953N \\
& {P330E} & F275W& F336W& F350LP& F438W& F475W& F547M& F555W& F600LP& F606W& F621M& F625W&  F775W\\
& & & & & & & & & & & F814W& F845M& F850LP \\
\hline
\multirow{2}{*} & {G191B2B} & & & & & & & & F275W& F336W& F475W& F625W& F775W \\
{15113} & {GD153}   & & & & & & & & F275W& F336W& F475W& F625W& F775W \\
& {GD71}   & & & & & & & & F275W& F336W& F475W& F625W& F775W \\
\hline
\multirow{2}{*}{15398} & {GD153} & & & & & & F218W& F225W& F275W& F336W& F438W& F606W& F814W \\
& {GRW70} & & & & & & F218W& F225W& F275W& F336W& F438W& F606W& F814W \\
\hline
\multirow{2}{*}{15399} & {GD153} & & & & & & & F218W& F225W& F275W& F336W& F606W& F814W \\
& {P330E} & & & & & & & F218W& F225W& F275W& F336W& F606W& F814W \\
\hline
\multirow{2}{*} & {GD153} & F218W& F225W& F275W& F336W& F438W& F475W& F547M& F555W& F606W& F621M& F625W& F657N\\
& & & & & & & & & & & F775W& F814W& F845M \\
& {GD71} & F218W& F225W& F275W& F336W& F438W& F475W& F547M& F555W& F606W& F621M& F625W&  F657N\\
{15582}  & & & & & & & & & & & F775W& F814W& F953N \\
& {P330E} & F275W& F336W& F350LP& F438W& F475W& F547M& F555W& F600LP& F606W& F621M& F625W&  F775W\\
& & & & & & & & & & & F814W& F845M& F850LP \\
& {GRW70} & F218W& F225W& F275W& F336W& F438W& F475W& F547M& F555W& F606W& F621M& F625W& F657N\\
& & & & & & & & & & & F775W& F814W& F953N \\
\hline
\multirow{2}{*}{15583} & {GD153} & & & & & &  F218W& F225W& F275W& F336W& F438W& F606W& F814W \\
& {GRW70} & & & & & & F218W& F225W& F275W& F336W& F438W& F606W& F814W \\
\hline
\hline
%\end{scriptsize}
\end{longtable*}
}

A \texttt{Python} pipeline based on \texttt{Photutils} and \texttt{WFC3\_tools}\footnote{https://github.com/spacetelescope/wfc3tools} was developed to perform photometry on the thousands of images available. 
Below we provide a description of the pipeline steps that produce photometric catalogs for each standard star and filter.

{1)} The PAM-corrected {\it \_flc} images were divided by the exposure time to convert total counts (e$^-$) to count rates (e$^-$/s);

{2)} Standard stars in calibration programs are usually observed close to the center of the 512$\times$512 sub-array; therefore, a first attempt to detect the star near the center of the sub-array wass done by using a segmentation map. Images were smoothed with a 3$\times$3 pixel kernel with a Full Width Half Maximum (FWHM) of 1.8 pixels. A detection threshold of 30 and 100 connected pixels was found to work for all images to find most stars on the first try. 
If no sources were found, the detection parameters were adjusted, i.e. the threshold was set to 15 and the connection pixels to 75, and the segmentation map was created again. 
If the second try failed, the image was discarded; however, this happened in a very small fraction of data, $\le$ 2\%.
In the case where two or more sources were found, a method was devised to determine which of those was the standard star. The image header keywords {\it RA\_TARG} and {\it DEC\_TARG} were compared to the coordinates of the detected standard,  {\it RA} and {\it DEC}. 
Since the proper motion information was not included in the calibration proposals for pre-2015 data,
\texttt{astroquery} was used to query SIMBAD for the proper 
motion of the standard star and these were applied to {\it RA\_TARG } and {\it DEC\_TARG}.
The detected source with coordinates closest to the corrected target location 
was selected as the standard star;

{3)} The sky background and sky root mean square (RMS) were calculated as the sigma-clipped mean of the pixels in a circular annulus of 9 pixels in width, with an inner radius of 156 pixels centered
on the detected standard star. The sky background was then subtracted from the source count rates;

{4)} Aperture photometry was measured at different aperture radii, from 1 to 50 pixels, centered on the standard star;

{5)} Photometric errors were computed by following the prescription of \citet{stetson1987}, i.e. including Poisson, sky background and readout noises;

{6)} Outlier measurements were defined as those more than 5\% away from the median count rate value of the standard star on all the {\it \_flc} exposures for each filter and amplifier; these were clipped before the catalogs were finalized. 
This cleaning enabled the removal of images impacted by cosmic ray (CR) hits on the source Point-Spread Function (PSF) or of poor measurements.

\subsubsection{Scanned photometry}\label{sec:scan}
WFC3-UVIS spatial scan observations for two of the four WDs, namely GRW70 and GD153, 
were also included in the analysis to measure the sensitivity change of the 
UVIS1 and UVIS2 detectors with time. 
Spatial scans of bright sources, when compared to staring mode observations, 
are expected to yield higher precision photometry. Scans allow the collection of millions of source photons without causing saturation by spreading them across many pixels on the 
detector, and, thereby, reducing the Poisson noise. Averaging over a large number of pixels 
also helps to reduce noise originating from spatial effects such as bad pixels and flat-field errors. 
Indeed, it has been determined that sub-0.1\% photometric repeatability is possible 
with spatial scans \citep{shanahan2017b}.

\begin{table*}
\begin{center}
\caption{\footnotesize Program number of the regular calibration proposals, standard star name and filters for the WFC3-UVIS spatial scan observations included in this analysis. 
These data can be found on the MAST archive by using the following DOI: 10.17909/q0m5-n042.
\label{table:2}}
\begin{tabular}{ll p{1.0cm}p{1.0cm}p{1.0cm}p{1.0cm}p{1.0cm}p{1.0cm}p{1.0cm}}
\hline
Program & Star  & \multicolumn{7}{c}{Filters} \\
\hline
\hline
\multirow{2}{*}{14878} & {GRW70} & F218W& F225W& F275W& F336W& F438W& F606W& F814W \\
& {GD153} & F218W& F225W& F275W& F336W& F438W& F606W& F814W \\
\hline
\multirow{2}{*}{15398} &{GRW70} & F218W& F225W& F275W& F336W& F438W& F606W& F814W \\
& {GD153} & F218W& F225W& F275W& F336W& F438W& F606W& F814W \\
\hline
\multirow{2}{*}{15583} &{GRW70} & F218W& F225W& F275W& F336W& F438W& F606W& F814W \\
& {GD153} &  F218W& F225W& F275W& F336W& F438W& F606W& F814W \\
\hline
\hline
\end{tabular}
\end{center}
\end{table*}

Spatial scan data were collected during four calibration proposals between 
2017 and 2020; Table~\ref{table:2} lists the program numbers, the 
standard star names and the filters of the scan observations used in this analysis.
Program 14878 was exploratory and examined the viability of using 
spatial scans as a high-precision technique for studying temporal photometric stability. 
An optimal observing strategy, based on the results from this program, 
was established by \citet{shanahan2017b}, and all the observations included in 
this work were obtained following their prescriptions.
Data were acquired using either the UVIS1-C512A-SUB
%(near amplifier A on UVIS1) 
or the UVIS2-C512C-SUB
% (near amplifier C on UVIS2) 
sub-array. 
%In both cases, the single-lined, vertical 
%scan was placed in the middle of the sub-array.

As in the staring mode reduction process described above, raw scan images were processed 
through the {\it calwf3} pipeline version {\it 3.5.0}, by using the IMPHTTAB {\it 1681905hi\_imp.fits}, so that bias 
correction, dark subtraction, flat-fielding and gain conversion were performed.
However, unlike the staring mode data, scan images were not corrected for the charge transfer inefficiency effects since
these are minimal in the bright spatial scan trails centered within sub-arrays close to the readout amplifiers.
To further mitigate these effects, scans were executed in the vertical direction along the detector columns. In addition, scans were inclined by a 1$\deg$ angle to uniformly sample the pixel phase for each CCD.
%especially for scans in the vertical direction.

The {\it \_flt} image products were then processed by a multi-step reduction pipeline 
introduced and described fully in \citet{shanahan2017b}. In summary, this \texttt{Python} based pipeline utilizes 
various tools from scientific data analysis packages such as \texttt{astropy} and \texttt{Photutils}, 
%and \texttt{WFC3\_tools} 
and performed the following steps:

{1)} CR detection and repair - Longer exposure times and the spreading of source flux over a large area on 
the detector make the spatial scans more susceptible to CR hits compared to the staring mode observations. 
Building on a routine originally developed for CR identification in Space Telescope Imaging spectrograph (STIS) CCD images, this step identified
CR events in the data. The affected pixels were then repaired by interpolating from unaffected neighboring pixels;

{2)} Determining the scan location - Each image was designed to have the single-lined, vertical scan positioned at the 
center of the 512$\times$512 sub-array. However, to account for small shifts, an automated determination of the scan 
centroid location was performed for each image. A simultaneous determination of the scan direction was also performed. 
However, as mentioned before, the entire dataset considered in this work comprises vertical scans only;

{3)} Sky background subtraction - The sky region corresponding to each vertical scan was defined as all pixels excluding 
a 10-pixel wide strip bordering the sub-array and a 350$\times$75 pixel rectangular region centered on the scan. 
The sky background level and the sky RMS were calculated as the sigma-clipped mean and RMS of all the sky pixels. 
This background was then subtracted from the data and the errors were propagated accordingly; 

{4)} Scaling with pixel area maps - The sky subtracted image was scaled by applying the appropriate PAM
to account for geometric distortions of the detector;

{5)} Aperture photometry - The last step in this process is to perform aperture photometry on the sky-subtracted,
PAM corrected image to determine the sum of pixels in the scan. This was done using a 240$\times$36 pixel rectangular 
aperture placed at the scan centroid determined in step 2). The dimensions of the aperture were chosen such 
that it was large enough to contain the scan in its entirety, yet it was not too large to be affected by noise from the sky subtraction. 
In this regime of very high total source counts, the Poisson noise term should dominate and is therefore approximated as the measurement error. Finally, the photometric measurement was converted into source count rates (e$^-$/s) by dividing the sum of pixels by the image exposure time.

\begin{table*}
\begin{center}
\caption{ \footnotesize Star name and program numbers for the WFC3-IR observations used in this analysis.
These data can be found on the MAST archive by using the following DOI: 10.17909/04tn-rj35.
\label{table:3}}
%\def\arraystretch{1.25}
%\begin{tabular}{ l | l}
\begin{tabular}{l p{1.0cm}p{1.0cm}p{1.0cm}p{1.0cm}p{1.0cm}p{1.0cm}p{1.0cm}p{1.0cm}p{1.0cm}p{1.0cm}p{1.0cm}}
%\hline
\hline
Stars & \multicolumn{11}{c}{Program} \\
\hline
\hline
\multirow{2}{*} {GD153} & 11451 & 11552 & 11926 & 12334 & 12699 & 12702 & 13089 & 13092 & 13575 & 13579 & 13711 \\
                      &       & 14021 & 14384 & 14386 & 14544 & 14883 & 14992 & 14994 & 15113 & 15582 & 16030 \\
\hline
\multirow{2}{*} {GD71}  & 11926 & 11936 & 12333 & 12334 & 12357 & 12699 & 12702 & 13711 & 14024 & 14384 & 14883 \\
                      &       &       &       &       &       &       &       & 14992 & 15113 & 15582 & 16030 \\
\hline
                {GRW70} &     &        &        &       & 11557 & 12333 & 12698 & 13088 & 13575 & 15582 & 16030 \\
\hline
\multirow{2}{*}  {P330E} & 11451 & 11926 & 12334 & 12699 & 13089 & 13573 & 13575 & 14021 & 14328 & 14384 & 14883 \\
                      &       &       &       &       &       &       &       &       &       & 14992 & 16030 \\
\hline
              {G191B2B} &       &       &       &       &       & 11926 & 12334 & 13094 & 13576 & 13711 & 15113 \\
\hline
\hline
\end{tabular}
\end{center}
\end{table*}

\subsection{WFC3-IR}\label{redu_ir}
Observations for four CALSPEC standard WDs and for the G-type standard star P330E were collected with WFC3-IR between June 2009 and August 2020 during regular calibration and a few GO programs in all the 15 filters.

WFC3-IR is a 1014$\times$1014 detector, with a pixel scale of 0.13\arcsec and a total
field of view of 136\arcsec$\times$ 123\arcsec.
Sub-arrays of different sizes are available for the observations at the center of the detector. Standard star images were collected by using these sub-arrays, with the size determined from the exposure time used;this
ranged between 64 $\times$ 64 and 512 $\times$ 512 pixels (see Section 7.4 of the 
Instrument Handbook\footnote{https://hst-docs.stsci.edu/wfc3ihb} for more details on the different WFC3-IR sub-arrays).

As some of the programs were not designed for photometric calibration purposes, the number of observations for each target 
and filter varies. The list of programs in which data were taken for each star is presented in Table~\ref{table:3}.

The majority of the datasets used in this analysis were observed as part of photometric calibration programs, and typically feature long enough exposure times to exceed a $S/N \gtrsim$ 100.
Starting from 2017, the observations were designed to mitigate the effects of persistence, which is
critical to achieve high-precision photometry. In particular, 
frequent dithering of at least 10 pixels was used to place the star on a recently unused portion of the detector \citep{bajaj2019}.

Images were processed with {\it calwf3} version {\it 3.5.0} and the IMPHTAB {\it wbj1825ri\_imp.fits} was used.
New flat fields were delivered at the end of 2020 and were used in the image processing; these have errors $\lesssim$ 0.5\% \citep{mack2021}.

Images were grouped by target and filter, and those collected in the same visit were drizzled together, as images taken in the same visit typically have very precise relative astrometry. 
Drizzling reduced the number of discrepant artifacts in the images, such as CRs, hot and bad pixels.

Source finding was performed on the drizzled images by using the DAOFIND algorithm \citep{stetson1987} as implemented in the \texttt{Python} package \texttt{Photutils}. The FWHM was set to the width of the WFC3-IR PSF, i.e. $\approx$ 1.2 pixels. Though the FWHM varies slightly for different filters, a parametrization with respect to wavelength was unnecessary to achieve satisfactory results.  
Due to the highly undersampled nature of the WFC3-IR PSF, many spurious objects would often be detected on the drizzled image.
In some cases, due to a larger sub-array and longer exposure times, other sources also appeared in the images, 
leading to extra detections. To dispense of the superfluous detections, an initial pass of aperture photometry was 
performed on the drizzled images. The measured count rates of all sources were then compared to synthetic count rates 
of the standard stars computed with {\it Pysynphot} by using the latest SEDs and the total system throughput curves. 
The object that reported the closest count rates to the synthetic ones was used to record an approximate position 
of the standard star in each drizzled image. %This proved to be the most successful out of many source detection methods
%that were implemented, as the synthetic flux values and photometric performance are consistent enough (within a few percent) to ensure an accurate selection of the target.
This position was transformed from the drizzled to the {\it \_flt} 
image coordinate system via the \verb+all_pix2world()+ and \verb+all_world2pix()+ methods of 
the WCS package within \texttt{astropy} (\cite{astropy}).  
The position was then re-centered in each {\it \_flt} image using a 2D gaussian fitting to 
the central-most pixels of the PSF, ensuring that the small aperture used in the photometry is placed correctly. 

Aperture photometry was then performed on the PAM-corrected  {\it \_flt} images with an aperture radius of 3 pixels 
($\approx$ 0.4\arcsec), a background annulus ranging from 15 to 30 pixels centered on the standard star, 
and using a $\sigma$-clipped median to calculate the sky background. 
Unlike the analysis in \citet{kalirai2011}, the PAM multiplication was necessary, as the placement of the 
stars on the images spanned much of the total detector area. 
%Thus, the PAM was sliced to contain only the pixels used in each sub-array image, 
%and multiplied by the data array before photometry was performed. 
Aperture photometry was perforemed by using \texttt{Photutils} and the \texttt{wfc3\_photometry} package \citep{bradley_photutils}. 
Since the {\it \_flt} images for the IR channel are already in units of $e^{-}/s$, the exposure time corrections was not required.  
In \citet{kalirai2011}, the aperture radius of 3 pixels was used, but was described as not optimal for minimizing 
the dispersion of the measurements. However, in repeating this analysis with more data, we found that the 3-pixel 
aperture minimizes the standard deviation for GD153 (the most observed star of the set) flux measurements in 
both the $F110W$ and $F160W$ filters.

\begin{figure*}
\begin{center}
\includegraphics[width=0.55\textwidth, height=0.75\textheight, angle=90]{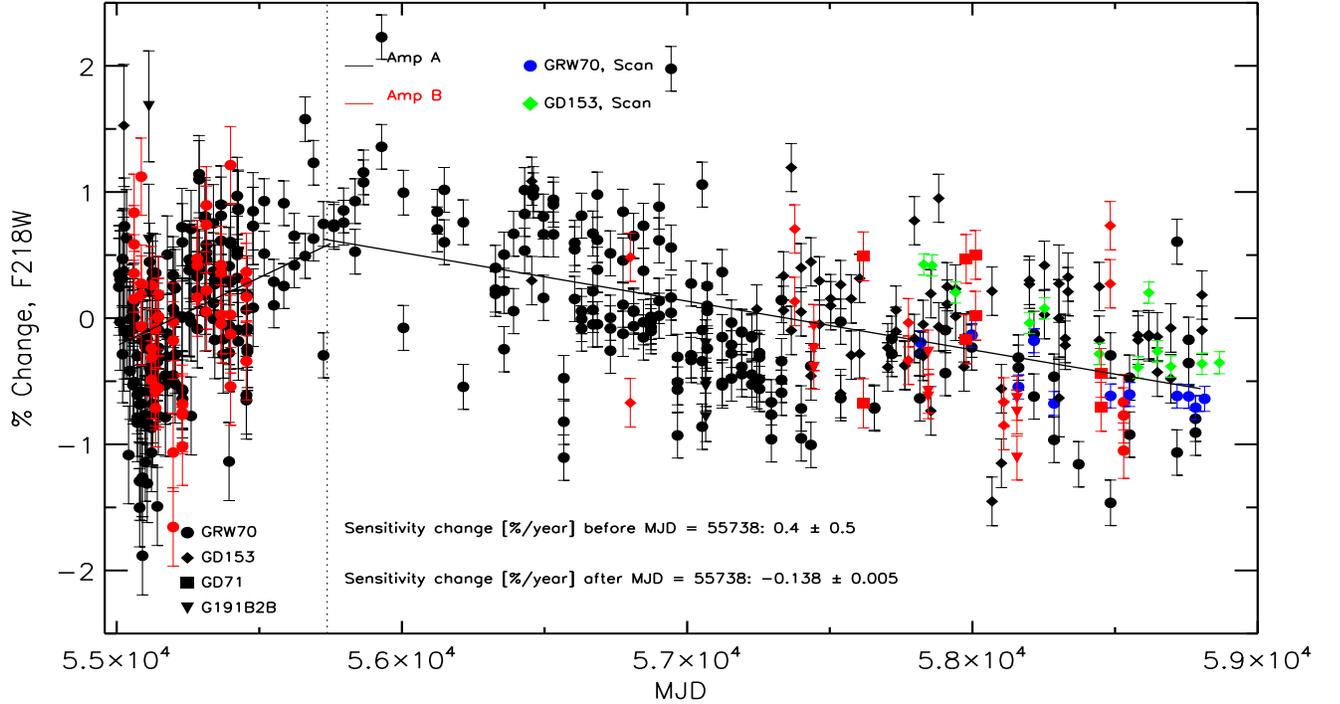}
\caption{Aperture photometry performed with a 10-pixel radius on {\it \_flc} images collected with the 
F218W filter and the UVIS1-C512A-SUB (Amp A, black) and UVIS1-C512B-SUB (Amp B, red) sub-arrays for four CALSPEC white dwarfs, namely GRW70 (filled circle), GD153 (diamond), GD71 (square), 
and G191B2B (triangle), is plotted as a percent change values versus the $MJD$ of the observations.
Photometry performed on scan images in the same filter and for GRW70 (blue) and GD153 (green) is also shown. The solid lines show the fit to the photometry of all
the stars before and after {\bf $MJD =$ 55738}, indicated by a vertical dotted line. 
The sensitivity change rates in \%/yr derived from fitting the data are labeled in the figure. \label{fig:f218ascan}}
\end{center}
\end{figure*}

\begin{figure*} 
\begin{center}
\includegraphics[width=0.55\textwidth, height=0.75\textheight, angle=90]{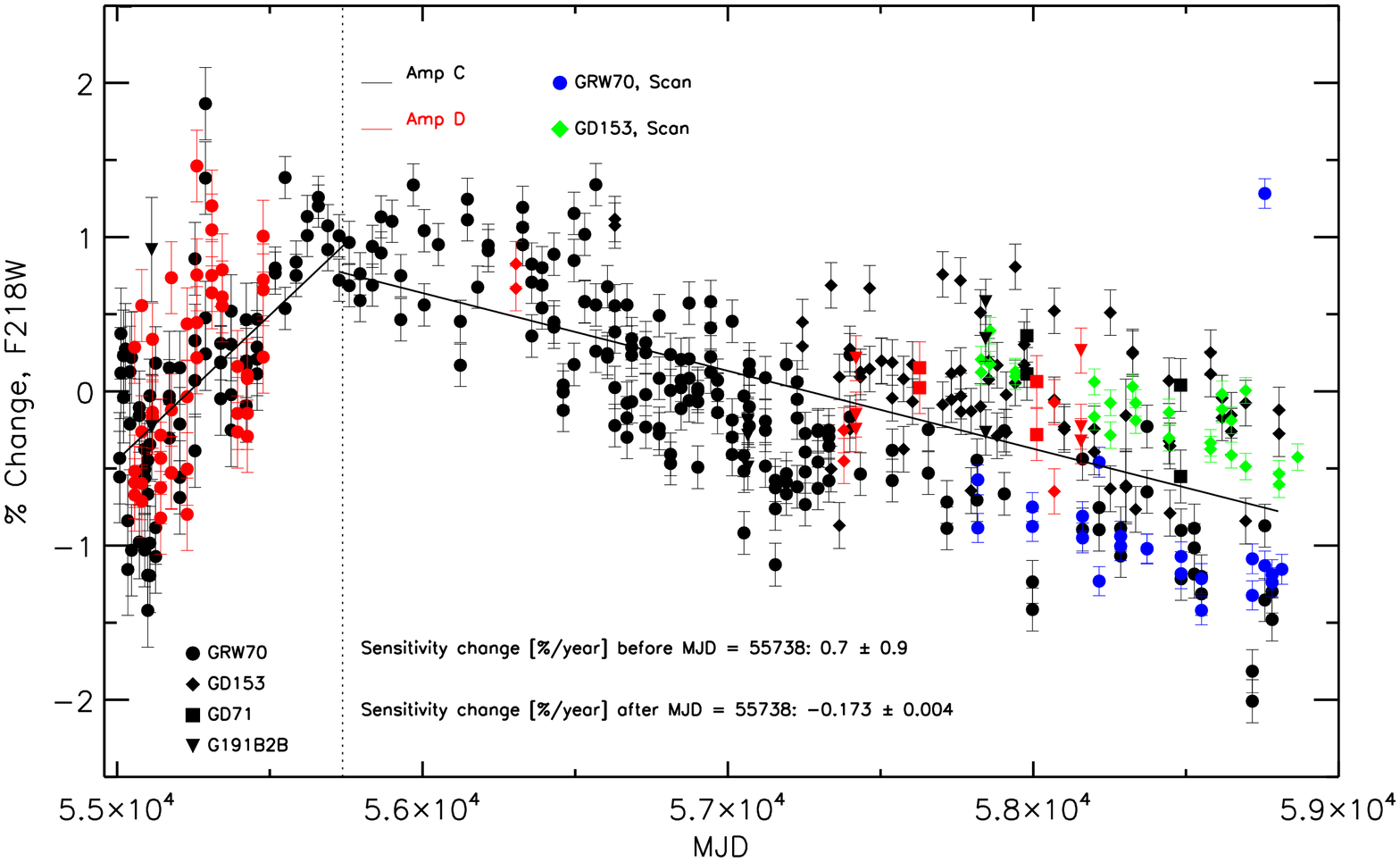}
\caption{Same as Fig.~\ref{fig:f218ascan} but for data collected with the UVIS2-C512C-SUB (Amp C, black) 
and the UVIS2-C512D-SUB (Amp D, red) sub-arrays. \label{fig:f218cscan}}
\end{center}
\end{figure*}

\begin{figure*}
\begin{center}
\includegraphics[width=0.55\textwidth, height=0.75\textheight, angle=90]{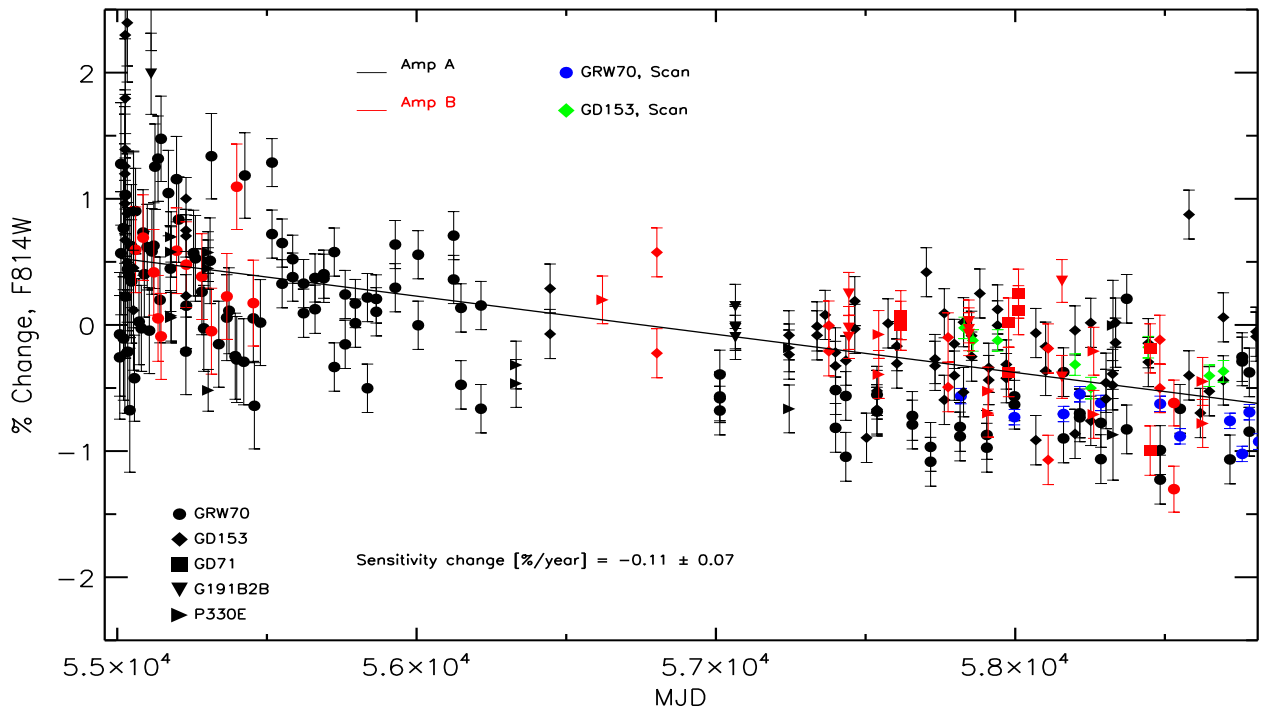}
\caption{Same as Fig.~\ref{fig:f218ascan} but for the $F814W$ filter. In this case, photometry for the CALSPEC G-type standard P330E (horizontal triangle) is also included. \label{fig:f814ascan}}
\end{center}
\end{figure*}

\begin{figure*}
\begin{center}
\includegraphics[width=0.55\textwidth, height=0.75\textheight, angle=90]{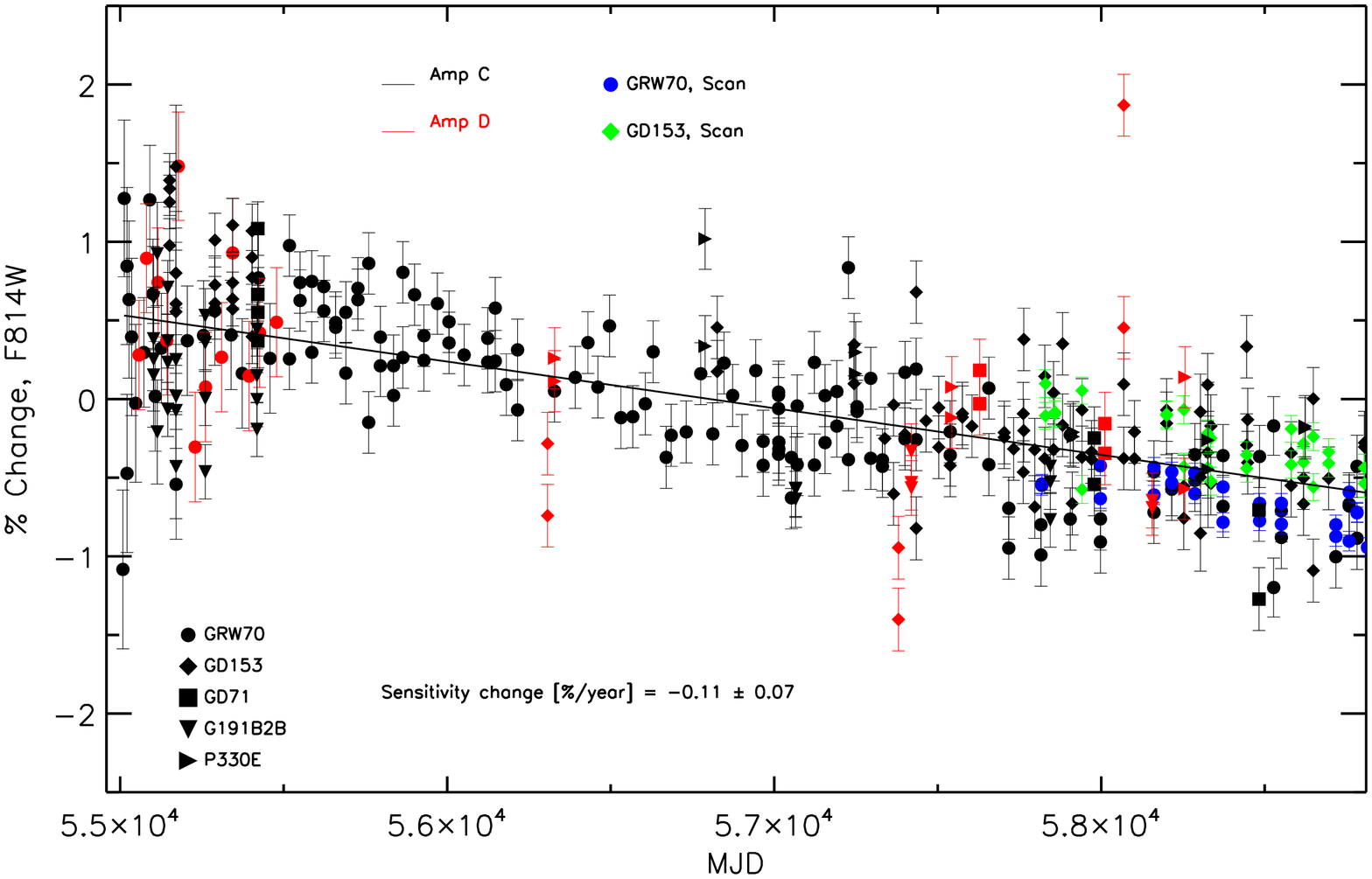}
\caption{Same as Fig.~\ref{fig:f814ascan} but for data collected with  the UVIS2-C512C-SUB (Amp C, black) 
and the UVIS2-C512D-SUB (Amp D, red) sub-arrays. \label{fig:f814cscan}}
\end{center}
\end{figure*}

\section{Data Analysis}\label{sec:analysis}
\subsection{WFC3-UVIS}
Aperture photometry with a 10-pixel radius for each standard star and filter was 
normalized to the mean value over the full time interval and the percent change of the 
count rates was plotted as a function of the Modified Julian Day ($MJD$) of the observation. 
The spatial scan photometry for GRW70 and GD153, when available,
was normalized to the staring mode photometry at 10 pixel for the same stars in the same 
time interval, $\sim$ 55700 -- 58800, i.e. $\sim$ 2016.8-- 2019.8, and the percent change 
values were overplotted. 
This normalization was done by manually shifting the scan count rates to match the mean count rate of the staring mode of each observed standard star over the same time interval. The WFC3 team is currently testing new 'enrectangled' energy corrections to be applied to synthetic count rates in order to compare them to the  standard star observed count rates. This process is described in Section~\ref{sec:scan} and more details can be found in Marinelli et al. (2022, in prep.).

As shown in Figs.~\ref{fig:f218ascan}, \ref{fig:f218cscan}, 
\ref{fig:f814ascan}, and \ref{fig:f814cscan}, small offsets between the photometry of different standard stars is present and is currently under investigation by the WFC3 team. We used these data to determine the best 
sensitivity change slopes for each filter and chip and then corrected the count rates of each 
standard star over time. The corrected count rates were then used to determine the final normalized photometry in all filters. Since most measurements were collected on Amps A and C,
mean count rates on Amps B and D were normalized to the mean count rates on Amp A and C, respectively, 
in order to correct for small ($\lesssim$ 2\%) errors in the flat-field across each chip \citep{mack2015}.

\begin{table*}
\begin{center}
\begin{scriptsize}
\caption{\footnotesize Slopes and their 1-$\sigma$ dispersion of the sensitivity changes 
for UVIS1 (Amp A) and UVIS2 (Amp C) 42 full-frame filters. The UV filters have two slopes, one for $MJD \le$55738 
and one for later times, while the redder filters have only one slope (see text for more details). \label{table:4}}
\vspace{0.3cm}
\begin{tabular}{lcccccc}
\hline
\hline
Filter & Pivot  & Slope1/$\sigma$ ($MJD \le$ 55738) & Slope2/$\sigma$ ($MJD >$ 55738) & Pivot  & Slope1/$\sigma$ ($MJD \le$ 55738) & Slope2/$\sigma$ ($MJD >$ 55738) \\
         &  (\AA) &  (\%/yr)  &    (\%/yr)  & &   (\%/yr)  &    (\%/yr)\\
\hline
\multicolumn{4}{c}{UVIS1 (Amp A)} & \multicolumn{3}{l}{UVIS2 (Amp C)} \\
\hline
F200LP& 4971.86 &  \ldots & -0.092/0.674  & 4875.10 &  \ldots &  -0.100/2.057  \\
F218W & 2228.04 &  0.394/0.532 & -0.137/0.006  & 2223.72 &  0.685/0.863 &  -0.173/0.004  \\
F225W & 2372.05 &  0.228/0.479 & -0.158/0.005  & 2358.39 &  0.552/0.790 &  -0.192/0.003  \\
F275W & 2709.69 &  0.120/0.564 & -0.135/0.005  & 2703.30 &  0.337/0.806 &  -0.173/0.004  \\
F280N & 2832.86 &  0.023/0.627 & -0.138/0.007  & 2829.98 &  0.337/0.806 &  -0.173/0.004  \\
F300X & 2820.47 &  0.023/0.627 & -0.138/0.007  & 2805.84 &     \ldots   &  -0.040/0.068  \\
F336W & 3354.49 & \ldots & -0.029/0.075  & 3354.66 & \ldots &  -0.040/0.068  \\
F343N & 3435.15 & \ldots & -0.029/0.080  & 3435.19 & \ldots &  -0.049/0.076  \\
F350LP& 5873.87 & \ldots & -0.092/0.199  & 5851.15 & \ldots &  -0.144/0.479  \\
F373N & 3730.17 & \ldots & -0.120/0.269  & 3730.17 & \ldots &  -0.067/0.287  \\
F390M & 3897.24 & \ldots & -0.120/0.269  & 3897.00 & \ldots &  -0.067/0.287  \\
F390W & 3923.69 & \ldots & -0.162/0.295  & 3920.72 & \ldots &  -0.025/0.017  \\
F395N & 3955.19 & \ldots & -0.053/0.950  & 3955.15 & \ldots &  -0.025/0.017  \\
F410M & 4108.99 & \ldots & -0.167/0.318  & 4108.88 & \ldots &  -0.034/0.357  \\
F438W & 4326.23 & \ldots & -0.152/0.063  & 4325.14 & \ldots &  -0.111/0.074  \\
F467M & 4682.58 & \ldots & -0.231/0.277  & 4682.60 & \ldots &  -0.226/0.289  \\
F469N & 4688.10 & \ldots & -0.048/0.492  & 4688.10 & \ldots &  -0.180/0.317  \\
F475W & 4773.10 & \ldots & -0.140/0.134  & 4772.17 & \ldots &  -0.061/0.099  \\
F475X & 4940.72 & \ldots & -0.133/0.474  & 4937.41 & \ldots &  -0.192/0.893  \\
F487N & 4871.38 & \ldots & -0.116/0.446  & 4871.38 & \ldots &  -0.061/0.099  \\
F502N & 5009.64 & \ldots & -0.123/0.422  & 5009.64 & \ldots &  -0.133/0.322  \\
F547M & 5447.50 & \ldots & -0.121/0.128  & 5447.24 & \ldots &  -0.135/0.133  \\
F555W & 5308.43 & \ldots & -0.181/0.154  & 5307.91 & \ldots &  -0.054/0.207  \\
F600LP& 7468.12 & \ldots & -0.148/0.185  & 7453.66 & \ldots &  -0.075/0.339  \\
F606W & 5889.17 & \ldots & -0.213/0.068  & 5887.71 & \ldots &  -0.171/0.075  \\
F621M & 6218.85 & \ldots & -0.116/0.155  & 6219.16 & \ldots &  -0.139/0.220  \\
F625W & 6242.56 & \ldots & -0.155/0.169  & 6241.96 & \ldots &  -0.187/0.191  \\
F631N & 6304.29 & \ldots & -0.000/1.903  & 6304.28 & \ldots &   0.000/1.533  \\
F645N & 6453.59 & \ldots & -0.000/1.903  & 6453.58 & \ldots &  -0.001/1.534  \\
F656N & 6561.37 & \ldots & -0.031/0.373  & 6561.36 & \ldots &  -0.023/0.301  \\
F657N & 6566.63 & \ldots & -0.031/0.373  & 6566.60 & \ldots &  -0.023/0.301  \\
F658N & 6584.02 & \ldots & -0.031/0.373  & 6583.92 & \ldots &  -0.012/1.322  \\
F665N & 6655.88 & \ldots & -0.031/0.373  & 6655.84 & \ldots &   0.000/1.525  \\
F673N & 6765.94 & \ldots & -0.031/0.373  & 6765.91 & \ldots &   0.000/1.525  \\
F680N & 6877.60 & \ldots & -0.135/0.476  & 6877.41 & \ldots &  -0.000/4.050  \\
F689M & 6876.75 & \ldots & -0.135/0.476  & 6876.50 & \ldots &  -0.252/0.581  \\
F763M & 7614.37 & \ldots & -0.126/0.470  & 7612.74 & \ldots &  -0.271/0.545  \\
F775W & 7651.36 & \ldots & -0.062/0.162  & 7648.30 & \ldots &  -0.092/0.158  \\
F814W & 8039.06 & \ldots & -0.110/0.066  & 8029.32 & \ldots &  -0.108/0.072  \\
F845M & 8439.06 & \ldots & -0.126/0.197  & 8437.27 & \ldots &  -0.121/0.173  \\
F850LP& 9176.13 & \ldots & -0.035/0.147  & 9169.94 & \ldots &   0.012/0.181  \\
F953N & 9530.58 & \ldots & -0.016/0.090  & 9530.50 & \ldots &  -0.016/0.090  \\
\hline
\hline
\end{tabular}
\end{scriptsize}
\end{center}
\end{table*}

It is worth noting that error bars of the individual data points in Figs.~\ref{fig:f218ascan}, \ref{fig:f218cscan}, 
\ref{fig:f814ascan}, \ref{fig:f814cscan} represent uncertainties on the standard star 
aperture photometry. These were calculated following the recipe of \citet{stetson1987}, and include the Poisson and readout noise, and the sky brightness error. The figures show that these uncertainties are underestimated; for instance, cosmic rays were not removed from the standard star images; although outlier measurements ($>$ 5\%) were excluded from the final photometric tables (see Sec.~\ref{sec:obs}), a few measurements could still
be contaminated by cosmics. Detector artifacts, such as defective or unstable pixels, or uncertainties in the actual length of the exposure time for short exposures (e.g. the shutter shading effect) can also affect the measurements, while not being included in the final error estimate. 
Before 2010, for instance, standard star images were collected with very short exposure times ($<$ 1s); for these short times, the shutter vibration can affect the actual duration of the exposures, leading to fainter measured magnitudes on the image \citep{hilbert2009,sabbi2009,sahu2014,sahu2015}. 
This issue is reflected in the larger scatter of the standard star measurements at earlier epochs, i.e. for $MJD \lesssim$ 55300 (Figs.~\ref{fig:f218ascan}, \ref{fig:f218cscan}, 
\ref{fig:f814ascan}, \ref{fig:f814cscan}).

%In addition, a slight offset between GRW70 and the other standard stars measurements is present, and the WFC3 team is currently investigating this.}

\begin{figure*}
\begin{center}
\includegraphics[width=0.55\textwidth, height=0.75\textheight, angle=90]{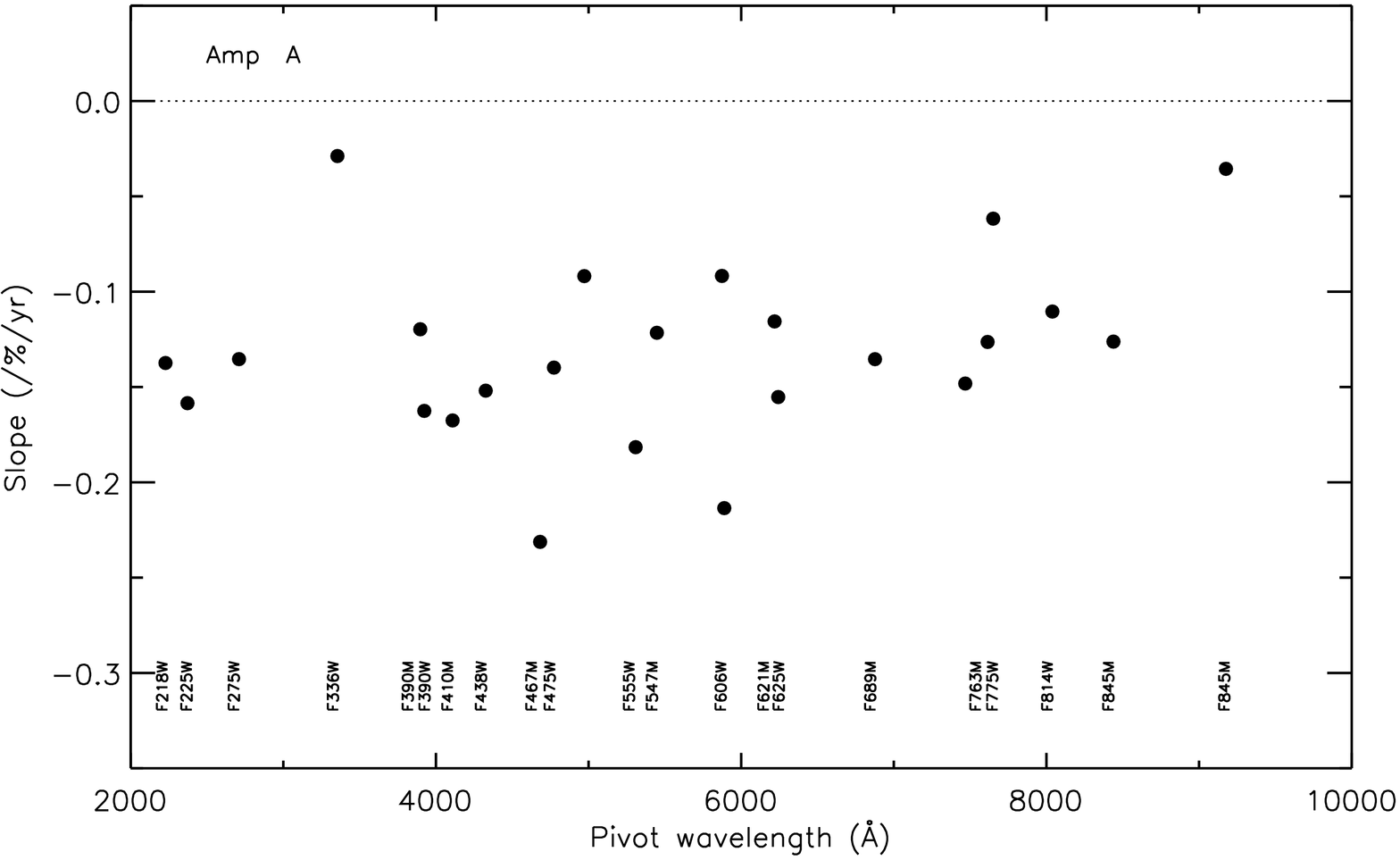}
\caption{Measured sensitivity change slopes for the wide- and medium-band filters for the UVIS1
detector (Amp A) as a function of pivot wavelength. Note that for the three UV filters the slope 
is the one calculated after $MJD =$ 55738 (see text and Table~\ref{table:4} for more details). \label{fig:slopeA}}
\end{center}
\end{figure*}

\begin{figure*}
\begin{center}
\includegraphics[width=0.55\textwidth, height=0.75\textheight, angle=90]{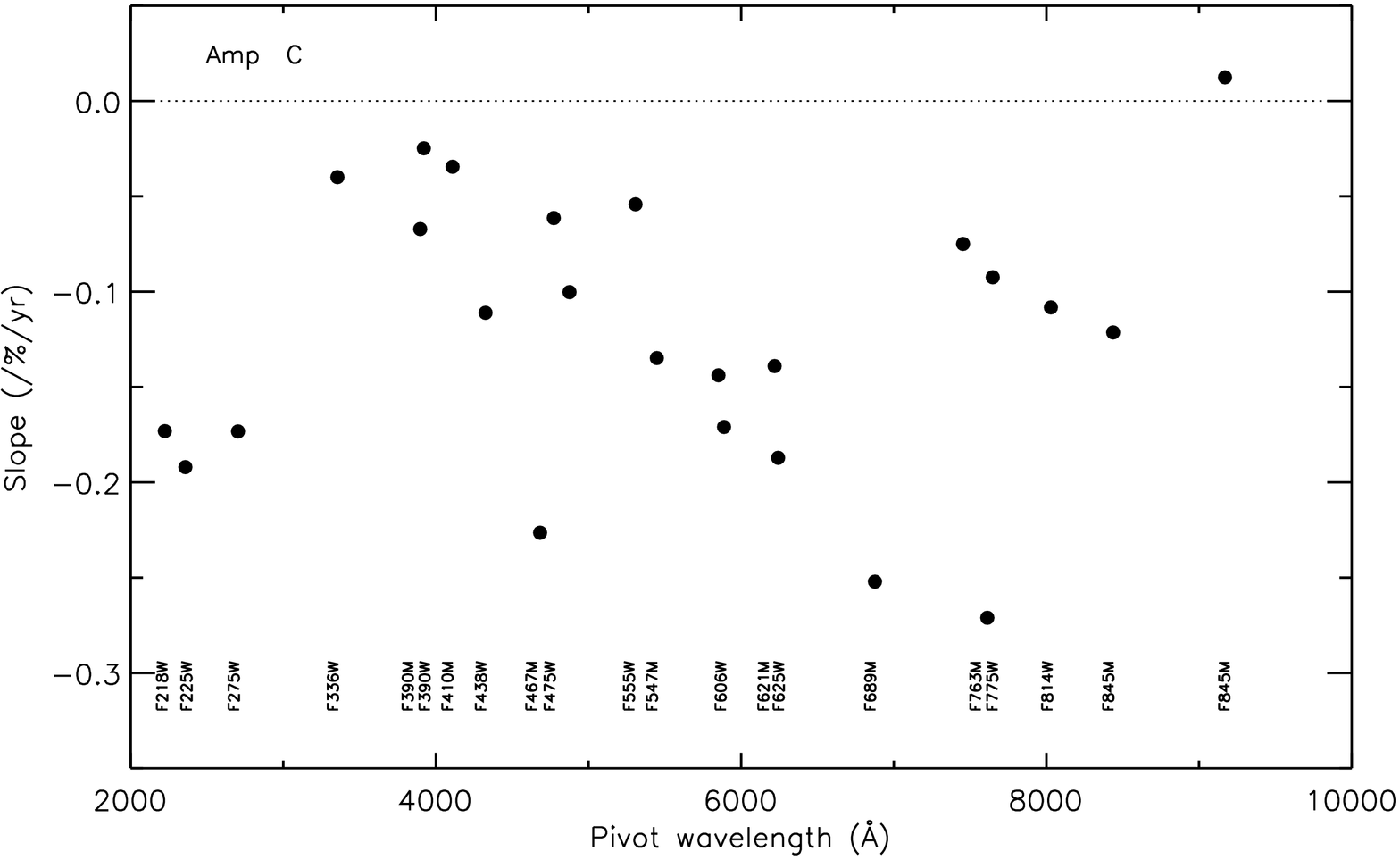}
\caption{Same as Fig.~\ref{fig:slopeA} but for the UVIS2 detector (Amp C). \label{fig:slopeC}}
\end{center}
\end{figure*}

Plots of the time-dependent sensitivity evolution are displayed for the UV filter $F218W$
%$F606W$, 
and the red filter $F814W$ in Figs.~\ref{fig:f218ascan}, 
\ref{fig:f218cscan}, 
%\ref{fig:f606ascan}, \ref{fig:f606cscan},  
\ref{fig:f814ascan}, and \ref{fig:f814cscan}, respectively.
Figs.~\ref{fig:f218ascan} and \ref{fig:f218cscan} show that in the case of the $F218W$ filter, the
sensitivity of the UVIS1 and UVIS2 detectors increased with time for the first 2 years of WFC3 life, from $MJD =$ 55008 to $\approx$ 55738 (2009 to $\approx$ 2011), and later decreased. The same is true for the other UV filters ($F225W$, $F275W$ and $F280N$). %see figures in the Appendix). 
This effect was already observed in \citet{shanahan2017a} and \citet{khandrika2018}, and
it is also present in other instruments with UV capabilities on board \emph{HST}, such as STIS \citep{carlberg2017}.

\begin{figure*}
\centering
\includegraphics[width=1.05\textwidth, height=0.40\textheight]{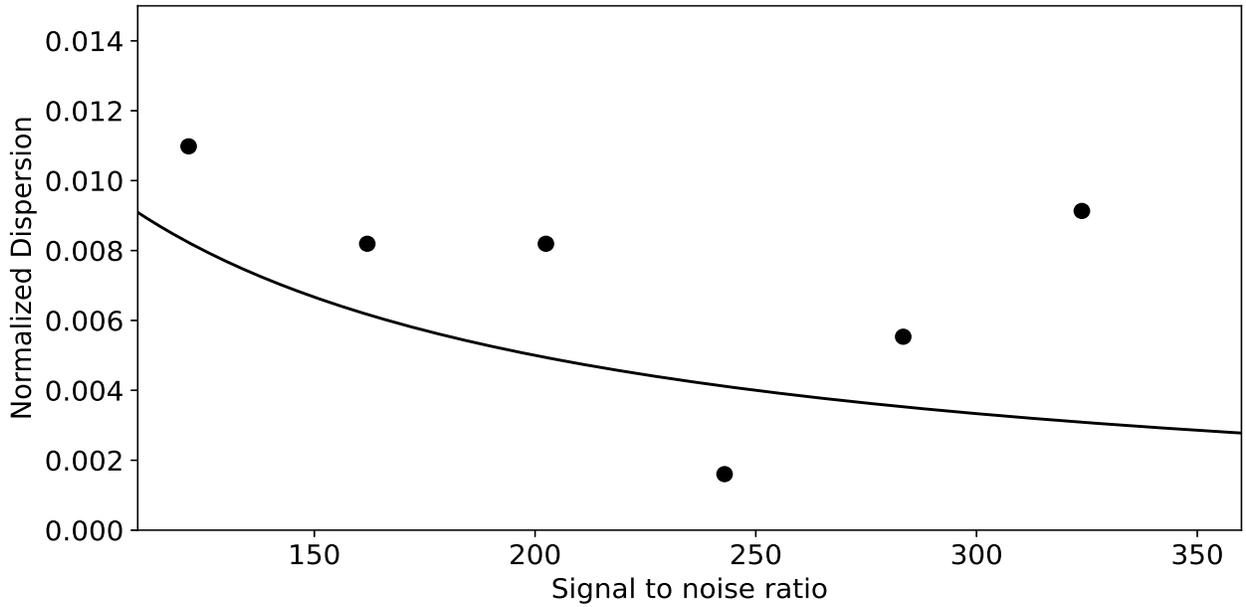} %gd153_snr_disp.ps}
\caption{The normalized standard deviation (in percent) of WFC3-IR photometric measurements of GD153 plotted versus the signal to noise ratio.
The solid line represents the expected normalized standard deviation ($1/S/N$). \label{fig:gd_153_disp}}
\end{figure*}

In order to calculate the sensitivity changes over time we performed a first least-square linear fit by including all the measurements of the five standard stars. In the case of the UV filters, we 
performed two different fits, one for $MJD \le$ 55738 and a second for all data acquired 
through $MJD \approx$ 58800, to take into account the change from an increase 
to a decrease of the sensitivity. 
We then performed a 2.5 $\sigma$-clipping of the outlier measurements and a second least-square fit 
that resulted in the final slope values. These are indicated as sensitivity change rates (\%/year)
in Figs.~\ref{fig:f218ascan}, \ref{fig:f218cscan}, 
%\ref{fig:f606ascan}, \ref{fig:f606cscan},  
\ref{fig:f814ascan}, \ref{fig:f814cscan}.
The final slope values with their uncertainties for UVIS1 and UVIS2 are listed in Table~\ref{table:4}
and are plotted as a function of the filter pivot wavelength in Figs.\ref{fig:slopeA} and \ref{fig:slopeC}.

It is worth noticing that errors on the slopes for some filters are quite large, in particular for the UV filters in the first two years of observations (i.e. when the UV sensitivity change rate was positive).
Moreover, uncertainties are larger for a few narrow-, medium-band or long-pass filters, where a limited number ($<$ 20) of standard star measurements were available.
In some cases, there were not enough measurements available to calculate a reliable slope. Therefore, we assumed that the sensitivity change rates 
for these filters were the same as those derived for filters similar in wavelength, but with a much larger data sample.
For instance, the slope for filter $F373N$
(pivot wavelength 3730\AA) was assumed to be the same as for $F390M$ (pivot wavelength 3897\AA, see Table~\ref{table:4}).

\begin{table}
%\centering
%\def\arraystretch{1.3}
\caption{\footnotesize The normalized standard deviation of the WFC3-IR photometric measurements of the standard stars, in percent.  
The entries labeled with "*" have less than three data points, and should not be considered as meaningful. \label{table:5}}
\begin{tabular}{ l c c c c c}
\hline
Filter & GD153 &	GD71 &	P330E &	GRW70 & G191B2B \\
\hline
\hline
F098M &	0.99 &	1.69 &	1.07 &	1.17 &	1.76 \\
F105W &	1.84 &	1.92 &	1.16 &	1.19 &	1.41 \\
F110W &	0.96 &	1.2  &	1.01 &	0.63 &	1.31 \\
F125W &	1.14 &	1.16 &	1.03 &	1.58 &	0.72 \\
F126N &	1.95 &	0.91 &	1.05 &	0.65 &	0.22*\\
F127M &	1.19 &	1.03 &	0.98 &	0.83 &	1.36 \\
F128N &	1.77 &	0.94 &	1.26 &	0.59 &	0.1* \\
F130N &	2.3  &	0.84 &	1.16 &	0.39 &	0.13 \\
F132N &	2.49 &	0.96 &	1.19 &	0.29 &	0.01*\\
F139M &	1.69 &	0.93 &	1.0  &	0.99 &	0.99 \\
F140W &	0.94 &	1.22 &	0.88 &	0.72 &	0.89 \\
F153M &	1.59 &	0.75 &	0.71 &	0.6 & 	0.72 \\
F160W &	0.85 &	0.96 &	0.92 &	0.72 &	0.71 \\
F164N &	2.13 &	1.13 &	1.14 &	0.47 &	0.18*\\
F167N &	2.04 &	1.04 &	1.07 &	0.63 &	0.05*\\
\hline
\hline
\end{tabular}
\end{table}

Once all the slopes were finalized, we used them to normalize the 10-pixel radius aperture photometry 
for each standard star and filter to the reference epoch $MJD =$ 55008 (June 26, 2009), 
corresponding to the time at which the first WFC3 observations were collected.
A weighted mean of all measurements was then calculated after a 2.5 $\sigma$-clipping of the outliers. 
This mean was used to define the value of the photometry at 10 pixels in units of $e^{-}/s$, i.e. in count rates, for each standard star and filter at the reference epoch. 

The slopes were then used to derive inverse sensitivities at six different $MJD$ values spaced by 2 years, namely 55008, 56468, 57198, 57928, 58658, 559388, for each filter. 
{\it calwf3} pipeline then calculates inverse sensitivities at any observing epoch by interpolating over the six provided values (more details are in the following sections).
Therefore, the fact that the two least-square fit lines in the case of the $F218W$ filter (Figs.\ref{fig:f218ascan} and \ref{fig:f218cscan}), for example, do not perfectly coincide at the established 
inversion epoch, $MJD =$ 55738, does not affect {\it calwf3} inverse sensitivity calculation.

\begin{figure*}
\begin{center}
\includegraphics[width=0.65\textwidth, height=0.75\textheight, angle=90]{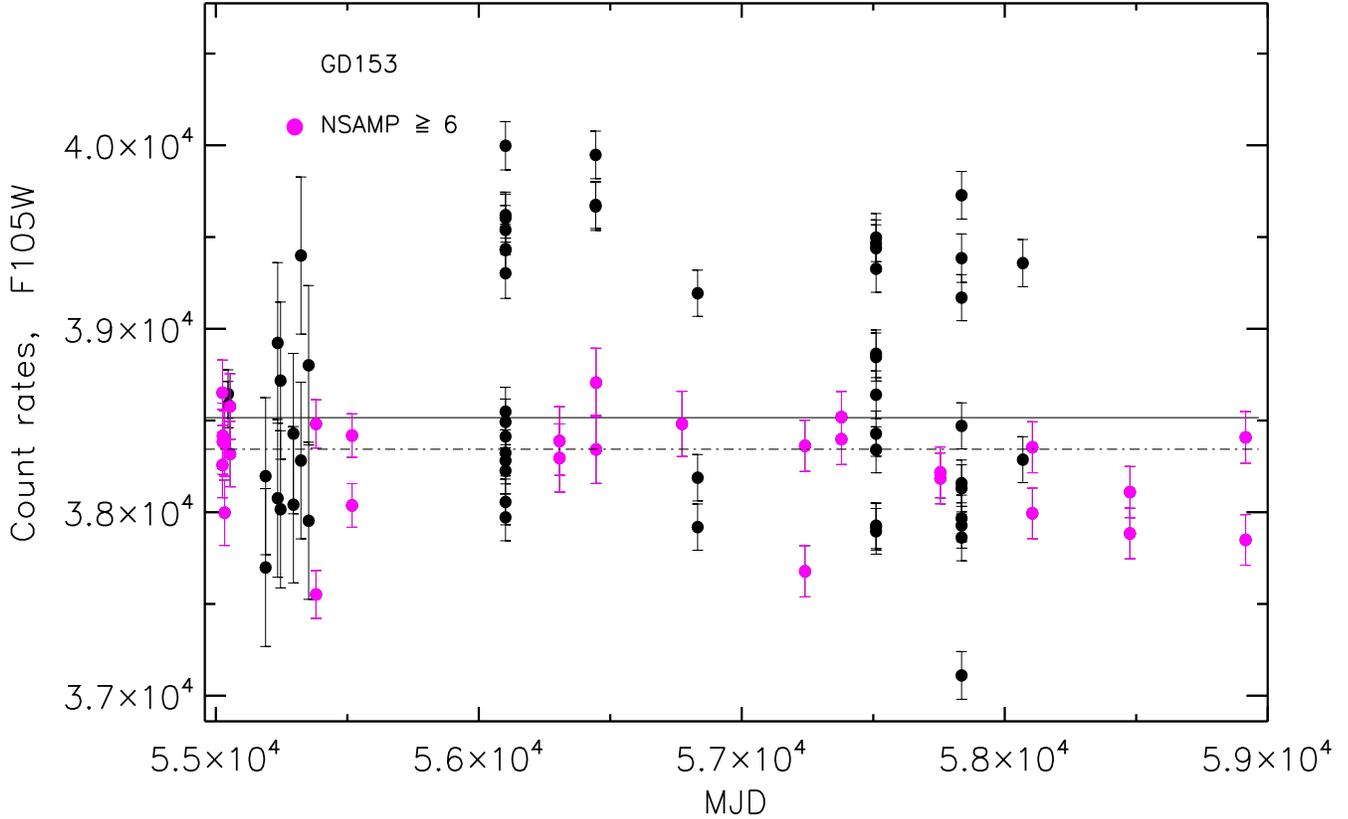}  %{samp_flux.ps}
%\vspace{0.5cm}
\caption{GD153 count rates for the WFC3-IR $F105W$ filter are shown with black filled circles, while 
measurements from exposures collected with a number of samples (NSAMP) greater than six are marked 
with magenta filled circles. Error bars are also shown. 
The solid line indicates the mean of all measurements, while the dotted-dashed line indicates the 
mean of the NSAMP$\ge$ 6 measurements. \label{fig:samp_flux}}
\end{center}
\end{figure*}

\subsection{WFC3-IR}
In order to investigate the repeatability of WFC3-IR photometry and to look for possible 
systematics affecting it, we calculated the 3-$\sigma$ clipped standard deviation of the flux measurements 
of the five standard stars in each filter normalized by the median flux measurement.
This normalized standard deviation is presented in percent and listed in Table~\ref{table:5}: while 
the percent deviation is $\gtrsim$ 1\% for most filters, the $S/N$ of many observations
is often substantially larger than 100, even including noise imparted from calibration (as reported in the error array of the {\it \_flt} images).  
Fig.~\ref{fig:gd_153_disp} shows how the dispersion of photometry evolves with the $S/N$ of the exposures:
 notably, the actual standard deviations are consistently higher than predicted for all $S/N$ levels.  

A factor differentiating this analysis from \citet{kalirai2011} is the usage of updated flat fields \citep{mack2021}.
However, by using the new flat fields the scatter of the standard star measurements was not 
substantially reduced. This may be partially due to the clustering of the standard star 
observations near the center of the detector, as the WFC3-IR sub-arrays are all centered. 
The flat-field error in the center of the detector was already below the half percent level \citep{dahlen2013}, 
and thus the improvement in the new flat fields pixel to pixel variation was minimal.  

In some cases, the inclusion of images collected with different observing strategies imparted a higher 
dispersion on the photometry. Several observations of GD153 in the $F105W$ filter, for example, were collected for the WFC3-IR grism calibration and only included a small number of reads (NSAMP) per exposure; this 
resulted in noisier data, possibly due to the behavior of the first read of the WFC3-IR integrations (see Fig.~\ref{fig:samp_flux}). 
Removing low sample exposures from the analysis increased the precision of the photometry for a small subset of the filters, though not to the level predicted by the $S/N$. For instance, the GD153 $F105W$ filter images with less than 
six reads (NSAMP $\le$ 6) have a clipped standard deviation of $\approx$ 2\%, while those with more reads had a much smaller dispersion, i.e. $\approx$ 0.7\%.  In addition, the difference between the means of the two populations is $\approx$ 1.3\% (Fig.~\ref{fig:samp_flux}). However, removing images collected with less than six reads did not always yield a more precise result; for some stars and filter combinations the dispersion of the measurements increased.
The WFC3 team is currently working at better understanding this issue.

The WFC3-IR detector is also affected by persistence, i.e. the residual signal of a large incident light level that can last on the images from minutes to days \citep{long2011,long2013,gennaro2018}.
As noted in \citet{bajaj2019}, the effects of persistence significantly
lower the precision of WFC3-IR observations. This is likely due to the dependence of persistence signals on time from the stimulus (the exposures that caused the persistence), and fluence of the previous exposures causing the persistence. Additionally, longer term persistence (from observations up to days before) 
can sometimes still affect the standard star observations \citep{ryan2015}, though this effect is generally 
smaller than the self-persistence (persistence from observations in the same visit). 
The excess flux from persistence is thus not well constrained, and it is virtually indistinguishable from the real flux. 
The variability of persistence is one of the causes of the lower than expected precision of WFC3-IR observations. 
Because the effects of persistence on precision photometry were not initially well understood, 
many of the earlier observations of the standard stars dithered infrequently, and sometimes only by a few pixels. 
While this may maximize observational efficiency, it incurred a loss of precision. Frequent and large dithers can mitigate much 
of the effect of the persistence and lead to substantially better precision, and are therefore used in photometric 
calibration programs since 2017 \citep{bajaj2019}.

However, the WFC3-IR detector also exhibits longer term behaviors, where even the first observations in a visit 
(which should be unaffected by persistence) show photometric offsets compared to previous visits \citep{bajaj2019}. 
In some cases, these offsets are present across a visit. The visit-to-visit variation is distinct from the Poisson error, 
as Poisson errors manifest randomly. This effect is also detected in WFC3-IR spatial scan data, where Poisson noise terms are
effectively close to zero \citep{som2021}. 

A portion of the non-repeatability of the WFC3-IR detector may be then attributed to varying observation 
configurations (e.g. different sample sequences, number of samples, and exposure time). 
A substantial detection or correction of systematic behavior as a function of these observation characteristics 
would likely require additional, extensive processing in the calibration pipelines. 
This instability between visits is not currently well-understood 
and the WFC3 team will further investige this issue.

As shown in Fig.~\ref{fig:samp_flux}, standard star data collected over a baseline longer than 10 years
do not support a change of sensitivity of the WFC3-IR detector with time. 
The overall stability of the detector appears to remain similar to the results found in \citet{kalirai2011}
and \citet{bajaj2019}, with a typical dispersion of $\approx$ 1\% and no significant 
consistent trends.
However, the lack of precision and the non-repeatability of the photometric measurements might 
ultimately limit the ability to detect small sensitivity losses (BA20).
Specifically, the visit-to-visit variation of the photometry substantially reduces the precision of any 
time-dependent measurement of the sensitivity. Thus, the standard star measurements are unable 
to support the findings seen in other studies, such as \citet{kozhurina2020} and \citet{bohlin2019}.
The first analysis detected sensitivity losses of the order of 2\% over 10 years for the $F160W$ filter 
by using observations of the core of the globular cluster $\omega$ Cen; the second analysis found
sensitivity losses of $\approx$ 0.17 and 0.08 \%/yr for the $G102$ abd $G141$ grism, respectively,
by using observations of the four CALSPEC standard WDs.
The WFC3 team currently has a calibration program to measure WFC3-IR 
sensitivity losses via spatial scanning, since this observation strategy allows for extremely small Poisson noise terms.  
However, preliminary analysis showed uncertainties much larger than the Poisson noise would predict within a visit, 
and from visit to visit \citep{som2021}. This effect is not persistence related but appears consistent with the 
visit-to-visit variability of the standard star measurements.  
%However, these observations were designed to limit 
%persistence effects, and have a much more unified observation strategy (compared to the standard star observations), reducing parameter space in the analysis of systematics.
Another technique currently used tby the WFC3 team to verify for WFC3-IR sensitivity losses is observing globular clusters in regions farther away
from the core where stellar crowding is less of a concern. A consistent observing strategy between epochs is used in these
calibration programs, and should yield more precise measurements of sensitivity losses.

Since no time-dependent correction was applied to 
the photometry, we calculated a weighted mean of all measurements after a 1.0 $\sigma$-clipping of the outliers for each standard star and filter.
The mean was used to define the value of the photometry for each standard star and filter at 3 pixels in units of $e^{-}/s$, i.e. in count rates, at the reference epoch, $MJD =$ 55008 (June 26, 2009).

\begin{figure}
\includegraphics[width=0.5\textwidth, height=0.65\textheight]{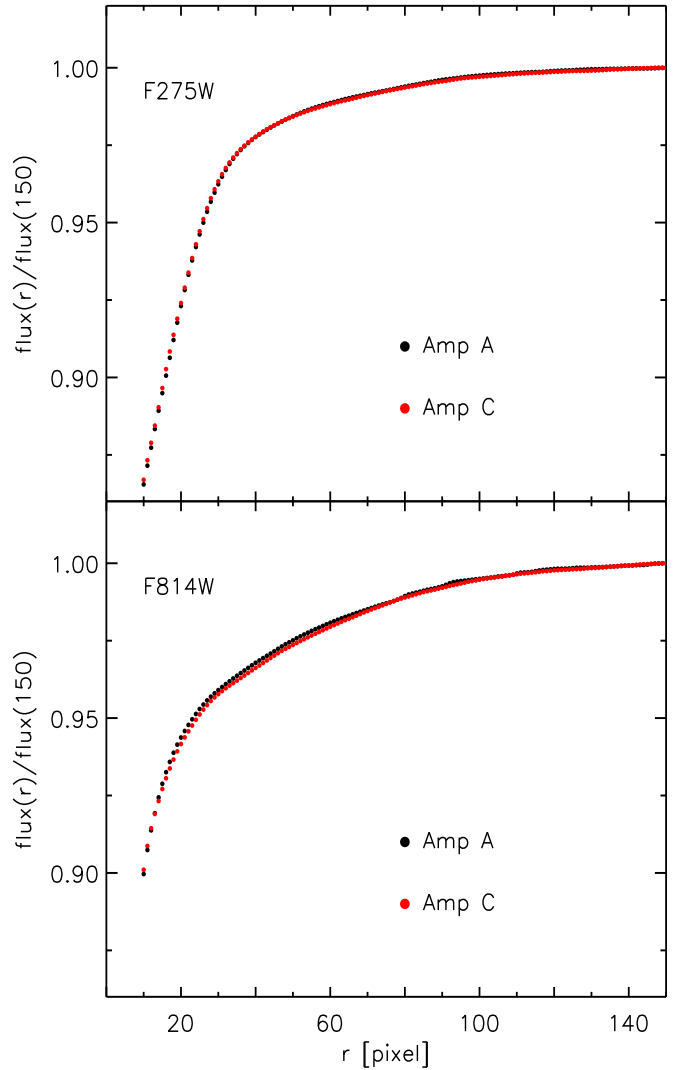}
\caption{Encircled energy fraction, $flux(r)/flux(150)$, as a function of the aperture radius in pixels, $r$, for Amp A (UVIS1, black filled circles) and Amp C (UVIS2, red) for the $F275W$ (top panel) and the $F814W$ filter (bottom). Photometry was measured on the combined {\it \_drc} images for the standard star GRW70. See text for more details. \label{fig:ee}} 
\end{figure}

\section{Encircled energy corrections}\label{sec:energy}

To calculate new inverse sensitivities at {\it infinity}, the radius enclosing all of the light emitted by a point source, we first needed to apply encircled energy (EE) corrections (or fractions) to the standard star photometry computed using an aperture radius of 10 pixels for WFC3-UVIS and 3 pixels for WFC3-IR. Uncertainties in the EE corrections are carried over to the uncertainties in the inverse sensitivities. Therefore, in the case of WFC3-UVIS, we applied the new sensitivity change slopes to improve the EE corrections for a subset of filters. For WFC3-IR, new EE corrections were not calculated since the sensitivity changes with time are not well characterized yet for this detector. Instead, the EE solutions from \citet{hartig09} were used to correct the standard star photometry from a 3 pixel radius aperture to {\it infinity}. 

\begin{figure}
\begin{center}
\includegraphics[width=0.5\textwidth, height=0.35\textheight]{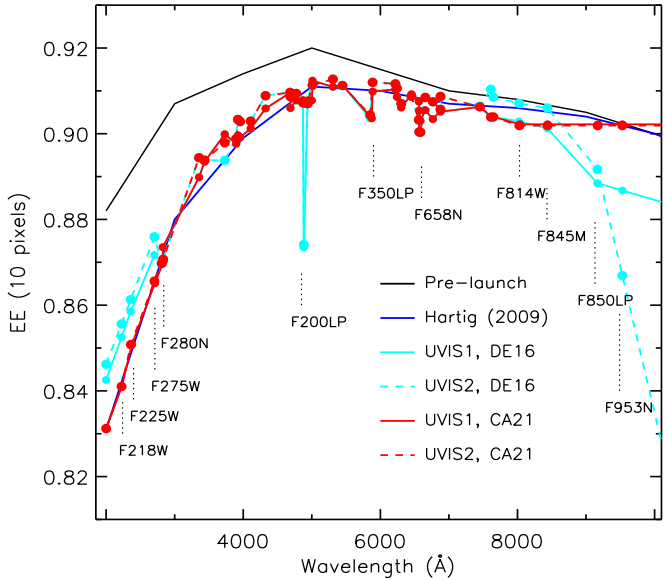}
\caption{Old (DE16, cyan) and new (CA21, red) EE corrections for UVIS1 (solid line) and UVIS2 (dashed line), as a function of wavelength. The EE
model values from \citet{hartig2009} are shown with a blue solid line and the pre-launch values with a black solid line. \label{fig:ee2020} }
\end{center}
\end{figure}

For WFC3-UVIS, the derived slopes were used to correct the science arrays of the standard star {\it \_flc} images prior to combining them with \texttt{AstroDrizzle}, and we used this new procedure to recompute EE curves for the $F275W$ and $F814W$ filters. $F275W$ was selected because the EE values in the DE16 (and DE17) solutions differ by $\approx$ 1\% from the original in-flight EE calculation by \citet{hartig2009} for both UVIS1 and UVIS2. $F814W$ was selected because the EE correction for UVIS2 in the DE16 solution differs from UVIS1 by $\approx$ 0.5\% or more for the reddest filters \citep[see Fig.~\ref{fig:ee2020}]{hartig2009}. 

\begin{table*}
\begin{center}
\caption{\footnotesize \texttt{AstroDrizzle} parameters with non-default settings used for this analysis. \label{table:6}}
\begin{tabular}{lll}
\hline
\hline
Name & \footnotesize Description & Value \\
\hline
\texttt{skymethod} & \footnotesize Equalize sky background between the input frames & \texttt{match}  \\
\texttt{skystat}  & \footnotesize Use the sigma-clipped mean background & \texttt{mean}  \\
\texttt{driz\_sep\_bits} & \footnotesize For single images, set DQ values considered to be {\it good} data & \texttt{64, 16}  \\
\texttt{combine\_type} & \footnotesize Combine images using the median & \texttt{median}  \\
\texttt{combine\_nhigh} & \footnotesize Set the number of high value pixels to reject for the median  & \texttt{1} \\
\texttt{driz\_cr\_snr} & \footnotesize $S/N$ to be used in detecting CRs, performed in two iterations & \texttt{3.5 3.0} \\
\texttt{driz\_cr\_scale} & \footnotesize Scaling factors applied to the derivative for detecting CRs  & \texttt{2.0 1.5} \\
\texttt{final\_bits} & \footnotesize For the final stack, set DQ values considered to be {\it good} data & \texttt{64, 16} \\
\hline
\hline
\end{tabular}
\end{center}
\end{table*}

EE corrections for these two filters were calculated using all in-flight observations collected for the standard star GRW70 from the reference epoch ($MJD$ = 55008) until about $MJD$ = 58800. In particular, each {\it \_flc} image was multiplied by an inverse sensitivity ratio, i.e. the time-dependent inverse sensitivity value for each image divided by the value at the reference epoch. This was performed using the task \texttt{phot\_eq}\footnote{https://drizzlepac.readthedocs.io/en/latest/photeq.html}, which scales the {\it \_flc} science array values by their respective inverse sensitivity ratio. In this way, all {\it \_flc} images were corrected to have approximately equal count rates, in preparation for combining the individual frames.

The flux scaled {\it \_flc} images were then processed using \texttt{Astrodrizzle} to create the combined {\it\_ drc} image that is used for the EE fraction calculation. The {\it \_drc} images were produced by combining many individual {\it \_flc} images, significantly improving the $S/N$ of the standard star and reducing the overall noise, thus enhancing the visibility of the PSF wings. This is very important to achieve more precise photometric measurements at larger aperture radii. For $F275W$, the Amp A (UVIS1) drizzled image was derived from 229 {\it \_flc} images with a total exposure time of 1,012 seconds, while the Amp C (UVIS2) drizzled image was derived from 185 {\it \_flc} images, totalling 939 seconds. For $F814W$, the Amp A drizzled image was derived from 117 {\it \_flc} images, with a total exposure time of 497 seconds, while the Amp C drizzled image was derived from 134 {\it \_flc} images, totalling 736 seconds.

Whereas the \texttt{Astrodrizzle} algorithm uses pointing information from the  {\it \_flc} image header to align images on the sky, we devised a new approach to align the images in detector coordinates. This ensured that the drizzled PSF did not rotate as the nominal \emph{HST} orientation varied over the years, which would change the position of the diffraction spikes and structures in the PSF wings. We achieved this by modifying the following astrometry header keywords in each {\it \_flc} image before drizzling:

\begin{itemize}
\item \texttt{CRPIX1} and \texttt{CRPIX2} were modified to match the $X,Y$ position of the centroid of the standard star in each image;

\item \texttt{CRVAL1} and \texttt{CRVAL2} were set to match that of the reference image in order to remove any proper motion applied to the RA and DEC of the standard star over the 10 years;

\item The linear terms of the CD matrix (\texttt{CD1\_1}, \texttt{CD1\_2}, \texttt{CD2\_1}, \texttt{CD2\_2}) were set to the value in the reference image in order remove any orientation and plate scale changes with date.
\end{itemize}

Once the astrometry header keywords were updated, we combined the {\it \_flc} images for each detector using the \texttt{AstroDrizzle} parameter values listed in Table~\ref{table:6}. By aligning each star in detector coordinates, we were able to accurately flag and reject artifacts such as cosmic rays and unstable hot pixels, while not affecting any PSF structure. Additionally, by not rotating the images on the sky, the {\it \_flc} frames have minimal pixel resampling. 

Photometry was performed on the {\it \_drc} images for both filters using aperture radii in the range 1 -- 150 ({\it infinity}) pixels. The sky value used for background subtraction was computed as the 3-$\sigma$ clipped mean value in an annulus with radii ranging from 160 -- 200 pixels. The EE correction for each filter was then estimated as the ratio of the flux (in units of $e^{-}$/s) at different aperture radii and the flux at {\it infinity}, defined at a radius of 150 pixels ($\approx$ 6\arcsec) for WFC3-UVIS.

After accounting for changes in sensitivity, we find improved agreement in the EE correction values for UVIS1 and UVIS2. For $F275W$ the fraction of flux included in a 10-pixel aperture radius is 86.5$\pm$0.1\% for UVIS1 and 86.6$\pm$0.1\% for UVIS2, as shown in the top panel of Fig.~\ref{fig:ee}. This differs by $\approx$ 1\% from the EE corrections from DE16 which were 87.2\% and 87.6\% for UVIS1 and UVIS2. Following the results for the F275W filter, we corrected the EE fractions for the other UV filters, namely $F218W$, $F225W$, and $F280N$, scaling them by the difference between the new and old $F275W$ values (see solid and dashed red lines and the marked filter names in Fig.~\ref{fig:ee2020}). For $F814W$, the new EE fraction is 90.2$\pm$0.1\% for UVIS1 and 90.2$\pm$0.1\% for UVIS2, as shown in the bottom panel of Fig.~\ref{fig:ee}. The value for UVIS1 agrees with the previous DE16 value of 90.3\% to within the measurement uncertainty, while the value for UVIS2 differs by $\approx$ 0.5\% with the DE16 value of 90.7\%. The new EE corrections for both F275W and F814W agree very well with the values derived from the 2009 optical model \citep[see Fig.~\ref{fig:ee2020}]{hartig2009}.

For filters with pivot wavelengths longer than $F814W$, namely $F845M$, $F850LP$ and $F953N$ (also marked in Fig.~\ref{fig:ee2020}), we assumed the same EE correction value as derived for the $F814W$ filter. For $F775W$, the DE16 EE values for UVIS2 were $\approx$ 0.5\% larger than for UVIS1, so we adopted the UVIS1 values for both detectors. The DE16 EE fractions for a few long-pass and narrow-band filters (marked in the figure) are in large disagreement with the 2009 model values; therefore, we used interpolated EE fractions for these filters based on the values for the two filters closest in wavelength (see Fig.~\ref{fig:ee2020}).

New aperture correction files, {\it wfc3uvis1\_aper\_007\_syn.fits} and {\it wfc3uvis2\_aper\_007\_syn.fits}, were created for use in {\it STsynphot} and are shown as solid and dashed red lines in Fig.~\ref{fig:ee2020}. For comparison the DE16 aperture correction files, {\it wfc3uvis1\_aper\_005\_syn.fits} and {\it wfc3uvis2\_aper\_005\_syn.fits}, are shown as solid and dashed cyan lines. The 2009 EE model values, {\it wfc3\_uvis\_aper\_002\_syn.fits}, are shown as a solid blue line, while the pre-launch EE values, {\it wfc3\_uvis\_aper\_001\_syn.fits}, are shown as a solid black line. It is worth noting how well the new UVIS1 and UVIS2 aperture correction files agree with one another and with the model values from \citep{hartig2009}.

\begin{figure*}
\begin{center}
\includegraphics[width=0.55\textwidth, height=0.75\textheight, angle=90]{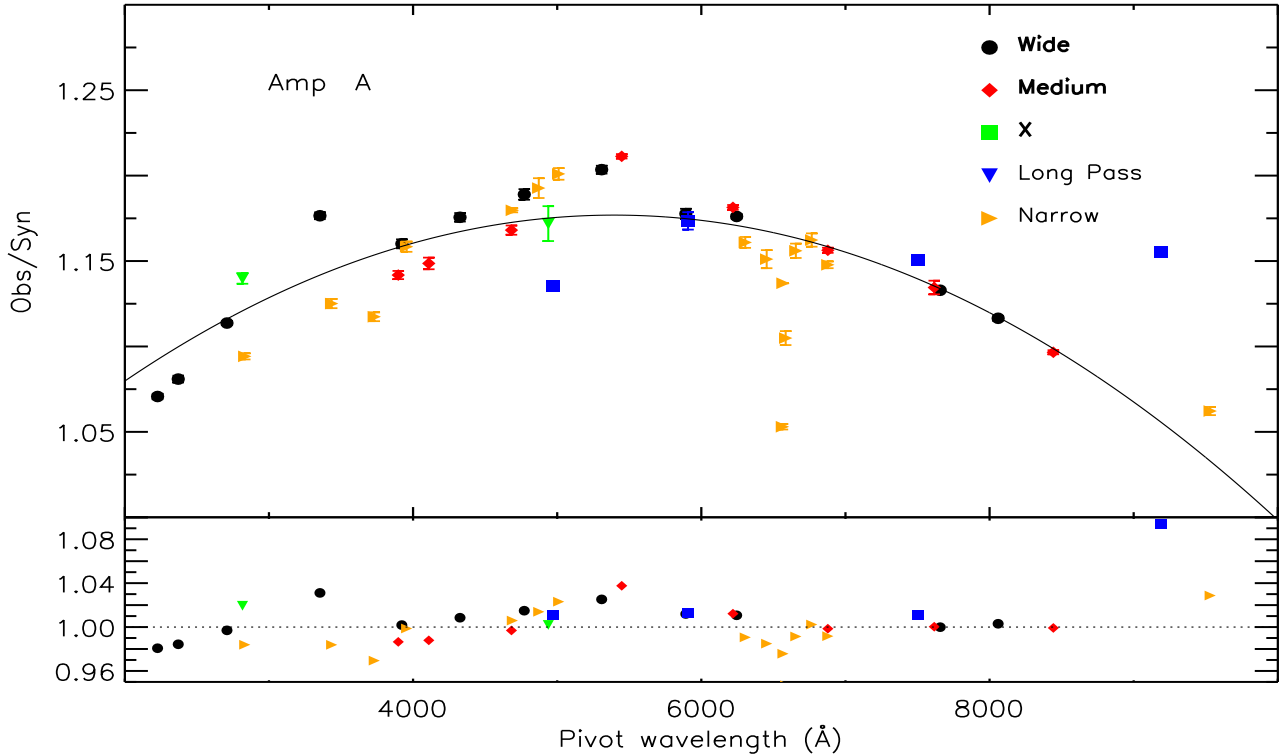}
\vspace{0.5cm}
\caption{Top - Observed over synthetic count rates for the 42 UVIS1 (Amp A) full-frame filters: wide (black), medium (red), X (green), long Pass (blue), and narrow (orange). These were calculated as a weighted mean over the standard stars. Error bars are displayed.
The solid black line is a quadratic polynomial least-square fit to the data of the wide, medium and X filters only.
Bottom - The residual ratios after the polynomial fit are shown. The dotted line shows a residual ratio of 1.0 (see text for more details). \label{fig:corA}}
\end{center}
\end{figure*}

\begin{figure*}
\begin{center}
\includegraphics[width=0.55\textwidth, height=0.75\textheight, angle=90]{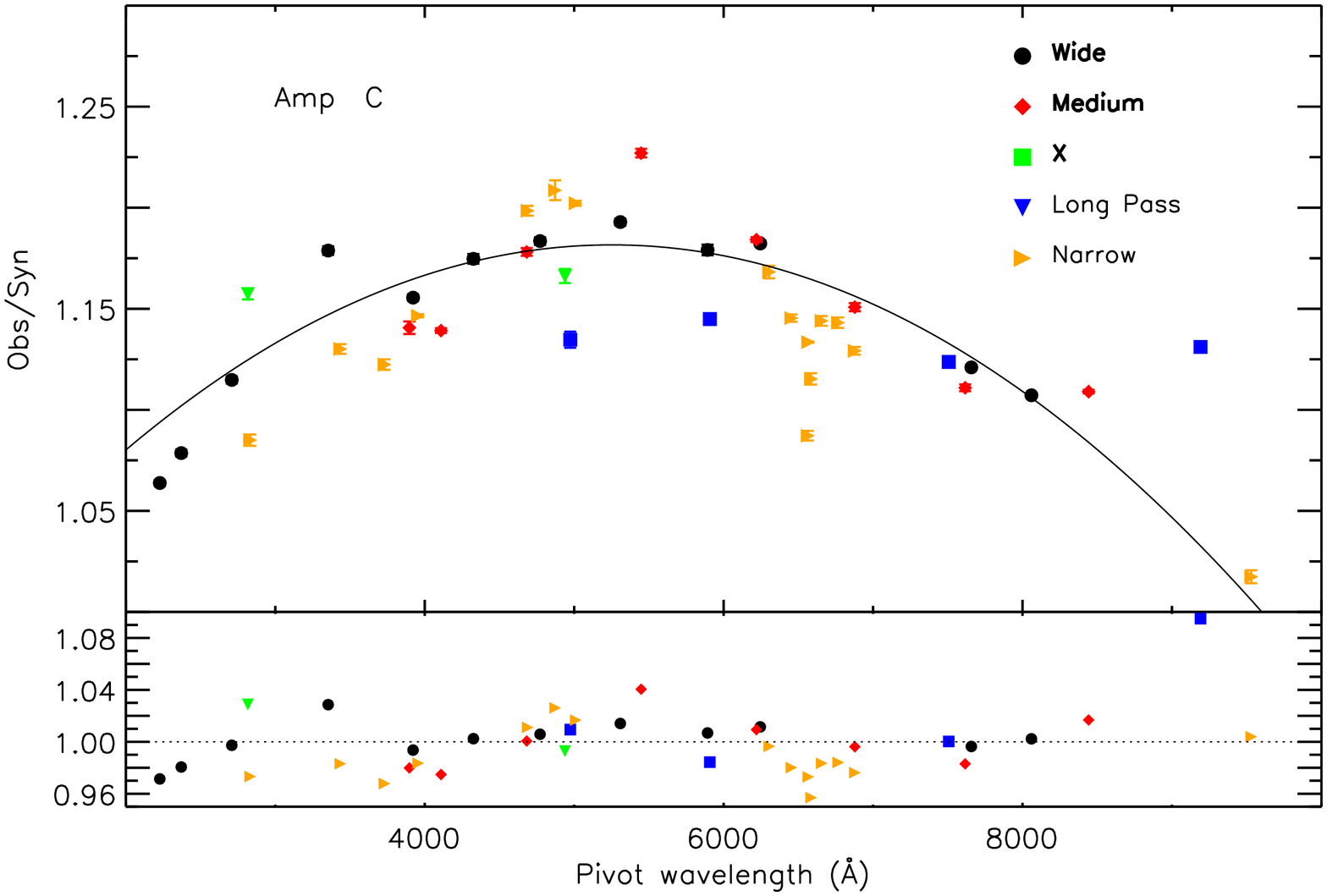}
\vspace{0.5cm}
\caption{Same as Fig.~\ref{fig:corA} but for the 42 UVIS2 (Amp C) full-frame filters. \label{fig:corC}}
\end{center}
\end{figure*}

\section{New in-flight corrections and filter curves}\label{sec:corr}
We used the new EE fractions to correct WFC3-UVIS standard star photometry
from a 10 pixel aperture radius to {\it infinity}. We obtained mean count rates for each standard star as observed with the two detectors through all the 42 full-frame filters at the reference epoch $MJD =$ 55008. 

\begin{figure*}
\begin{center}
\includegraphics[width=0.55\textwidth, height=0.75\textheight, angle=90]{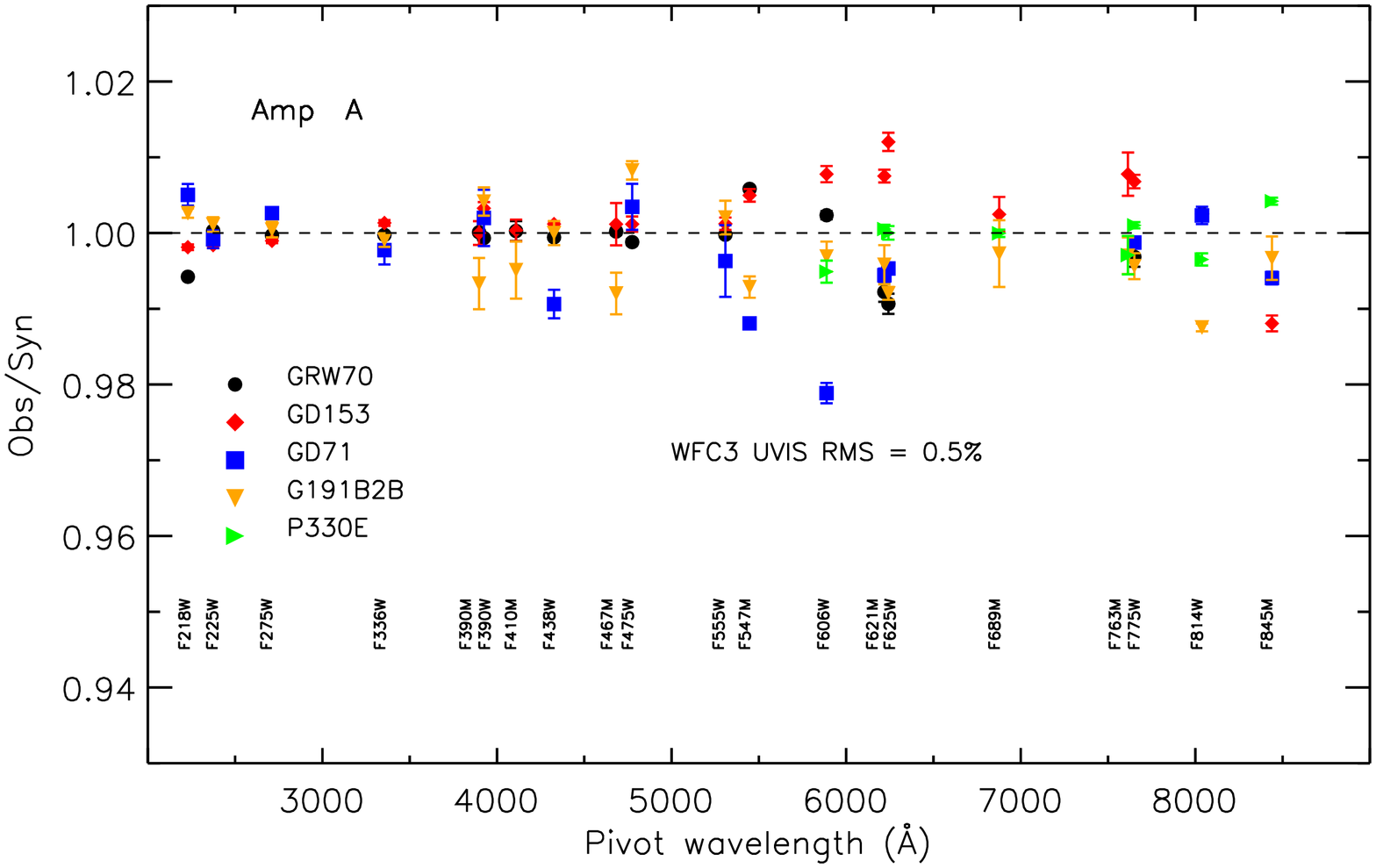}
\caption{Observed over synthetic count rates for UVIS1 (Amp A) wide- and medium-band filters for the five standard stars used in the calibration as a function of the pivot wavelength. 
Error bars are displayed. Note that ratios for P330E were calculated for $\lambda >$ 6000\AA~ only. \label{fig:ratioA}}
\end{center}
\end{figure*}

\begin{figure*}
\begin{center}
\includegraphics[width=0.55\textwidth, height=0.75\textheight, angle=90]{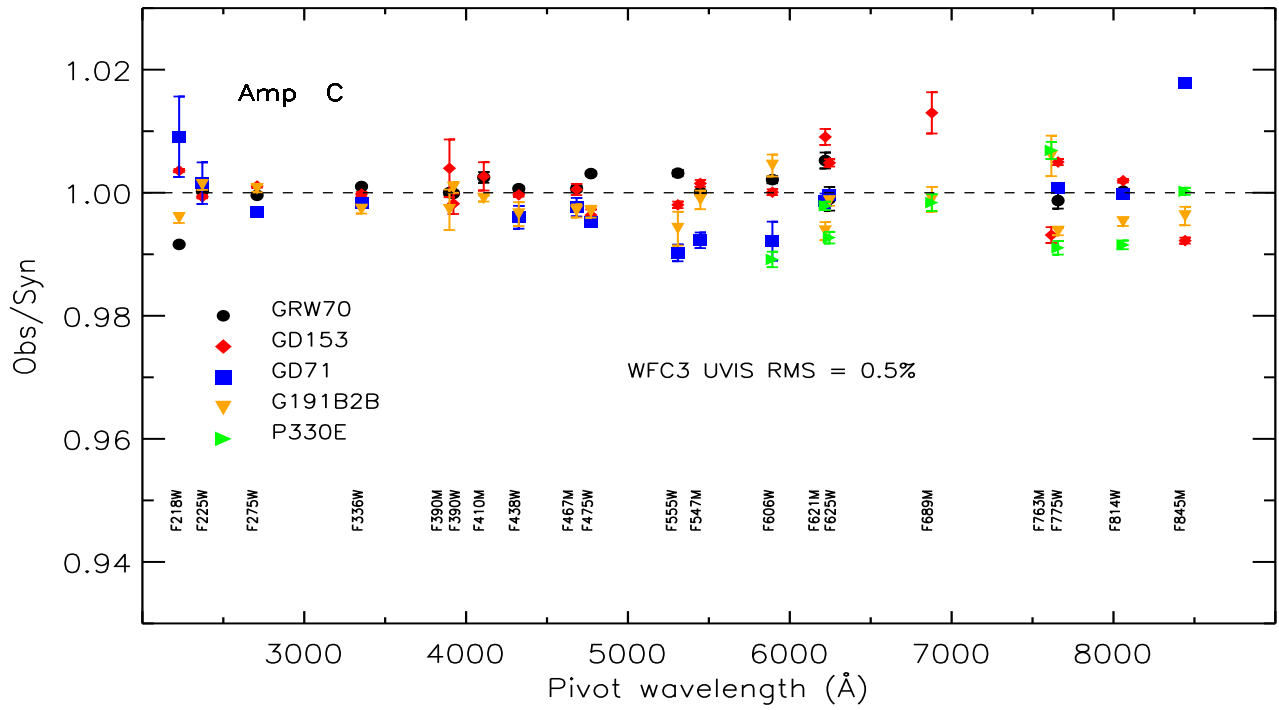}
\caption{Same as Fig.~\ref{fig:ratioA} but for UVIS2 (Amp C) filters. \label{fig:ratioC}}
\end{center}
\end{figure*}

\begin{table}
%\begin{center}
\begin{scriptsize}
\caption{\footnotesize Files used in the synthetic simulations performed with {\it Pysynphot.} \label{table:7}}
\begin{tabular}{lll}
\hline
\hline
Component & Description \\
\hline
\multicolumn{2}{c}{Simulations to derive the new in-flight corrections for WFC3-UVIS}  \\
\hline
wfc3\_uvis\_cor\_003\_syn.fits & Original in-flight correction, all entries set to 1.0 \\
wfc3uvis1\_aper\_007\_syn.fits & New aperture correction for UVIS1 \\
wfc3uvis2\_aper\_007\_syn.fits & New aperture correction for UVIS2 \\
wfc3\_uvis\_FXXXX\_002/003\_syn.fits & Pre-launch filter curves (TV3) \\
gd153\_stiswfcnic\_002.fits             & New CALSPEC SED \\
gd71\_stiswfcnic\_002.fits               & New CALSPEC SED \\
gd191b2b\_stiswfcnic\_002.fits       & New CALSPEC SED \\
grw\_70d5824\_stiswfcnic\_002.fits & New CALSPEC SED \\
p330e\_stiswfcnic\_002.fits             & New CALSPEC SED \\
\hline
\multicolumn{2}{c}{Simulations to derive the new filter curves for WFC3-UVIS}  \\
\hline
wfc3uvis1\_cor\_005\_syn.fits & New in-flight correction for UVIS1 \\
wfc3uvis2\_cor\_005\_syn.fits & New in-flight correction for UVIS2 \\
wfc3uvis1\_aper\_007\_syn.fits & New aperture correction for UVIS1 \\
wfc3uvis2\_aper\_007\_syn.fits & New aperture correction for UVIS2 \\
wfc3\_uvis\_FXXXX\_002/003\_syn.fits & Pre-launch filter curves (TV3) \\
gd153\_stiswfcnic\_002.fits             & New CALSPEC SED \\
gd71\_stiswfcnic\_002.fits               & New CALSPEC SED \\
gd191b2b\_stiswfcnic\_002.fits       & New CALSPEC SED \\
grw\_70d5824\_stiswfcnic\_002.fits & New CALSPEC SED \\
p330e\_stiswfcnic\_002.fits             & New CALSPEC SED \\
\hline
\multicolumn{2}{c}{Simulations to derive the final synthetic count rates for WFC3-UVIS}  \\
\hline
wfc3uvis1\_cor\_005\_syn.fits & New in-flight correction for UVIS1 \\
wfc3uvis2\_cor\_005\_syn.fits & New in-flight correction for UVIS2 \\ wfc3uvis1\_aper\_007\_syn.fits & New aperture correction for UVIS1 \\
wfc3uvis2\_aper\_007\_syn.fits & New aperture correction for UVIS2 \\
wfc3uvis1\_FXXXX\_008\_syn.fits & New filter curves for UVIS1 \\
wfc3uvis2\_FXXXX\_008\_syn.fits & New filter curves for UVIS2 \\
gd153\_stiswfcnic\_002.fits             & New CALSPEC SED \\
gd71\_stiswfcnic\_002.fits               & New CALSPEC SED \\
gd191b2b\_stiswfcnic\_002.fits       & New CALSPEC SED \\
grw\_70d5824\_stiswfcnic\_002.fits & New CALSPEC SED \\
p330e\_stiswfcnic\_002.fits             & New CALSPEC SED \\
\hline
\multicolumn{2}{c}{Simulations to derive the new filter curves for WFC3-IR}  \\
\hline
wfc3\_ir\_cor\_004\_syn.fits & Original in-flight correction \\
wfc3\_ir\_aper\_002\_syn.fits & Original aperture correction \\
wfc3\_ir\_FXXXX\_004/005\_syn.fits & 2012 filter curves \\
gd153\_stiswfcnic\_002.fits             & New CALSPEC SED \\
gd71\_stiswfcnic\_002.fits               & New CALSPEC SED \\
gd191b2b\_stiswfcnic\_002.fits       & New CALSPEC SED \\
grw\_70d5824\_stiswfcnic\_002.fits & New CALSPEC SED \\
p330e\_stiswfcnic\_002.fits             & New CALSPEC SED \\
\hline
\multicolumn{2}{c}{Simulations to derive the final synthetic count rates for WFC3-IR}  \\
\hline
wfc3\_ir\_cor\_004\_syn.fits & Original in-flight correction \\
wfc3\_ir\_aper\_002\_syn.fits & Original aperture correction \\
wfc3\_ir\_FXXXX\_007\_syn.fits & New filter curves \\
gd153\_stiswfcnic\_002.fits             & New CALSPEC SED \\
gd71\_stiswfcnic\_002.fits               & New CALSPEC SED \\
gd191b2b\_stiswfcnic\_002.fits       & New CALSPEC SED \\
grw\_70d5824\_stiswfcnic\_002.fits & New CALSPEC SED \\
p330e\_stiswfcnic\_002.fits             & New CALSPEC SED \\
\hline
\hline
\end{tabular}
%\end{center}
\end{scriptsize}
\end{table}

WFC3-UVIS filter curves were first calculated during three thermal vacuum (TV) tests performed at NASA Goddard by using the CASTLE apparatus. This system illuminated the detector with a monochromatic flux source and aperture photometry was derived on the images. The filter curves resulting from the third test, TV3, were delivered and presented in \citet{brown2008}. These curves were updated after WFC3 was installed on \emph{HST}, and the first in-flight correction and inverse sensitivities were derived by \citet{kalirai09b}.

Different in-flight corrections for UVIS1 and UVIS2 were later delivered by DE16, when WFC3 chips were independently calibrated. 

In order to update the in-flight corrections and derive new filter curves we
%We then proceeded to calculate new in-flight corrections for UVIS1 and UVIS2 
used {\it Pysynphot\footnote{https://pysynphot.readthedocs.io/en/latest/}} (Lim et al.\ 2015) to predict the count rates for each filter and standard star as observed with UVIS1 and UVIS2 at the reference epoch.
As input for the {\it Pysynphot} simulations, we used no in-flight correction (the {\it wfc3\_uvis\_cor\_003\_syn.fits} file has all entries set to 1.0), the filter curves from TV3 ({\it wfc3\_uvis\_FXXX\_002/003\_syn.fits}), and the new aperture correction files we calculated ({\it wfc3uvis1/2\_aper\_007\_syn.fits}), all listed in Table~\ref{table:7}. The simulations also used other components, such as the \emph{HST} Optical Telescope Assembly (OTA), the pick-off mirror, the mirrors' reflectivity, the inner and outer window, and the quantum efficiency (QE) of each detector.

We also used new SEDs of the three \emph{HST} primary WDs provided by the CALSPEC database ({\it \_stiswfcnic\_002})\footnote{Note that the latest version of the three \emph{HST} primary WD SEDs is {\it \_stiswfcnic\_003}}; these were calculated with the 
Non-Local Thermal-Equilibrium (NLTE) code from TMAP \citep{rauch2013} and TLUSTY \citep{hubeny2017}.
The models were normalized to an absolute flux level defined by the flux of 3.47$\times$10$^{-9}$ erg cm$^{-2}$ s$^{-1}$ \AA$^{-1}$ for Vega at 0.5556$\mu$m, as reconciled with the MSX mid-IR absolute flux measures (for more details please see BO20). 
We used new SEDs for GRW70 and P330E as well, as delivered in the CALSPEC database by BO20 and listed in the same table.

We then derived the ratio of observed over synthetic count rates for each star, detector and filter:
in the case of UV and bluer filters, i.e. for wavelengths $\lambda <$ 6000\AA, 
we calculated a weighted mean of the ratios by using the four standard WDs, while for longer
wavelengths we used all five stars in the calculations, i.e. we also included the G-type star P330E, 
when measurements were available. We followed this strategy since photometric measurements for
P330E have a much lower $S/N$ in the bluer filters and a significant 
color term ($\approx$ 1 to 8\%) is present when observing red sources with UV filters, 
i.e. the response of the detector and filter for red stars is different compared to the response for blue stars (CA18).
Figs.~\ref{fig:corA} and \ref{fig:corC} show the ratios of observed over synthetic count rates for all filters and 
Amp A (UVIS1) and Amp C (UVIS2), respectively. The ratio values for all filters are larger than 1.0, i.e. 
the throughputs were underestimated before WFC3 launch. A very similar result was found by 
\citet[see their Fig.~5]{kalirai09b} and DE16 (see their Fig.~8) based on observations collected in 2009 and 6 
years of standard star photometry, respectively. The pre-launch throughput values were 
measured during the TV3 campaign and were systematically underestimated, on average by 5--10\% and up to 20\% for wavelengths around $\lambda \sim$ 5000\AA. A possible explanation 
provided by Kalirai et al. is that the TV3 calibration error was due to problems with the CASTLE 
apparatus (see also \citealt{brown2008}).

The residuals of the observed over synthetic ratios after applying the new in-flight corrections are larger 
for the narrow-band filters (see bottom panel in Figs.~\ref{fig:corA} and \ref{fig:corC}), as expected, due to the availability of many less standard star measurements in these filters compared to the others,  the lower $S/N$, and in some cases the presence of absorption lines. 
For example, the ratio and the residual for the $F656N$ filter are systematically lower (by $\approx$ 10 and 5\%) 
compared to the other filters, probably due to the presence of a $H_{\alpha}$ line in the standard WD SEDs. 

The long-pass filters also show slightly larger residuals due to few measurements available.
Therefore, we only used the wide-, medium-band and the extremely-wide (X) filters 
to derive the new in-flight corrections. 

A least-square fit with a quadratic polynomial (Amp A: $0.93 + 9\cdot 10^{-5} \cdot \lambda - 8\cdot 10^{-9} \cdot \lambda^{2}$, Amp C: $0.92 + 1\cdot 10^{-4}\cdot \lambda - 1\cdot 10^{-8}\cdot \lambda^{2}$, where the wavelength is in units of \AA) resulted the best method to reproduce the data points and 
it is shown with a solid line in Figs.~\ref{fig:corA} and \ref{fig:corC}. The bottom panel of the figures shows the 
residual ratios for each filter after the fit. The ratios between observed and synthetic count rates have a mean value of 1.00 with a dispersion $\approx$ 0.02. 

We then created a new in-flight correction file for each detector, {\it wfc3uvis1\_cor\_005\_syn.fits} and {\it wfc3uvis2\_cor\_005\_syn.fits}, 
by using the derived polynomials (Table~\ref{table:7}).
New synthetic count rates were thus calculated with the new in-flight corrections and the same SEDs, filter curves and aperture corrections.
The ratio of observed and new synthetic count rates was used to derive a multiplicative scalar correction to be applied to each filter curve. New filter curves were created, and named
as {\it wfc3uvis1\_FXXXX\_008\_syn.fits} and {\it wfc3uvis2\_FXXXX\_008\_syn.fits}, and used to calculate 
the final synthetic count rates for each detector, 
filter and standard star. These new filter curves provided count rates as observed at the reference epoch.
Time-dependent filter curves were also created,  {\it wfc3uvis1\_FXXXX\_mjd\_008\_syn.fits} and
{\it wfc3uvis2\_FXXXX\_mjd\_008\_syn.fits}, by using the sensitivity change rates and calculating the filter curve for six different epochs spaced by two years each. The {\it wfc3uvis1,2\_FXXXX\_mjd\_008\_syn.fits} files thus have seven different throughput columns, one for the reference epoch and other six for different increasing $MJD$ values, until $MJD$ = 59388 (June 23, 2021). 

To generate synthetic count rates for any star and any filter as measured by WFC3-UVIS at different epochs ($MJDs$), {\it Pysynphot} or the more recently delivered 
{\it STsynphot}\citep{synphot2018}, interpolate between two of the six consecutive $MJD$ values included in the filter curve tables. If the requested epoch is outside the current lifetime of WFC3, the values will be extrapolated in the future or in the past. However, the extrapolation to $MJD$ values before the reference epoch, i.e. before WFC3 was launched, is not reliable and should not be used in simulations.

Figs.~\ref{fig:ratioA} and ~\ref{fig:ratioC} show the observed over synthetic count rates for the five standard stars and the wide- and medium-band filters obtained by using the new filter curves in the {\it Pysynphot} simulations.
The ratio values cluster around 1.0, as expected, with a RMS of 0.5\% for both detectors and including all filters.

\begin{figure*}
\begin{center}
\includegraphics[width=0.55\textwidth, height=0.75\textheight, angle=90]{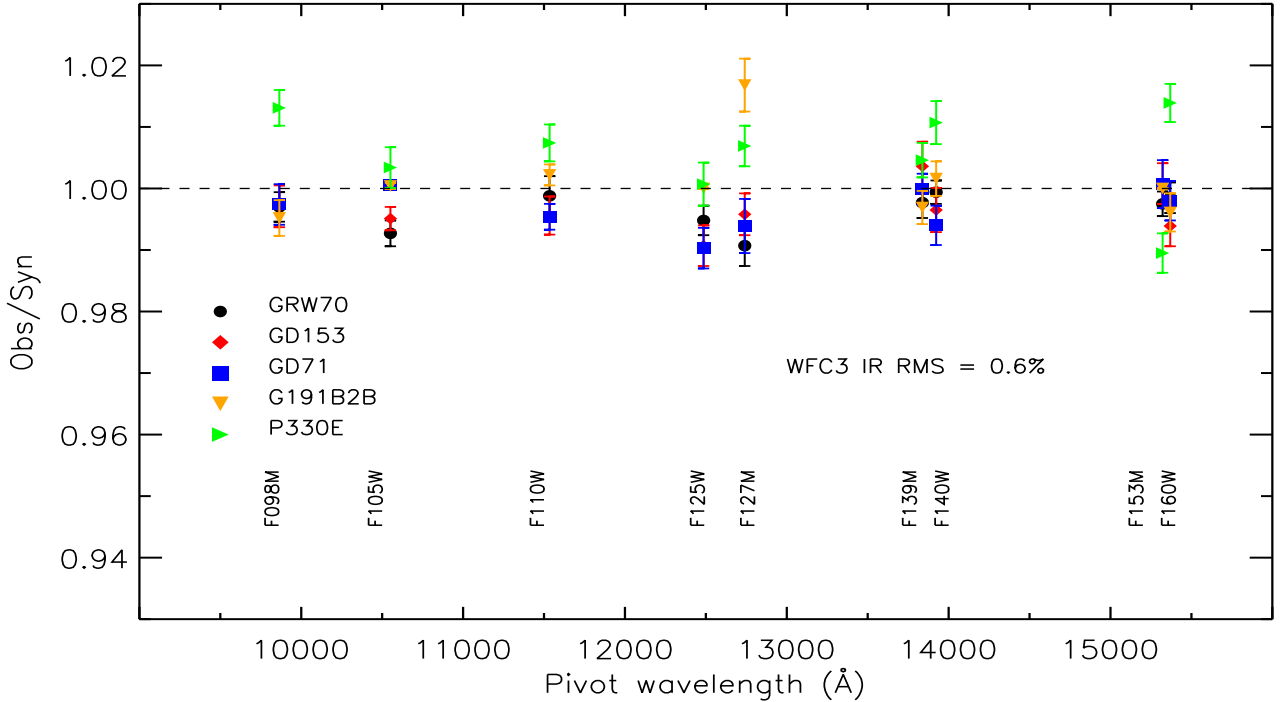}
\caption{Observed over synthetic count rates for WFC3 IR wide- and medium-band 
filters for the five standard stars used in the calibration as a function of the pivot wavelength. \label{fig:ratioIR}}
\end{center}
\end{figure*}

In the case of WFC3-IR, we performed {\it Pysynphot} simulations by using
the new standard star SEDs, the original in-flight correction ({\it 004}) from \citet{kalirai09a}, the 2012 filter curves ({\it 004/005}), and the original aperture correction ({\it 002}), as listed in Table~\ref{table:7}. We also used all the other components, such as the \emph{HST} OTA, the pick-off mirror, the mirror reflectivity, the inner and outer window, and the QE for the WFC3-IR detector.

We thus derived the ratio of observed over synthetic count rates for each standard star and filter and calculated a weighted mean of the ratios by using the four WDs and the G-type star P330E.
The ratio of observed and new synthetic count rates was then used to derive a multiplicative 
scalar correction to be applied to each filter curve. New filter curves were created, and named as {\it wfc3\_ir\_FXXXX\_007\_syn.fits}, and used to calculate the final synthetic count rates 
for each filter and standard star. These new filter curves provided count rates as observed at the 
same reference epoch as WFC3-UVIS, i.e. $MJD$ = 55008.
Filter curves including the same time-dependent columns ($MJD$) as the WFC3-UVIS filter curves were also created, 
{\it wfc3\_ir\_FXXXX\_mjd\_007\_syn.fits}; however, no time dependence was introduced for WFC-IR inverse sensitivities and the different $MJD$ columns all contain the same values.
Fig.~\ref{fig:ratioIR} shows the observed over synthetic count rates for the five standard stars and the wide- and medium-band filters obtained with the new WFC3-IR filter curves.
The ratio values cluster around 1.0, as expected, with a RMS of 0.6\%, including all filters.

\subsection{Quad filters}\label{quad}
The quad filters are made of a 2$\times$2 mosaic of elements occupying a single 
filter slot, with each quadrant providing a different bandpass; therefore, the five quad filter sets generate 20 different narrow- and medium-band filters. 
The readout amplifier for each quad filter is listed in Table~\ref{table:9}.

In the case of the quad filters, only the standards GD153, G191B2B 
and P330E were observed and not enough measurements were collected
to determine slopes for the sensitivity changes with time. Therefore, we used the available photometry to calculate the weighted mean count rates for each standard star in each filter and assumed the same 
reference epoch as for the other WFC3-UVIS filters, i.e. $MJD$ = 55008.

Synthetic count rates were calculated by using the new SEDs for the standard stars and the original in-flight
correction ({\it wfc3\_uvis\_cor\_003\_syn.fits}), the original aperture correction 
file ({\it wfc3\_uvis\_aper\_002\_syn.fits}), and the original filter curves ({\it wfc3\_uvis\_FQXXX\_004\_syn.fits} or
{\it wfc3\_uvis\_FQXXX\_005\_syn.fits}). We then obtained a
weighted mean of the ratios by using the two WDs and the G-type star P330E and 
derived multiplicative scalar factors to create new filter curves. 
These are named as {\it wfc3\_uvis\_FQXXXX\_008\_syn.fits} or {\it wfc3uvis2\_FQXXXX\_mjd\_008\_syn.fits}. As in the case of WFC3-IR, no time dependence is included in the quad filter curves, and the different $MJD$ columns have all the same throughput values.
It is worth noting that only one in-flight, one aperture
correction and one filter curve file, named as {\it \_uvis}, is available for the quad filters, irregardless of the quadrant (amplifier) in which the filters fall.

\section{Calculating the inverse sensitivities}\label{sec:sensi}
Updated synthetic count rates and photometry for the five standard stars for both
WFC3-UVIS and IR were used to derive new inverse sensitivities. We followed the 
method presented in \citet{bohlin2014a, bohlin2020} and DE16.
The point source mean flux over a passband can be defined in wavelength units, 
$erg$ $s^{-1}$ $cm^{-2}$ \AA$^{-1}$, as:

\begin{equation}
<F_{\lambda}> =  \frac{\int F_{\lambda }\cdot \lambda \cdot R\cdot d\lambda}{\int \lambda \cdot R\cdot d\lambda} = S_{\lambda}\cdot N_{e}
\end{equation}

or in frequency units, $erg$ $s^{-1}$ $cm^{-2}$ $Hz^{-1}$, as:

\begin{equation}
<F_{\nu}> =  \frac{\int F_{\nu }\cdot \nu \cdot R\cdot d\nu}{\int \nu \cdot R\cdot d\nu} = S_{\nu}\cdot N_{e}
\end{equation}

where $R$ is the system throughput, $S_{\lambda}$ and $S_{\nu}$ are the instrument sensitivities, $N_e$ is the instrumental count rate in $e^{-}/s$, and the integrals are calculated over the passband \citep{koornneef1986,rieke2008}.

The detector count rate, $N_e$, can be measured or calculated as:

\begin{equation}
N_{e} = \frac{A}{hc}\int F_{\lambda }\cdot \lambda \cdot R \cdot d\lambda = A \int \frac{ F_{\nu}}{h \nu} \cdot R \cdot d\nu 
\end{equation}

where $A$ is the telescope collecting area, $h$ is the Planck constant, $c$ is the speed of light.

The instrument sensitivities $S$ are then defined by dividing the mean flux of Eqs. (1) and (2) by the detected count rate, $N_{e}$, and are expressed in units of 
$erg$ $cm^{-2}$ \AA$^{-1}$ $e^{-1}$: 

\begin{equation}
S_{\lambda} = \frac{<F_{\lambda}>}{N_e} = \frac{hc}{A\cdot \int \lambda \cdot R\cdot d\lambda }
\end{equation}

 or $erg$ $cm^{-2}$ $Hz^{-1}$ $e^{-1}$: 

\begin{equation}
S_{\nu} = \frac{<F_{\nu}>}{N_e} = \frac{h}{A\cdot \int \nu^{-1} \cdot R\cdot d\nu }
\end{equation}

We refer to $S$ as the 'inverse sensitivity' since a more sensitive detector will have larger count rates, $N_e$, for the same source flux $F_{\lambda}$ or $F_{\nu}$.

$S_{\lambda}$ at {\it infinity} is provided in the image header for UVIS1 and UVIS2 as
the PHFTLAM1 and PHTFLAM2 keywords, respectively. The PHOTFLAM keyword is set to the value of 
PHTFLAM1, except for the UV filters (see below).
The ratio of the UVIS2 and UVIS1 inverse sensitivities ($S_{2}/ S_{1}$ or PHTFLAM2/PHTFLAM1) 
is indicated in the image header by the keyword PHTRATIO (see Section~\ref{sec:photratio}).

In the case of the WFC3-IR detector, $S_{\nu}$ is also provided in the image header as the PHOTFNU keyword.

For observations collected with UV filters, namely $F218W$, $F225W$, $F275W$ and $F200LP$, 
the value of the UVIS1 inverse sensitivity is modified ($S_{1}^{'}$), such that the ratio of the inverse sensitivities, PHTRATIO ($S_{2}/S_{1}^{'}$), is equal to the ratio of the observed count rates, $C_{1} / C_{2}$ (DE17).
This tweak is necessary since the response functions of UVIS1 and UVIS2 are significantly 
different in the UV regime, and WFC3 processing pipeline, {\it calwf3}, needs PHTRATIO to flux scale the UVIS2 detector to UVIS1. However, the equivalency of the modified inverse sensitivity ratio, $S_{2}/S_{1}^{'}$, to the count rate 
ratio, $C_{1} / C_{2}$, only holds for hot stars, i.e. $T_{eff} \gtrsim$ 30,000K, since cooler stars have a largely different SEDs in the UV, and the response of the detector + filter system is different for these sources. Therefore, 
magnitude offsets for the UV filters as a function of the source color need be applied 
to magnitudes measured on UVIS2 to transform the photometry to the UVIS1 photometric system.
These corrections are currently available in CA18 (see \S 9 for more details).
%however, new corrections will be provided in the future based on the new inverse sensitivities presented in CA21 and this manuscript.

Inverse sensitivities at {\it infinity} were also derived for the 15 WFC3-IR filters and indicated in the image header as the PHOTFLAM and PHOTFNU keywords. 

The new inverse sensitivities for UVIS1 and UVIS2 at the reference epoch, $MJD =$ 55008 (June 26, 2009), 
are listed in Tables~\ref{table:8} (42 full-frame filters) and Table~\ref{table:9} (20 quad filters).
Table~\ref{table:10} lists the new inverse sensitivities for the 15 WFC3-IR filters at the same reference epoch. 
Inverse sensitivities are also provided at the WFC3 Photometric Calibration web pages for WFC3-UVIS\footnote{\scriptsize https://www.stsci.edu/hst/instrumentation/wfc3/data-analysis/photometric-calibration/uvis-photometric-calibration} and WFC3-IR\footnote{\scriptsize https://www.stsci.edu/hst/instrumentation/wfc3/data-analysis/photometric-calibration/ir-photometric-calibration}. 

Inverse sensitivities can also be computed for any observation epoch by using {\it STsynphot} and 
the new set of filter curves, in-flight and aperture corrections: example tutorials are provided at the same web pages or at the STScI WFC3 Software Library on GitHub\footnote{https://github.com/spacetelescope/WFC3Library}. 

\LTcapwidth=\textwidth
{ \footnotesize
%\begin{table}[!h]
\begin{longtable*}{lcccccccc}
%\centering
%\begin{scriptsize}
\caption{\footnotesize New inverse sensitivity values (PHOTFLAM) and ZPs in different 
photometric systems for UVIS1 (Amp A) and UVIS2 (Amp C) 
42 full-frame filters calculated at the reference epoch $MJD =$ 55008 (June 26, 2009). Errors are also listed. \label{table:8}}\\
%\vspace{0.3cm}
%\begin{tabular}{lcccccccc}
\hline
\hline
Filter & Pivot  & PHOTBW  & ZP$_{AB}$ & ZP$_{Vega}$ &  ZP$_{ST}$  &  ZP$_{ERR}$ & PHOTFLAM & PHOTFLAM$_{ERR}$	\\
       & (\AA)  & (\AA)   & (Mag)     & (Mag)      & (Mag)        & (Mag)    & (erg cm$^{-2}$ \AA$^{-1}$ e$^{-1}$) & (erg cm$^{-2}$ \AA$^{-1}$ e$^{-1}$) \\
\hline
\hline
\endfirsthead
\hline
\multicolumn{9}{c}{UVIS1 (Amp A)}  \\
\hline
%\endfirsfoot
F200LP& 4971.86 & 1742.20 & 27.3356 & 26.8857 & 27.1261 & 0.0128 & 5.1234e-20 & 6.0032e-22 \\
F218W & 2228.04 &  128.94 & 22.9368 & 21.2726 & 20.9843 & 0.0072 & 1.4664e-17 & 9.6609e-20 \\
F225W & 2372.05 &  177.43 & 24.0631 & 22.4257 & 22.2467 & 0.0015 & 4.5849e-18 & 6.2529e-21 \\
F275W & 2709.69 &  164.43 & 24.1569 & 22.6759 & 22.6294 & 0.0017 & 3.2227e-18 & 5.1180e-21 \\
F280N & 2832.86 &  200.69 & 20.9180 & 19.5016 & 19.4871 & 0.0085 & 5.8231e-17 & 4.5543e-19 \\
F300X & 2820.47 &  316.56 & 24.9638 & 23.5611 & 23.5234 & 0.0024 & 1.4147e-18 & 3.1311e-21 \\
F336W & 3354.49 &  158.42 & 24.6908 & 23.5260 & 23.6269 & 0.0018 & 1.2860e-18 & 2.1606e-21 \\
F343N & 3435.15 &   86.71 & 23.8868 & 22.7517 & 22.8745 & 0.0016 & 2.5716e-18 & 3.6774e-21 \\
F350LP& 5873.87 & 1490.06 & 26.9647 & 26.8116 & 27.1173 & 0.0050 & 5.1653e-20 & 2.4005e-22 \\
F373N & 3730.17 &   18.34 & 21.9076 & 21.0354 & 21.0742 & 0.0090 & 1.3499e-17 & 1.1206e-19 \\
F390M & 3897.24 &   65.48 & 23.6216 & 23.5457 & 22.8834 & 0.0052 & 2.5506e-18 & 1.2257e-20 \\
F390W & 3923.69 &  291.27 & 25.3725 & 25.1735 & 24.6489 & 0.0032 & 5.0170e-19 & 1.4587e-21 \\
F395N & 3955.19 &   26.29 & 22.6678 & 22.7115 & 21.9616 & 0.0024 & 5.9616e-18 & 1.3191e-20 \\
F410M & 4108.99 &   57.03 & 23.5959 & 23.7699 & 22.9726 & 0.0038 & 2.3495e-18 & 8.2162e-21 \\
F438W & 4326.23 &  197.31 & 24.8367 & 25.0015 & 24.3252 & 0.0060 & 6.7593e-19 & 3.7819e-21 \\
F467M & 4682.58 &   68.42 & 23.6935 & 23.8567 & 23.3539 & 0.0062 & 1.6536e-18 & 9.5492e-21 \\
F469N & 4688.10 &   19.97 & 21.8160 & 21.9825 & 21.4790 & 0.0029 & 9.2985e-18 & 2.5187e-20 \\
F475W & 4773.10 &  421.30 & 25.7039 & 25.8094 & 25.4058 & 0.0055 & 2.4984e-19 & 1.2504e-21 \\
F475X & 4940.72 &  660.68 & 26.1558 & 26.2131 & 25.9327 & 0.0017 & 1.5379e-19 & 2.3980e-22 \\
F487N & 4871.38 &   21.71 & 22.2269 & 22.0479 & 21.9731 & 0.0039 & 5.8987e-18 & 2.1052e-20 \\
F502N & 5009.64 &   26.96 & 22.3262 & 22.4190 & 22.1332 & 0.0050 & 5.0899e-18 & 2.3595e-20 \\
F547M & 5447.50 &  206.24 & 24.7550 & 24.7583 & 24.7440 & 0.0100 & 4.5959e-19 & 4.2627e-21 \\
F555W & 5308.43 &  517.49 & 25.8097 & 25.8379 & 25.7425 & 0.0028 & 1.8324e-19 & 4.6668e-22 \\
F600LP& 7468.12 &  945.89 & 25.8820 & 25.5487 & 26.5560 & 0.0070 & 8.6611e-20 & 5.5311e-22 \\
F606W & 5889.17 &  657.20 & 26.0872 & 26.0039 & 26.2454 & 0.0129 & 1.1529e-19 & 1.3885e-21 \\
F621M & 6218.85 &  185.65 & 24.6124 & 24.4620 & 24.8889 & 0.0070 & 4.0217e-19 & 2.5967e-21 \\
F625W & 6242.56 &  451.28 & 25.5247 & 25.3736 & 25.8095 & 0.0094 & 1.7225e-19 & 1.4834e-21 \\
F631N & 6304.29 &   41.60 & 21.8849 & 21.7232 & 22.1910 & 0.0114 & 4.8259e-18 & 5.0616e-20 \\
F645N & 6453.59 &   41.45 & 22.2434 & 22.0478 & 22.6004 & 0.0039 & 3.3101e-18 & 1.1955e-20 \\
F656N & 6561.37 &   41.77 & 20.4221 & 19.8404 & 20.8151 & 0.0385 & 1.7137e-17 & 5.9545e-19 \\
F657N & 6566.63 &   41.00 & 22.6585 & 22.3324 & 23.0531 & 0.0043 & 2.1815e-18 & 8.7084e-21 \\
F658N & 6584.02 &  148.71 & 21.0271 & 20.6717 & 21.4275 & 0.0177 & 9.7468e-18 & 1.5697e-19 \\
F665N & 6655.88 &  42.19  & 22.7339 & 22.4901 & 23.1578 & 0.0096 & 1.9808e-18 & 1.7401e-20 \\
F673N & 6765.94 &  41.94  & 22.5877 & 22.3424 & 23.0473 & 0.0069 & 2.1931e-18 & 1.3993e-20 \\
F680N & 6877.60 &  112.01 & 23.8182 & 23.5546 & 24.3133 & 0.0140 & 6.8336e-19 & 8.9134e-21 \\
F689M & 6876.75 &  207.61 & 24.4777 & 24.1950 & 24.9725 & 0.0028 & 3.7238e-19 & 9.6694e-22 \\
F763M & 7614.37 &  229.42 & 24.2260 & 23.8366 & 24.9421 & 0.0068 & 3.8296e-19 & 2.3862e-21 \\
F775W & 7651.36 &  419.72 & 24.8714 & 24.4800 & 25.5981 & 0.0048 & 2.0930e-19 & 9.1984e-22 \\
F814W & 8039.06 &  666.76 & 25.1272 & 24.6985 & 25.9612 & 0.0075 & 1.4980e-19 & 1.0373e-21 \\
F845M & 8439.06 &  260.30 & 23.8216 & 23.3150 & 24.7610 & 0.0091 & 4.5246e-19 & 3.8212e-21 \\
F850LP& 9176.13 &  470.53 & 23.8557 & 23.3253 & 24.9769 & 0.0066 & 3.7086e-19 & 2.2782e-21 \\
F953N & 9530.58 &   71.19 & 20.4250 & 19.8019 & 21.6285 & 0.0111 & 8.1018e-18 & 8.2727e-20 \\
\hline
\multicolumn{9}{c}{UVIS2 (Amp C)}  \\
%\endsecondhead
\hline
%\endsecondfoot
F200LP& 4875.10 & 1725.22 & 27.3803 & 26.9000 & 27.1282 & 0.0127 & 5.1134e-20 & 5.9509e-22 \\
F218W & 2223.72 &  124.92 & 23.2115 & 21.5463 & 21.2548 & 0.0106 & 1.1430e-17 & 1.1093e-19 \\
F225W & 2358.39 &  173.15 & 24.2791 & 22.6377 & 22.4501 & 0.0012 & 3.8015e-18 & 4.2937e-21 \\
F275W & 2703.30 &  165.58 & 24.2223 & 22.7373 & 22.6897 & 0.0021 & 3.0488e-18 & 5.9952e-21 \\
F280N & 2829.98 &  202.41 & 20.9303 & 19.5123 & 19.4972 & 0.0182 & 5.7693e-17 & 9.5770e-19 \\
F300X & 2805.84 &  316.95 & 25.0513 & 23.6394 & 23.5995 & 0.0117 & 1.3186e-18 & 1.4264e-20 \\
F336W & 3354.65 &  158.34 & 24.7185 & 23.5538 & 23.6547 & 0.0022 & 1.2535e-18 & 2.5211e-21 \\
F343N & 3435.19 &   86.65 & 23.9236 & 22.7885 & 22.9113 & 0.0042 & 2.4858e-18 & 9.6503e-21 \\
F350LP& 5851.15 & 1483.02 & 26.9356 & 26.7802 & 27.0798 & 0.0048 & 5.3469e-20 & 2.3475e-22 \\
F373N & 3730.16 &   18.29 & 21.9350 & 21.0628 & 21.1016 & 0.0051 & 1.3163e-17 & 6.1663e-20 \\
F390M & 3897.00 &   65.47 & 23.6375 & 23.5611 & 22.8992 & 0.0037 & 2.5138e-18 & 8.5689e-21 \\
F390W & 3920.72 &  291.16 & 25.3811 & 25.1779 & 24.6559 & 0.0017 & 4.9849e-19 & 7.8536e-22 \\
F395N & 3955.15 &  26.30  & 22.6700 & 22.7139 & 21.9638 & 0.0034 & 5.9495e-18 & 1.8615e-20 \\
F410M & 4108.88 &  56.96  & 23.5944 & 23.7683 & 22.9709 & 0.0029 & 2.3530e-18 & 6.3673e-21 \\
F438W & 4325.14 &  197.42 & 24.8343 & 24.9990 & 24.3223 & 0.0034 & 6.7777e-19 & 2.1424e-21 \\
F467M & 4682.60 &  68.37  & 23.6984 & 23.8616 & 23.3588 & 0.0024 & 1.6461e-18 & 3.7092e-21 \\
F469N & 4688.10 &  20.07  & 21.8199 & 21.9864 & 21.4828 & 0.0089 & 9.2649e-18 & 7.6216e-20 \\
F475W & 4772.17 &  421.76 & 25.6961 & 25.8017 & 25.3977 & 0.0048 & 2.5172e-19 & 1.1204e-21 \\
F475X & 4937.41 &  661.13 & 26.1519 & 26.2092 & 25.9273 & 0.0045 & 1.5455e-19 & 6.4132e-22 \\
F487N & 4871.38 &   21.84 & 22.2413 & 22.0624 & 21.9875 & 0.0165 & 5.8199e-18 & 8.8914e-20 \\
F502N & 5009.63 &   27.10 & 22.3215 & 22.4143 & 22.1285 & 0.0048 & 5.1120e-18 & 2.2581e-20 \\
F547M & 5447.24 &  206.18 & 24.7592 & 24.7625 & 24.7480 & 0.0050 & 4.5792e-19 & 2.1269e-21 \\
F555W & 5307.91 &  517.13 & 25.7962 & 25.8245 & 25.7288 & 0.0076 & 1.8556e-19 & 1.3004e-21 \\
F600LP& 7453.66 &  937.10 & 25.8573 & 25.5254 & 26.5271 & 0.0090 & 8.8952e-20 & 7.3298e-22 \\
F606W & 5887.71 &  656.93 & 26.0785 & 25.9954 & 26.2361 & 0.0079 & 1.1629e-19 & 8.5554e-22 \\
F621M & 6219.16 &  185.71 & 24.6065 & 24.4560 & 24.8831 & 0.0068 & 4.0434e-19 & 2.5181e-21 \\
F625W & 6241.96 &  451.09 & 25.5247 & 25.3736 & 25.8092 & 0.0049 & 1.7231e-19 & 7.8251e-22 \\
F631N & 6304.28 &   42.39 & 21.8900 & 21.7283 & 22.1961 & 0.0102 & 4.8033e-18 & 4.4718e-20 \\
F645N & 6453.58 &   42.24 & 22.2381 & 22.0425 & 22.5951 & 0.0050 & 3.3263e-18 & 1.5421e-20 \\
F656N & 6561.36 &   42.44 & 20.4568 & 19.8751 & 20.8497 & 0.0126 & 1.6600e-17 & 1.9100e-19 \\
F657N & 6566.60 &   41.07 & 22.6580 & 22.3319 & 23.0527 & 0.0051 & 2.1824e-18 & 1.0186e-20 \\
F658N & 6583.92 &  151.15 & 21.0376 & 20.6820 & 21.4379 & 0.0080 & 9.6562e-18 & 7.1081e-20 \\
F665N & 6655.84 &   42.26 & 22.7212 & 22.4775 & 23.1452 & 0.0062 & 2.0041e-18 & 1.1303e-20 \\
F673N & 6765.91 &   42.13 & 22.5625 & 22.3171 & 23.0220 & 0.0121 & 2.2447e-18 & 2.4847e-20 \\
F680N & 6877.41 &  112.06 & 23.7974 & 23.5339 & 24.2925 & 0.0065 & 6.9662e-19 & 4.1363e-21 \\
F689M & 6876.50 &  207.84 & 24.4682 & 24.1855 & 24.9630 & 0.0100 & 3.7566e-19 & 3.4228e-21 \\
F763M & 7612.74 &  228.87 & 24.2051 & 23.8160 & 24.9208 & 0.0087 & 3.9054e-19 & 3.1189e-21 \\
F775W & 7648.30 &  418.28 & 24.8610 & 24.4700 & 25.5868 & 0.0067 & 2.1149e-19 & 1.3204e-21 \\
F814W & 8029.32 &  663.97 & 25.1118 & 24.6841 & 25.9431 & 0.0056 & 1.5232e-19 & 7.8754e-22 \\
F845M & 8437.27 &  259.71 & 23.8125 & 23.3060 & 24.7515 & 0.0123 & 4.5641e-19 & 5.1436e-21 \\
F850LP& 9169.94 &  466.6  & 23.8099 & 23.2799 & 24.9297 & 0.0076 & 3.8736e-19 & 2.7041e-21 \\
F953N & 9530.50 &   72.85 & 20.3831 & 19.7601 & 21.5866 & 0.0122 & 8.4191e-18 & 9.5450e-20 \\
\hline
\hline
%\end{tabular}
%\end{scriptsize}
%\end{table}
\end{longtable*}
}

\begin{table*}
\begin{center}
\begin{scriptsize}
\caption{\footnotesize New inverse sensitivity values (PHOTFLAM) and ZPs in different 
photometric systems for the 20 quad filters calculated at the reference 
epoch $MJD =$ 55008 (June 26, 2009). Errors are also listed. \label{table:9}}
\vspace{0.3cm}
\begin{tabular}{lccccccccc}
\hline
\hline
Filter & Chip & Pivot  & PHOTBW  & ZP$_{AB}$ & ZP$_{Vega}$ &  ZP$_{ST}$  &  ZP$_{ERR}$ & PHOTFLAM & PHOTFLAM$_{ERR}$	\\
       &     & (\AA)  & (\AA)   & (Mag) & (Mag) & (Mag) & (Mag)  & (erg cm$^{-2}$ \AA$^{-1}$ e$^{-1}$) & (erg cm$^{-2}$ \AA$^{-1}$ e$^{-1}$) \\\hline
\hline
FQ232N & UVIS2 & 2432.22  &  263.50 & 20.4123  & 18.8028 & 18.6502 & 0.0064 & 1.2587e-16 & 7.4676e-19 \\
FQ243N & UVIS2 & 2476.32  &  193.97 & 20.7378  & 19.1082 & 19.0148 & 0.0129 & 9.0301e-17 & 9.5288e-18 \\
FQ378N & UVIS1 & 3792.41  &  32.14  & 22.7507  & 22.2919 & 21.9532 & 0.0136 & 6.0299e-18 & 3.8173e-19 \\
FQ387N & UVIS1 & 3873.66  &  15.01  & 21.3399  & 21.2738 & 20.5884 & 0.0136 & 2.0928e-17 & 2.8522e-19 \\
FQ422M & UVIS2 & 4219.21  &  38.33  & 22.6725  & 22.9269 & 22.1066 & 0.0185 & 5.1880e-18 & 5.3523e-20 \\
FQ436N & UVIS2 & 4367.16  &  22.82  & 21.6299  & 21.6775 & 21.1389 & 0.0081 & 1.2684e-17 & 3.1092e-19 \\
FQ437N & UVIS1 & 4371.04  &  21.60  & 21.2682  & 21.3942 & 20.7790 & 0.0369 & 1.7282e-17 & 5.5396e-19 \\
FQ492N & UVIS1 & 4933.44  &  35.18  & 22.8676  & 22.9380 & 22.6413 & 0.0073 & 3.1887e-18 & 1.5797e-20 \\
FQ508N & UVIS1 & 5091.05  &  42.37  & 22.8805  & 22.9579 & 22.7225 & 0.0389 & 2.9976e-18 & 1.5032e-19 \\
FQ575N & UVIS2 & 5757.69  &  42.20  & 20.5297  & 20.4709 & 20.6388 & 0.0537 & 2.0246e-17 & 7.3602e-19 \\
FQ619N & UVIS1 & 6198.52  &  36.45  & 21.9403  & 21.7985 & 22.2097 & 0.0160 & 4.7187e-18 & 6.4863e-20 \\
FQ634N & UVIS2 & 6349.21  &  43.00  & 21.9575  & 21.7809 & 22.2790 & 0.0307 & 4.4312e-18 & 7.6042e-20 \\
FQ672N & UVIS2 & 6716.38  &  70.00  & 20.3946  & 20.1585 & 20.8382 & 0.0960 & 1.5193e-17 & 1.6124e-18 \\
FQ674N & UVIS1 & 6730.68  &  39.20  & 20.6923  & 20.4535 & 21.1406 & 0.0403 & 1.6657e-17 & 6.8407e-18 \\
FQ727N & UVIS2 & 7275.23  &  63.22  & 21.5808  & 21.2474 & 22.1979 & 0.0676 & 4.7367e-18 & 2.6141e-19 \\
FQ750N & UVIS1 & 7502.50  &  28.12  & 21.5024  & 21.1309 & 22.1864 & 0.0561 & 4.6196e-18 & 2.9526e-19 \\
FQ889N & UVIS1 & 8892.15  &  55.49  & 21.0572  & 20.5360 & 22.1102 & 0.0181 & 5.0706e-18 & 5.8733e-19 \\
FQ906N & UVIS2 & 9057.76  &  57.30  & 20.9512  & 20.4312 & 22.0443 & 0.0500 & 5.3340e-18 & 2.4683e-19 \\
FQ924N & UVIS2 & 9247.59  &  46.28  & 20.7532  & 20.1576 & 21.8913 & 0.0500 & 6.3554e-18 & 2.9059e-18 \\
FQ937N & UVIS1 & 9372.42  &  54.80  & 20.6478  & 20.1671 & 21.8150 & 0.0045 & 7.2367e-18 & 4.2433e-19 \\
\hline
\hline
\end{tabular}
\end{scriptsize}
\end{center}
\end{table*}

\begin{table*}
\begin{center}
\begin{scriptsize}
\caption{\footnotesize New inverse sensitivity values (PHOTFLAM) and ZPs in different 
photometric systems for the 15 WFC3-IR filters calculated at the reference 
epoch $MJD =$ 55008 (June 26, 2009). Errors are also listed. \label{table:10}}
\vspace{0.3cm}
\begin{tabular}{lcccccccc}
\hline
\hline
Filter & Pivot  & PHOTBW  & ZP$_{AB}$    & ZP$_{Vega}$ &  ZP$_{ST}$  &  ZP$_{ERR}$ & PHOTFLAM & PHOTFLAM$_{ERR}$	\\
    &  (\AA)  & (\AA)   & (Mag) & (Mag) & (Mag) & (Mag)  & (erg cm$^{-2}$ \AA$^{-1}$ e$^{-1}$) & (erg cm$^{-2}$ \AA$^{-1}$ e$^{-1}$) \\\hline
\hline
F098M &  9864.72 &  500.85 & 25.6661 & 25.0900 & 26.9445 & 0.0080 & 6.0653e-20 & 4.3288e-22 \\ 
F105W & 10551.05 &  845.62 & 26.2637 & 25.6025 & 27.6882 & 0.0047 & 3.0507e-20 & 1.4230e-22 \\ 
F110W & 11534.46 & 1428.48 & 26.8185 & 26.0418 & 28.4364 & 0.0046 & 1.5318e-20 & 6.5390e-23 \\ 
F125W & 12486.06 &  866.28 & 26.2321 & 25.3117 & 28.0221 & 0.0078 & 2.2446e-20 & 1.6252e-22 \\ 
F126N & 12584.89 &  339.31 & 22.8491 & 21.9083 & 24.6563 & 0.0079 & 4.9671e-19 & 3.6832e-21 \\ 
F127M & 12740.29 &  249.56 & 24.6246 & 23.6432 & 26.4584 & 0.0122 & 9.4524e-20 & 1.0141e-21 \\ 
F128N & 12831.84 &  357.44 & 22.9561 & 21.8982 & 24.8055 & 0.0078 & 4.3480e-19 & 2.9263e-21 \\ 
F130N & 13005.68 &  274.24 & 22.9813 & 21.9849 & 24.8599 & 0.0081 & 4.1416e-19 & 2.5823e-21 \\ 
F132N & 13187.71 &  319.08 & 22.9325 & 21.9145 & 24.8413 & 0.0071 & 4.2096e-19 & 2.9310e-21 \\ 
F139M & 13837.62 &  278.02 & 24.4663 & 23.3663 & 26.4796 & 0.0043 & 9.3134e-20 & 3.2509e-22 \\ 
F140W & 13922.91 & 1132.38 & 26.4502 & 25.3528 & 28.4768 & 0.0068 & 1.4759e-20 & 9.1933e-23 \\ 
F153M & 15322.05 &  378.95 & 24.4469 & 23.1712 & 26.6815 & 0.0060 & 7.7161e-20 & 4.2952e-22 \\ 
F160W & 15369.18 &  826.25 & 25.9362 & 24.6622 & 28.1774 & 0.0086 & 1.9429e-20 & 1.5081e-22 \\ 
F164N & 16403.51 &  700.06 & 22.8921 & 21.4837 & 25.2748 & 0.0121 & 2.8257e-19 & 3.0682e-21 \\ 
F167N & 16641.60 &  645.24 & 22.9366 & 21.5504 & 25.3505 & 0.0095 & 2.6215e-19 & 2.8448e-21 \\ 
\hline
\hline
\end{tabular}
\end{scriptsize}
\end{center}
\end{table*}

\subsection{Sensitivity ratios for WFC3-UVIS}\label{sec:photratio}
We used the new WFC3-UVIS inverse sensitivities to calculate updated detector sensitivity ratio, PHTRATIO = PHTFLAM2/PHTFLAM1, values.
These values are used by the WFC3 processing pipeline, {\it calwf3}, to correct the fluxes measured on UVIS2 to the 
UVIS1 photometric system. Note that {\it calwf3} performs this correction by default; this is needed to allow users to apply only one inverse sensitivity value, PHTFLAM1, to calibrate the photometry
performed on the full WFC3-UVIS detector. The flux scaling is also needed to process the images with \texttt{AstroDrizzle} to create distorsion-free and CR corrected drizzled images (see \S 4 and \citealt{deustua2017b}).

However, if observations are done by using a UVIS2 sub-array, the flux correction can be avoided by setting FLUXCORR = OMIT in the image header and re-running the {\it calwf3} pipeline reduction on the raw images ({\it \_raw.fits}).
In this case, the PHTFLAM2 values in the image header must be used to calibrate the photometry.

\begin{figure}
%\begin{center}
%\hspace{-0.1cm}
\includegraphics[width=0.50\textwidth, height=0.35\textheight]{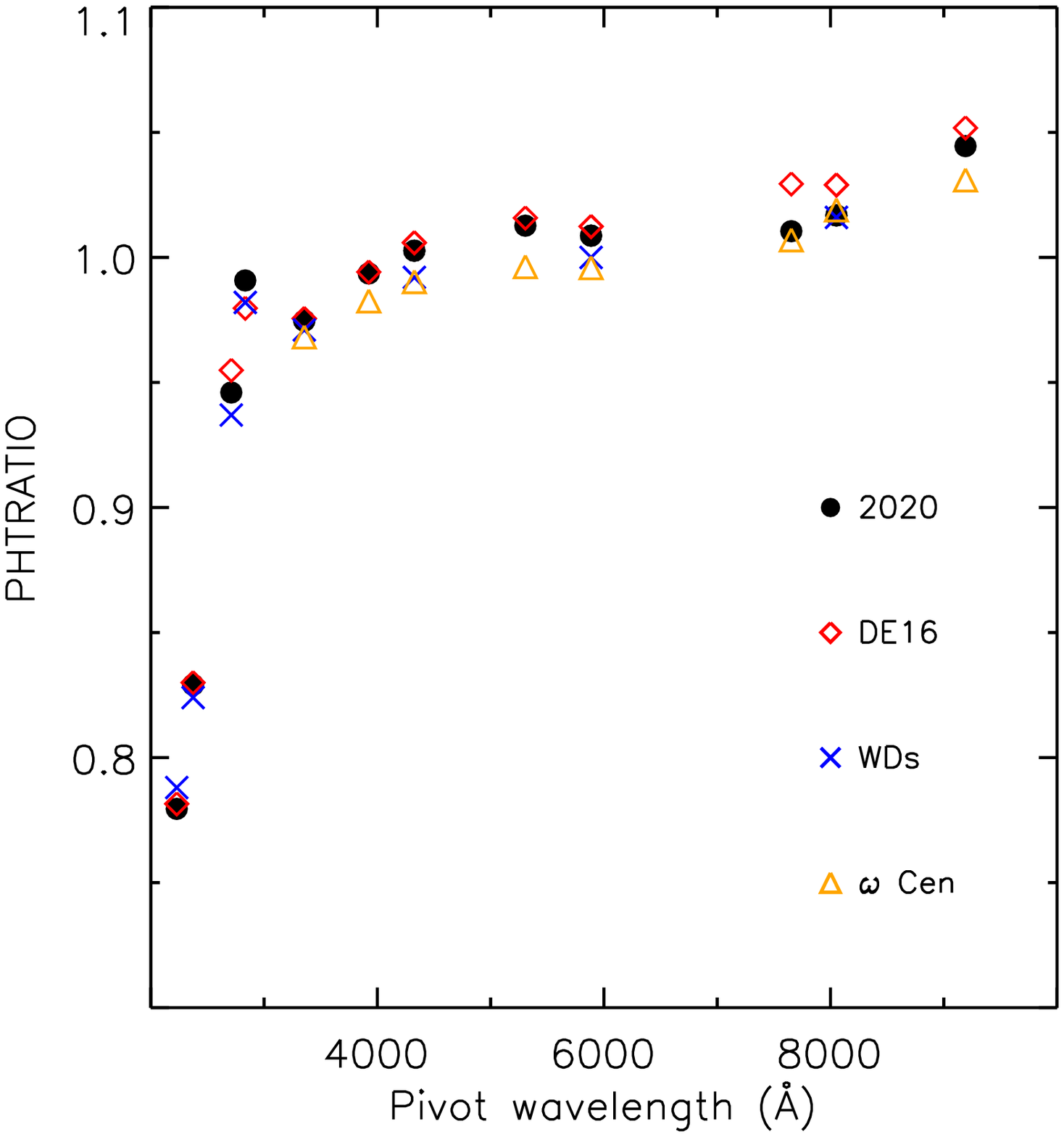}
\caption{Comparison of old (DE16, magenta diamond) and new (2020, filled black circle) 
synthetic detector sensitivity ratios, PHTRATIO, with observed values 
computed from WD (blue cross) and $\omega$ Cen (orange triangle) 
observations. See text for more details. \label{fig:photratio} }
%\end{center}
\end{figure}

\begin{table*}
\begin{center}
\begin{scriptsize}
\caption{\footnotesize PHTRATIO values (PHTFLAM2/PHTFLAM1) for the WFC3-UVIS 42 full-frame filters derived 
from synthetic photometry by using the new (syn\_2020) and the old (syn\_DE16) calibration {\it Pysynphot} files, 
and by using observations of standard WDs and $\omega$ Cen. \label{table:11}}
\vspace{0.3cm}
\begin{tabular}{lccccc}
\hline
\hline
Filter & Pivot  & PHTRATIO$_{syn\_2020}$ & PHTRATIO$_{syn\_DE16}$ & PHTRATIO$_{WDs}$  & PHTRATIO$_{\omega Cen}$   \\
\hline
\hline
F200LP& 4971.86 & 0.998 & 1.0023 & \ldots & \ldots   \\
F218W & 2228.04 & 0.779 & 0.7815 & 0.788  & \ldots   \\
F225W & 2372.05 & 0.829 & 0.8300 & 0.824  & \ldots   \\
F275W & 2709.69 & 0.946 & 0.9549 & 0.937  & \ldots   \\
F280N & 2832.86 & 0.991 & 0.9797 & 0.982  & \ldots   \\
F300X & 2820.47 & 0.932 & 0.9350 & \ldots & \ldots   \\
F336W & 3354.49 & 0.975 & 0.9756 & 0.971  & 0.968    \\
F343N & 3435.15 & 0.967 & 0.9690 & \ldots & \ldots   \\
F350LP& 5873.87 & 1.035 & 1.0316 & \ldots & \ldots   \\
F373N & 3730.17 & 0.975 & 0.9663 & \ldots & \ldots   \\
F390M & 3897.24 & 0.985 & 0.9870 & \ldots & \ldots   \\
F390W & 3923.69 & 0.994 & 0.9942 & \ldots & 0.983    \\
F395N & 3955.19 & 0.998 & 1.0031 & \ldots & \ldots   \\
F410M & 4108.99 & 1.001 & 1.0022 & \ldots & \ldots   \\
F438W & 4326.23 & 1.003 & 1.0059 & 0.992  & 0.990    \\
F467M & 4682.58 & 0.995 & 0.9979 & \ldots & \ldots   \\
F469N & 4688.10 & 0.996 & 1.0082 & \ldots & \ldots   \\
F475W & 4773.10 & 1.007 & 1.0126 & \ldots & \ldots   \\
F475X & 4940.72 & 1.005 & 1.0121 & \ldots & \ldots   \\
F487N & 4871.38 & 0.987 & 1.0028 & \ldots & \ldots   \\
F502N & 5009.64 & 1.004 & 1.0187 & \ldots & \ldots   \\
F547M & 5447.50 & 0.996 & 1.0158 & \ldots & \ldots   \\
F555W & 5308.43 & 1.013 & 1.0013 & \ldots & \ldots   \\
F600LP& 7468.12 & 1.027 & 1.0335 & \ldots & \ldots   \\
F606W & 5889.17 & 1.009 & 1.0124 & 1.000  & 0.996    \\
F621M & 6218.85 & 1.005 & 1.0140 & \ldots & \ldots   \\
F625W & 6242.56 & 1.000 & 1.0169 & \ldots & \ldots   \\
F631N & 6304.29 & 0.995 & 1.0049 & \ldots & \ldots   \\
F645N & 6453.59 & 1.005 & 1.0111 & \ldots & \ldots   \\
F656N & 6561.37 & 0.969 & 1.0008 & \ldots & \ldots   \\
F657N & 6566.63 & 1.000 & 1.0049 & \ldots & \ldots   \\
F658N & 6584.02 & 0.991 & 1.0069 & \ldots & \ldots   \\
F665N & 6655.88 & 1.012 & 1.0135 & \ldots & \ldots   \\
F673N & 6765.94 & 1.023 & 1.0202 & \ldots & \ldots   \\
F680N & 6877.60 & 1.019 & 1.0074 & \ldots & \ldots   \\
F689M & 6876.75 & 1.009 & 1.0139 & \ldots & \ldots   \\
F763M & 7614.37 & 1.020 & 1.0340 & \ldots & \ldots   \\
F775W & 7651.36 & 1.010 & 1.0294 & \ldots & 1.007    \\
F814W & 8039.06 & 1.017 & 1.0290 & 1.016  & 1.019    \\
F845M & 8439.06 & 1.009 &  1.0171 & \ldots & \ldots \\
F850LP& 9176.13 & 1.044 &  1.0518 & \ldots & 1.031    \\
F953N & 9530.58 & 1.039 &  1.0231 & \ldots & \ldots   \\
\hline
\hline
\end{tabular}
\end{scriptsize}
\end{center}
\end{table*}

Due to the new WFC3-UVIS inverse sensitivities being time-dependent, PHTRATIO now varies with the epoch
of observation. We calculated PHTRATIO by using the PHTFLAM1 and PHTFLAM2 values 
at the reference epoch, June 26 2009, and results are listed in Table~\ref{table:11}, with a comparison to
DE16 PHTRATIO values. 
We also compared the new values with PHTRATIO calculated by using the photometry of standard
star WDs dithered on the image and observed between 2010 and 2014 \citep{mack2015}, and
by using photometry from dithered images of $\omega$ Cen observed 
between 2009 and 2011 \citep{mack2013, mack2016}.
Note that these PHTRATIO values were calculated for a 10 pixel aperture radius and no time-dependent 
sensitivity correction was applied; however, the observations were mostly done at the beginning 
of WFC3 lifetime (2009 -- 2014) so the total sensitivity losses were $\approx$ 1\% at most, depending on
the filter. The observed PHTRATIO values derived from WD (blue crosses) and $\omega$ Cen (yellow triangles) 
photometry are shown in Fig.~\ref{fig:photratio}, where the new (2020, black filled circles) and 
old (DE16, magenta diamonds) synthetic PHTRATIO values are also
plotted for the same filters as a function of the pivot wavelength.
Table~\ref{table:11} and Fig.~\ref{fig:photratio} show that the
new PHTRATIO values agree very well with those obtained from both WD and $\omega$ Cen observations, 
and show improvements in several filters from the old PHTRATIO values which 
did not include any time-dependent sensitivity correction.
The new PHTRATIO values differ slightly from those calculated using either stepped WD or dithered $\omega$ Cen observations, with an average difference of $\approx$ 0.6\% and a dispersion of $\approx$ 0.7\% for ten wide-band filters and one narrow-band filter (see Table~\ref{table:11}).
These differences are most likely due to the new PHTRATIO values being derived from standard star photometry in the corner sub-arrays where the flat field is less accurate. Moreover, flat field 
differences between opposite corners of the WFC3-UVIS detectors can be as large as $\approx$ 3\% (see Table~3 in \citealt{mack2016}).

\section{Comparison to the old inverse sensitivity values\label{sec:comp}}
We compared the new WFC3-UVIS inverse sensitivities (converted in units of magnitudes as zero points, ZPs, in the ST photometric system) with those from the 2017 calibration. The latter ZPs were calculated by using the previous CALSPEC SEDs and six years of data available for the standard stars (2009 -- 2015): these measurements were simply averaged
without normalizing for the sensitivity changes with time. This resulted in different reference epochs 
per filter, due to the different observing cadence for each detector. Additionally, the 2017 calibration did not account for differences in the observed count rates 
due to flat field errors. 
In CA21, and as reported in this work, WFC3-UVIS inverse sensitivities were derived by using 
new CALSPEC SEDs for the standard stars and ten years of available data (2009 -- 2019). 
The photometric measurements of the standard stars were then normalized to a single reference epoch 
and weighted for photometric errors and number of measurements.
 
\begin{figure*}
\begin{center}
\includegraphics[width=0.55\textwidth, height=0.75\textheight, angle=90]{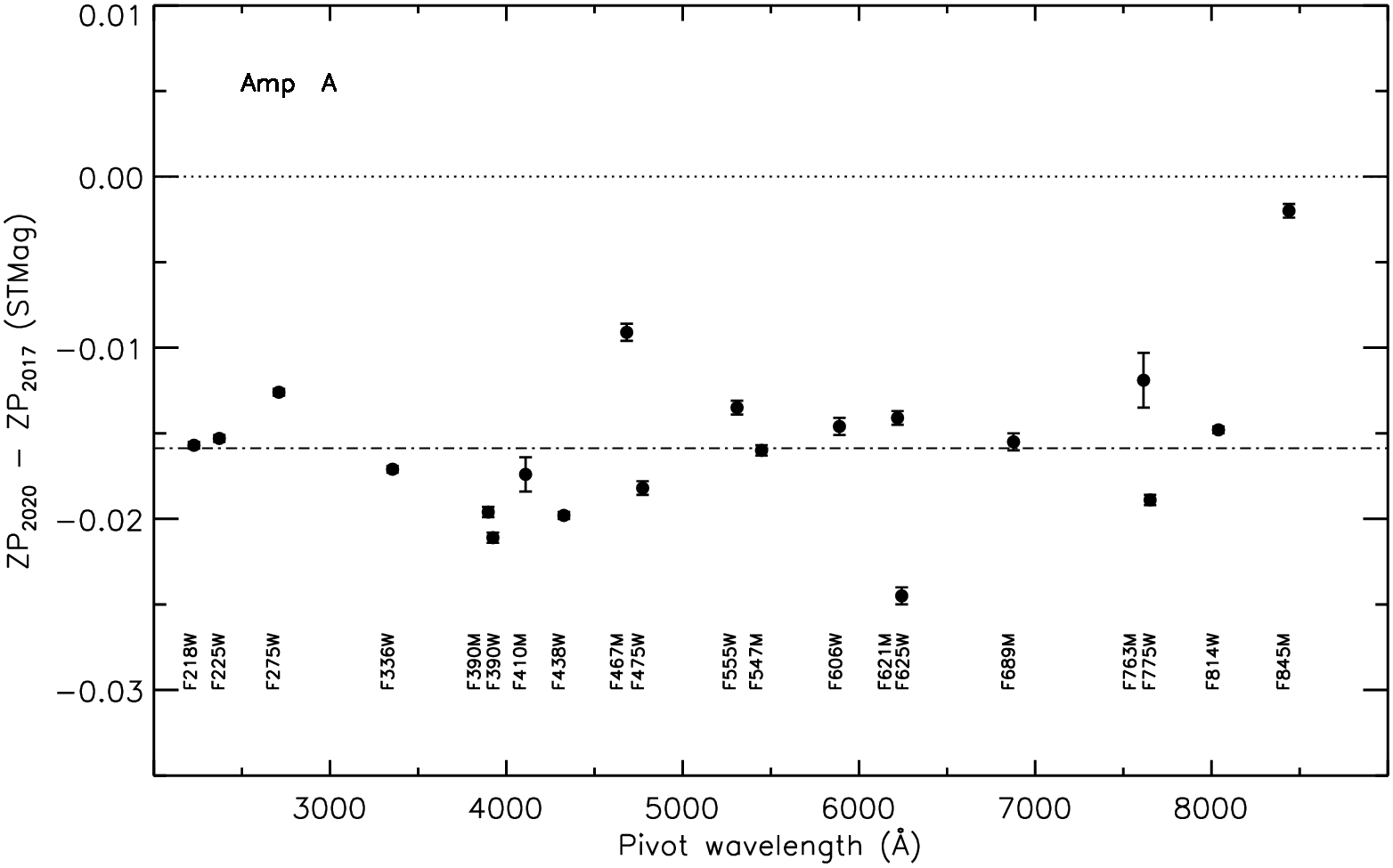}
%\vspace{-0.5cm}
\caption{Comparison between the new (2020) and old (2017) ZPs for WFC3 UVIS1 (Amp A) in the ST photometric system 
for the wide- and medium-band filters over the entire WFC3 UVIS wavelength range. Error bars are also shown. \label{fig:comp1}}
\end{center}
\end{figure*}

\begin{figure*}
\begin{center}
\includegraphics[width=0.55\textwidth, height=0.75\textheight, angle=90]{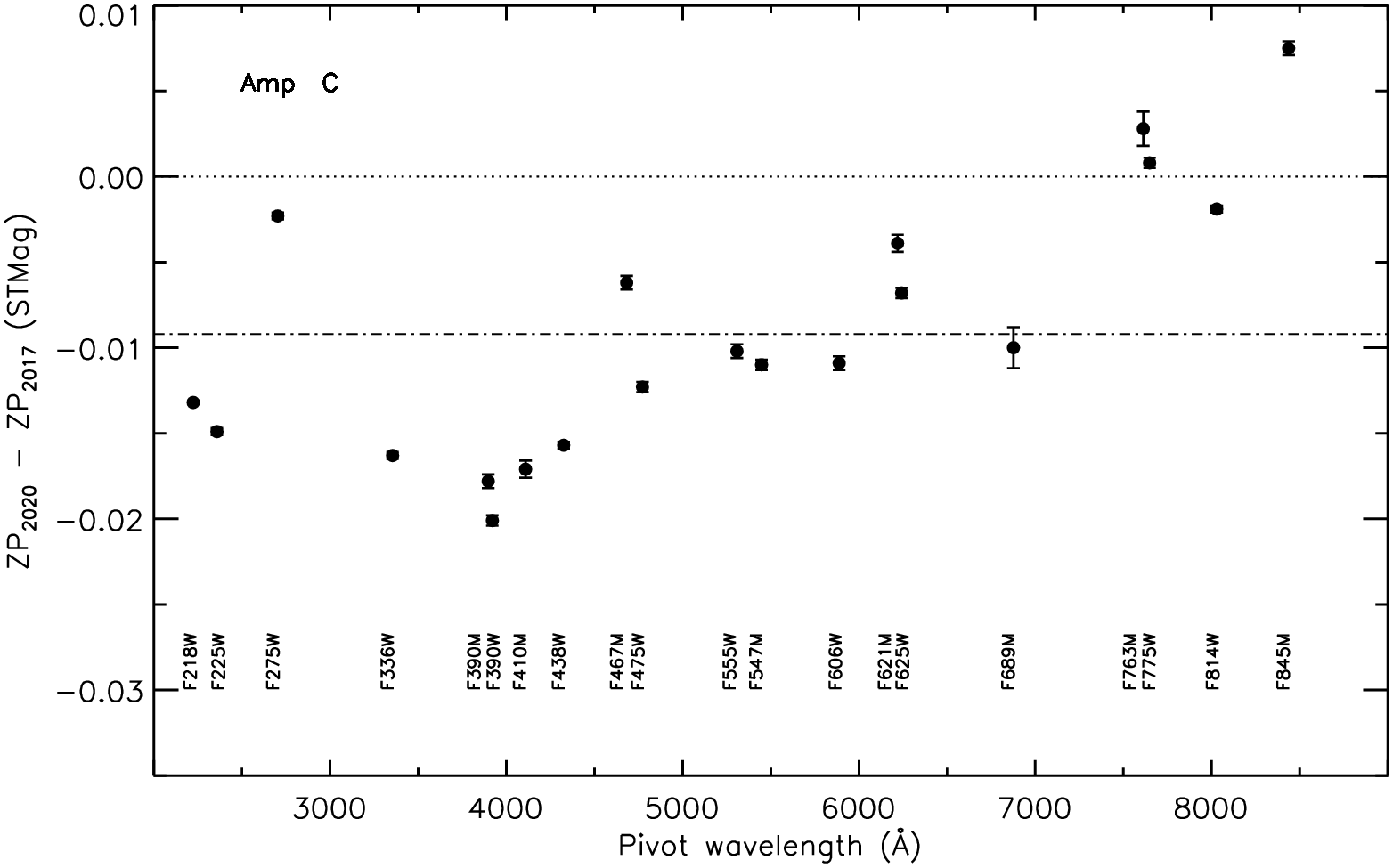}
%\vspace{-0.5cm}
\caption{Same as Fig.~\ref{fig:comp1} but for WFC3 UVIS2 (Amp C) detector. \label{fig:comp2}}
\end{center}
\end{figure*}

Fig.~\ref{fig:comp1} shows the comparison between new (2020) and old (2017) ZPs as a function of pivot wavelength for the wide- and medium-band UVIS1 filters. The ZPs differ on average by $\approx$ 1.5\%, with the new ZPs being brighter compared to the old ones. This is mostly due to the reference flux of Vega being $\approx$ 1\% brighter, and to the standard star photometry being corrected for losses in sensitivity. A similar comparison holds for the UVIS2 filters as shown in Fig.~\ref{fig:comp2}, where the ZPs differ on average by $\approx$ 0.9\%.
It is worth noting that the difference between old and new ZPs for the UVIS2 detector shows a 
trend between 4,000 $< \lambda <$ 8,000 \AA~ that is not clearly seen for UVIS1 values.
This difference between the two detectors may be due to how the five standard star measurements were 
collected between 2009 and 2019: for most filters, Amp A (UVIS1) has more observations 
at the beginning, 2009-2010, and throughout WFC3 lifetime compared to Amp C (UVIS2), 
which has less and more sparse measurements. Since the old ZPs were calculated by
simply averaging the photometry of the standard stars over the 2009 -- 2015 time interval, without applying any time
correction, the ZP values for UVIS2 resulted to be centered on later epochs compared to 
values for UVIS1.
 
\begin{figure*}
\begin{center}
\includegraphics[width=0.55\textwidth, height=0.75\textheight, angle=90]{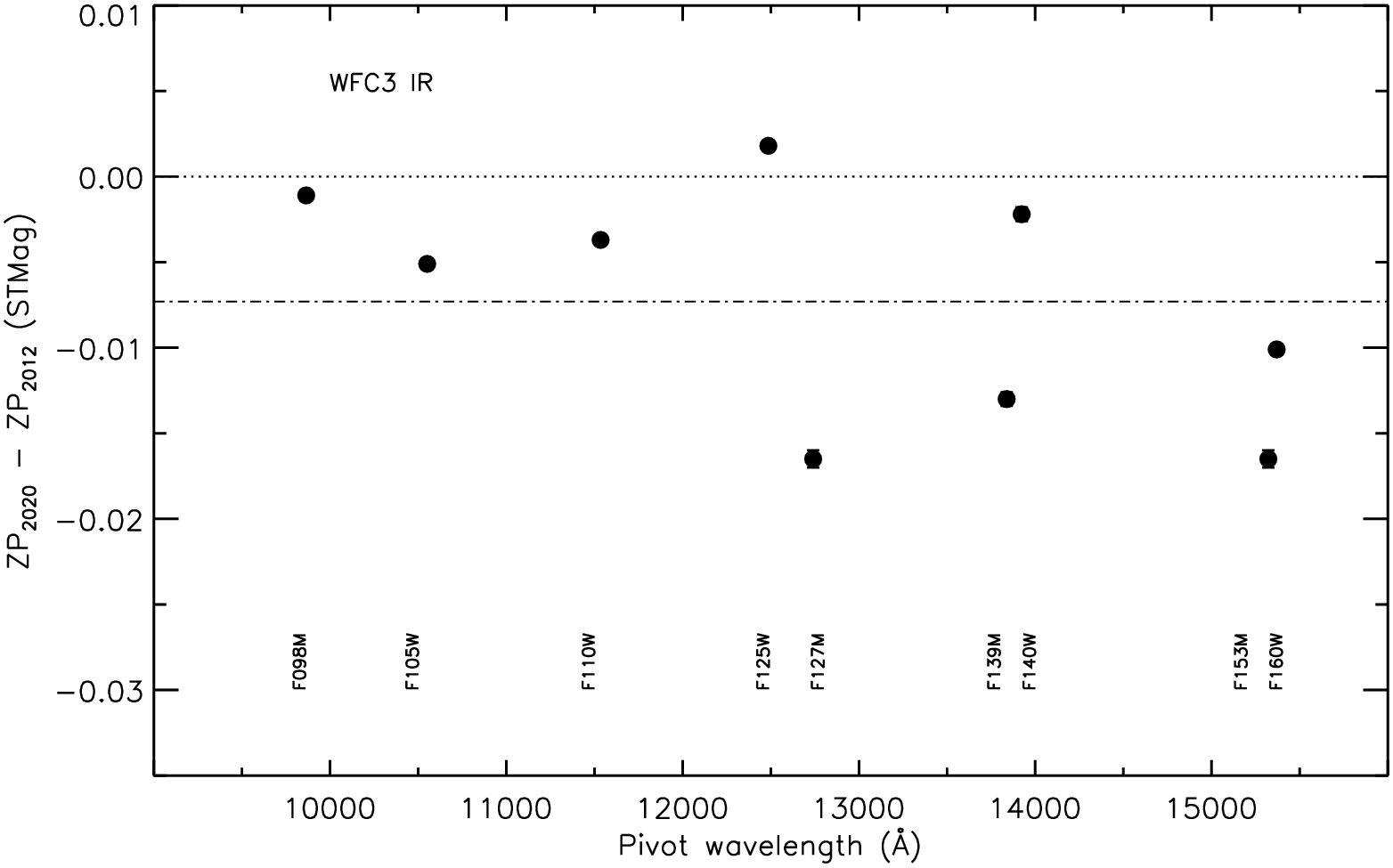}
%\vspace{-0.5cm}
\caption{Comparison between the new (2020) and old (2012) ZPs for WFC3-IR in the ST photometric system 
for the wide- and medium-band filters over the entire wavelength range. Error bars are also shown.  \label{fig:compIR}}
\end{center}
\end{figure*}

We performed a similar comparison for WFC3-IR inverse sensitivities; the new ZPs were based on updated CALSPEC SEDs and eleven years of data available for five standard stars (2009 -- 2020),
compared to the previous ZPs from \citet{kalirai2011}, based on only 1.5 years of photometry for four standard stars. Furthermore, the new ZPs were calculated by using updated pixel-to-pixel flat fields to
correct for spatial sensitivity residuals up to 0.5\% in the center of the detector and up to 2\% at the edges
\citep{mack2021} . Also, a new set of {\it delta} flat fields was used to correct for low-sensitivity artifacts known 
as blobs in six filters, namely $F098M$, $F105W$, $F110W$, $F125W$, $F140W$, and $F160W$ 
\citep{olszewski2021}. 
Fig.~\ref{fig:compIR} shows the comparison between new (2020) and old (2012) ZPs as a function of pivot
wavelength for the wide- and medium-band WFC3-IR filters. The ZPs differ on average by $\lesssim$ 1.0\%,
with the new ZPs being brighter compared to the old ones as expected. 
The comparison also shows that the difference is larger for redder filters and it is due to the new CALSPEC
models differing more at longer wavelengths  compared with the old ones.

\section{Testing the new time-dependent WFC3-UVIS inverse sensitivities\label{testing}}
\subsection{\omc photometry}
\begin{figure*}
\begin{center}
\includegraphics[width=0.55\textwidth, height=0.75\textheight, angle=90]{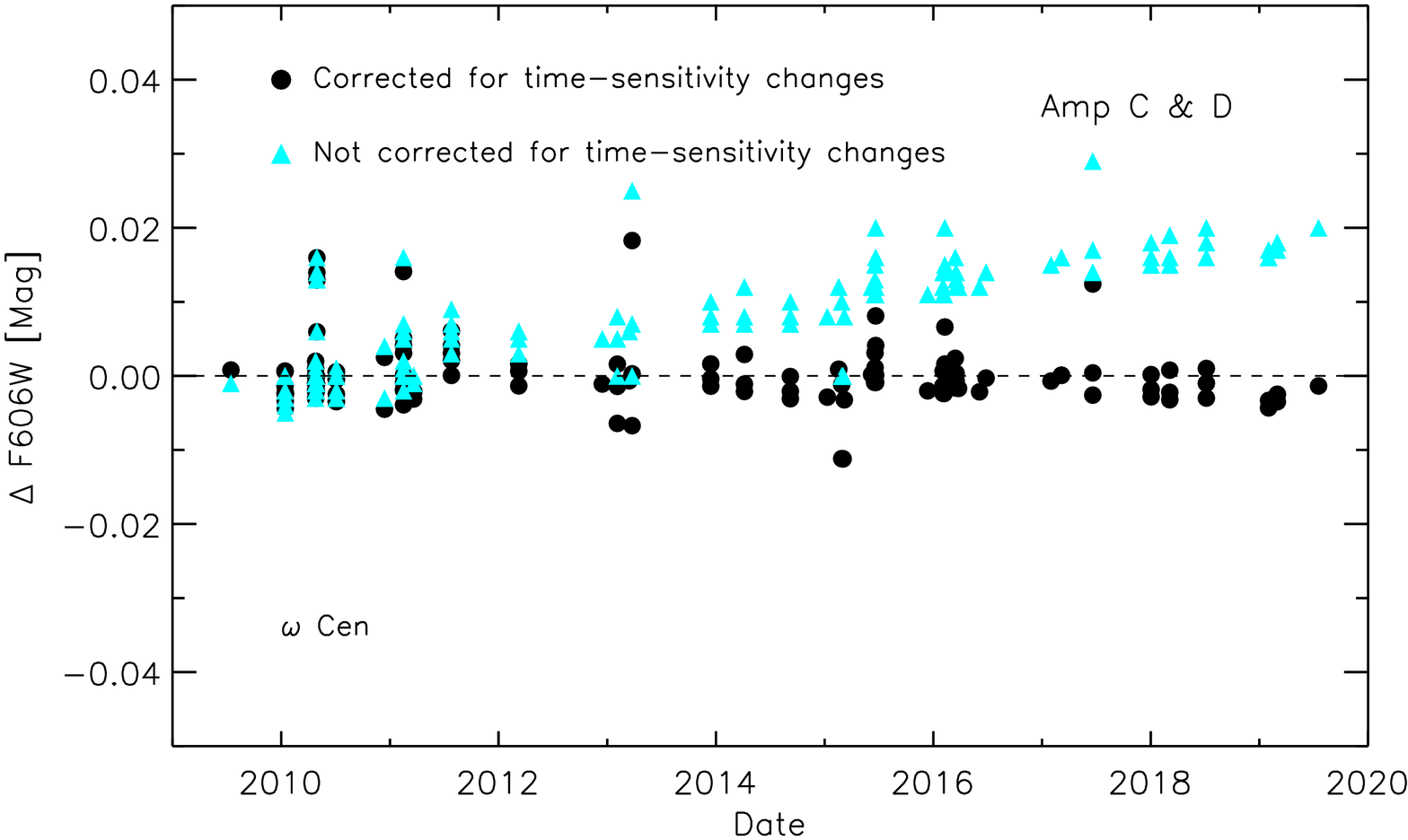}
%\vspace{-0.5cm}
\caption{Instrumental (cyan filled triangles) and calibrated (black filled circles) magnitude differences between a reference UVIS2 (Amp C \& D) $F606W$ image collected in 2009 for the Galactic globular cluster \omc and all subsequent 161 images.
The instrumental magnitude differences were not corrected for time sensitivity changes, while the calibrated magnitude differences were corrected by using the time-dependent
inverse sensitivity values provided in the image header as the PHTFLAM2 keyword.
 \label{fig:comp_ome}}
\end{center}
\end{figure*}

In order to verify the precision of the new WFC3-UVIS time-dependent photometric calibration we used data collected with the $F606W$ filter for the Galactic 
globular cluster \omc over a $\approx$ 10-year time interval.
Aperture photometry with a 5-pixel radius was performed on 162 {\it \_flc} images and the average instrumental 
(cyan filled triangles) and calibrated (black filled circles) magnitude difference for all stars measured in all images versus a reference image collected in 2009
is shown as a function of observing epoch in Fig.~\ref{fig:comp_ome}. 
Only photometry for stars 250 pixel away from the readout amplifiers (in this case Amp C and D) is shown to mitigate the possible effects of a non perfect CTE correction.
The instrumental magnitude differences (cyan triangles) were not corrected 
for the time sensitivity changes of the UVIS2 detector (Amp C and D) and show an increase of $\approx$ 0.02 mag over the 10-year time interval: this flux drop is expected from the calculated sensitivity loss rate of $0.02$\%/yr for UVIS2 and the $F606W$ filter (see Table~\ref{table:4}). The calibrated magnitude differences (black circles) were corrected by using the time-dependent
inverse sensitivity values provided in the image header as the PHTFLAM2 keyword: as the plot in Fig.~\ref{fig:comp_ome} shows, \omc magnitude differences cluster around 0, with a dispersion of $\approx$ 0.2\%, as it is expected in the absence of sensitivity changes with time of the detector.
We performed the same test for \omc UVIS1 observations and we obtained a similar result.

\subsection{Staring mode photometry}\label{sec:test_stare}
As a further validation of the time-dependent WFC3-UVIS photometric calibration, we downloaded from the Mikulski Archice for Space Telescopes (MAST) all available data in the $F814W$ filter for the standard WDs GD153 and GRW70 from 2009 until the end of 2021. These images were processed through the current {\it calwf3} pipeline, version {\it 3.6.2}. We performed aperture photometry with a 10-pixel radius and used the header PHOTFLAM keyword to correct the observed count rates for time sensitivity changes. We then compared the observed with the synthetic count rates derived by using {\it STsynphot} and the new SEDs, in-flight and aperture corrections and filter curves for $F814W$ and UVIS1 and UVIS2, as listed in Table~\ref{table:7}. 

\begin{figure}
\begin{center}
\includegraphics[width=0.55\textwidth, height=0.5\textheight]{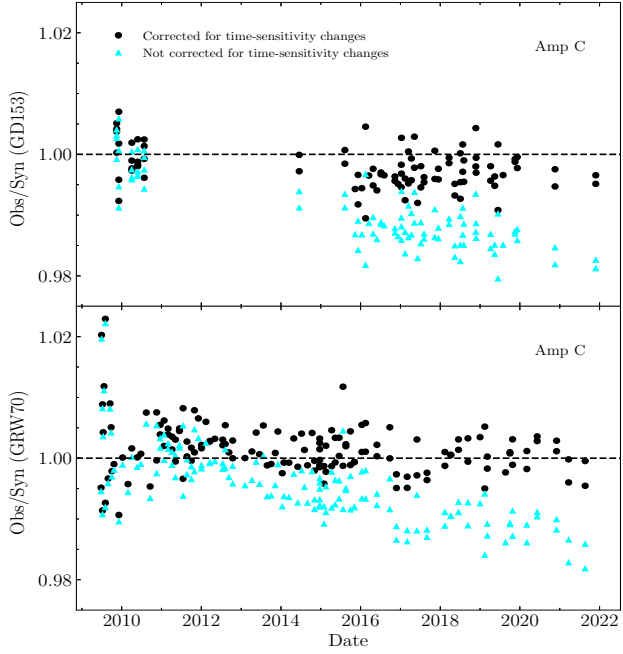}
%\vspace{-0.5cm}
\caption{Staring mode $F814W$ UVIS2 (Amp C) uncorrected (cyan filled triangles) and corrected (black filled circles) observed over synthetic count rate ratios for the standard star GD153 (top panel) and GRW70 (bottom) as a function of date. The observed count rates were corrected for sensitivity changes by using the time-dependent inverse sensitivity values provided in the image header as the PHTFLAM2 keyword. \label{fig:staring}}
\end{center}
\end{figure}

Fig.~\ref{fig:staring} shows the ratio of UVIS2 (Amp C) observed over synthetic count rates for GD153 (top panel) and GRW70 (bottom): the filled cyan triangles are the uncorrected ratios while the filled black circles indicate the corrected ones. The uncorrected ratios decrease with time for both stars, with a total loss of flux of $\approx$ 1.5\% over almost 13 years; this is compatible with the measured sensitivity loss rate of $\approx$ 0.1\%/yr for the $F814W$ filter and UVIS2. The corrected observed over synthetic count ratios cluster around 1 for both stars, with a dispersion of $\approx$ 0.33\% and 0.6\% for GD153 and GRW70, respectively, validating the current time-dependent inverse sensitivities values. 

We performed the same test for UVIS1 observations and we obtained a similar result.

\subsection{Scan mode photometry}\label{sec:test_scan}
We performed another validation test by using all available scan data for GD153 and GRW70 in the $F814W$ filter from 2017 until the end of 2021. The images were downloaded from MAST and processed with the current {\it calwf3} pipeline and aperture photometry was performed following the recipe presented in Section~\ref{sec:scan}. The count rates were corrected for sensitivity changes by using the PHOTFLAM keyword values provided in the image header.

To calculate synthetic count rates for the scan mode observations, 
a PSF was extrapolated from the latest $F814W$ EE curves and convolved with a scan line corresponding to the length of the spatial scan observations, thus creating a synthetic spatial scan with the same scale as the observed data. Aperture photometry was performed on the synthetic scan image, yielding a 'enrectangled' energy correction factor. Synthetic count rates were then produced by using {\it STsynphot} with the new SEDs for GD153 and GRW70, multiplied by the synthetic 'enrectangled' energy correction factor, the new in-flight and aperture corrections and filter curves for $F814W$ and UVIS1 and UVIS2, as listed in Table~\ref{table:7}.

\begin{figure}
\begin{center}
\includegraphics[width=0.5\textwidth, height=0.51\textheight]{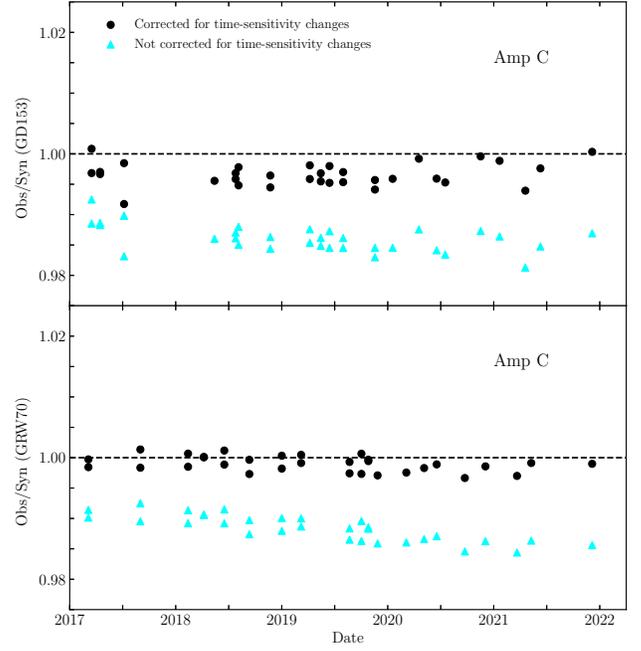}
%\vspace{-0.5cm}
\caption{Scan mode $F814W$ UVIS2 (Amp C) uncorrected (cyan filled triangles) and corrected (black filled circles) observed over synthetic count rate ratios for the standard star GD153 (top panel) and GRW70 (bottom) as a function of date. The scan observation count rates were corrected for sensitivity changes by using the time-dependent inverse sensitivity values provided in the image header as the PHTFLAM2 keyword. \label{fig:scan}}
\end{center}
\end{figure}

Fig.~\ref{fig:scan} shows the uncorrected (filled cyan triangles) and corrected (black filled circles) observed over synthetic count rates for UVIS2 (Amp C) for GD153 (top panel) and GRW70 (bottom): the uncorrected scan ratios are systematically lower than 1.0 by more than 1\%, due to the loss of sensitivity of the detector from 2009 to 2017, and maintain a decreasing trend until the end of 2021, for a total loss of $\approx$ 1.5\%. On the other hand, the corrected ratios cluster around 1 for both stars, with a dispersion of $\approx$ 0.15\% (GD153) and $\approx$ 0.1\% (GRW70), a factor of 2 and 6 lower, respectively, compared to the dispersion of the staring mode photometry for the same stars. 

We performed the same test for UVIS1 observations and, as in the case of the staring mode observations, we obtained a similar result.

\section{How to perform WFC3 photometric calibration \label{example}}
%The new WFC3 processing pipeline, currently version {\it calwf3\_v3.6.2}, 
%scales the UVIS2 detector to UVIS1. Therefore, users need to apply only one inverse sensitivity value, the header keyword PHTFLAM1 (same as the PHOTFLAM keyword) to calibrate their photometry. However, if observations are done by using a UVIS2 sub-array, the flux correction can be avoided by setting FLUXCORR = OMIT in the image header and re-running {\it calwf3} on the raw images. In this case, the PHTFLAM2 keyword value in the image header should be used to calibrate the photometry.

%Photometric measurements performed on images retrieved from MAST after October 15, 2020 or re-processed after this date with the current pipeline, {\it calwf3} version 3.6.2, should be calibrated by using the new time-dependent inverse sensitivities available in the image header. 

The PHOTFLAM inverse sensitivity keyword provided in the WFC3 image header can be used to convert the photometric measurements into fluxes in units of erg cm$^{-2}$ s$^{-1}$ \AA$^{-1}$.
For images retrieved from MAST after October 15, 2020, or re-processed after this date with the current pipeline, {\it calwf3} version 3.6.2, the PHOTFLAM inverse sensitivity keyword includes the time-dependent correction; this can than be used to scale all the photometric measurements collected at different epochs to the reference epoch of June 26, 2009.

Magnitudes in different photometric systems can also be obtained: WFC3 provides ZPs in three systems, namely AB, ST and Vega\footnote{https://www.stsci.edu/hst/instrumentation/wfc3/data-analysis/photometric-calibration/uvis-photometric-calibration} for the reference epoch.

%The AB photometric system \citep{oke74} is strictly speaking defined for monochromatic fluxes. 
%If the flux at frequency $\nu$ is denoted by $f_{\nu}$ and expressed
Magnitudes in the AB photometric system are based on a constant flux (flat spectrum) per unit frequency \citep{oke74}, with the ZP set so that VEGA has AB magnitude $\approx$ 0 in the Johnson $V$-band.  
Fluxes in the AB system are then defined in units of 
erg cm$^{-2}$ s$^{-1}$ Hz$^{-1}$, and the corresponding AB magnitude at frequency $\nu$ is defined as:

\begin{equation}
m(AB_{\nu}) = -2.5 \log (F_{\nu}) - 48.60
\end{equation}

%This corresponds to a normalization where an object with a flat spectrum has AB magnitude equal to its $V$ band magnitude \citep{oke1983}.

%To incorporate the idea of AB magnitudes for non-monochromatic use, say for a passband $X$, we use the extension as proposed 
%by \citet{fukugita1996} for a photon proportional detector system to define the quantity $f_{X}$:

%\begin{equation}
%f_{X} = \frac{\int f_{\nu} v^{-1}R d\nu }{\int \nu^{-1} R d\nu} = \frac{\int N_{\nu} R d\nu }{\int (h\nu)^{-1} R d\nu}
%\end{equation}

%where $R$ is the (telescope + instrument + filter) response function for passband $X$, $N_{\nu}$ is the count rate of photons per unit frequency and $h$ is Planck's constant. The numerator on rightmost side is the photon count rate in the band, so $f_{X}$ is directly proportional to the photon count rate. 

%The AB magnitude for passband $X$ is then given by:

%\begin{equation}
%m(AB_{X}) = -2.5 \log(f_{X}) - 48.60
%\end{equation}

where the zero point is set such as AB mag = 0 is for $F_{\nu}$= 3.96$\times$10$^{-20}$ erg cm$^{-2}$ s$^{-1}$ Hz$^{-1}$.

The Space Telescope (ST) photometric system is defined in the wavelength domain and magnitudes in this systems are:

\begin{equation}
m(ST_{\lambda}) = -2.5 log (F_{\lambda}) - 21.10
\end{equation}

where ST mag = 0 is $F_{\lambda}$ = 3.96$\times$10$^{-9}$ erg cm$^{-2}$ s$^{-1}$ \AA$^{-1}$.

A characteristic wavelength, {\it pivot wavelength}, can be defined 
to convert flux densities from the frequency to the wavelength domain as:

\begin{equation}
\lambda _{p} = \sqrt{\frac{c F_{\nu}}{F_{\lambda }}} = \sqrt{\frac{\int R \lambda d\lambda }{\int R \frac{d\lambda }{\lambda } } }
\end{equation}

where $R$ is the (telescope + instrument + filter) response function and $F_{\nu}$ and $F_{\lambda}$ are the fluxes in the frequency and wavelength domains.
We can then transform AB to ST magnitudes using the relation:

\begin{equation}
m(ST_{\lambda}) = m(AB_{\nu}) + 5 log( \lambda _{p}) - 18.70
\end{equation}

In the Vega photometric systems, magnitudes are calculated using the Vega
flux as a reference:

\begin{equation}
m(Vega) = -2.5 log (F/F_{Vega}) 
\end{equation}

In order to convert count rates measured on WFC3 images into magnitudes, 
users can use the inverse sensitivity provided as the header keyword PHOTFLAM. 
According to the chosen photometric system, the following
equations can be used to obtain the ZP at the epoch of the observation:

\begin{itemize}

\item ST photometric system:

\begin{equation}
ZP_{STMag} = -21.1 -2.5 log(PHOTFLAM)
\end{equation}

\item AB photometric system:

\begin{equation}
ZP_{ABMag} = -21.1 -2.5 log(PHOTFLAM) - 5 log(PHOTPLAM) + 18.70)
\end{equation}

where PHOTPLAM is the pivot wavelength keyword, also available in the image header.

\item Vega photometric system: the calculation of the ZP follows two steps. 
First, the user needs to calculate the flux of Vega as observed by the telescope, detector and filter, $FLAM_{Vega}$; to do this the new Vega SED should be used, 
{\it alpha\_lyr\_stis\_010.fits}, available from the CALSPEC database\footnote{https://archive.stsci.edu/hlsps/reference-atlases/cdbs/current\_calspec/} or from the CRDS database \footnote{https://hst-crds.stsci.edu/}.
Subsequently, the ZP can be calculated as:

\begin{equation}
ZP_{Vega} = -2.5 log(PHOTFLAM/FLAM_{Vega})
\end{equation}

\end{itemize}

The WFC3 team provides a Jupyter notebook that shows how to calculate the ZPs in the different photometric systems by using {\it synphot}, specifically {\it STSynphot}.
Another notebook is also provided and shows how to use the new 
time-dependent solutions to work with WFC3-UVIS data obtained at different epochs. 
These notebooks are available from the 
WFC3 photometric calibration 
web page\footnote{https://www.stsci.edu/hst/instrumentation/wfc3/data-analysis/photometric-calibration/uvis-photometric-calibration} 
and https://www.overleaf.com/project/619d593114f0b03d31bde4b4from the STScI WFC3 Software Library on 
Github\footnote{https://github.com/spacetelescope/WFC3Library}.

Users should note that the final photometry for both detectors will be in 
the UVIS1 system since WFC3 processing pipeline, {\it calwf3}, flux scales the UVIS2 detector to UVIS1 by multiplying UVIS2 by the ratio of the inverse sensitivities, PHTRATIO in the image header keyword. Therefore, users need to apply only one inverse sensitivity value, the header keyword PHOTFLAM (same as the PHTFLAM1 keyword) to calibrate their photometry.

However, UVIS1 and UVIS2 have significantly different quantum efficiencies in the UV regime ($\lambda \lesssim$ 4,000 \AA), and the modified PHTRATIO introduced by DE17 to match the
the count rate ratio of UVIS1 and UVIS2 for hot stars ($T_{eff} \ge$ 30,000 $K$) in the 
$F218W$, $F225W$, $F275W$ and $F200LP$ filters, does not work for cooler stars. 
Photometry for cooler stars measured on the UVIS2 detector in the UV filters thus needs to be corrected by applying a magnitude offset according to their UVIS2 color, if available, or temperature or spectral type. 

Offsets for magnitudes measured on the UVIS2 detector relative to UVIS1 were  calculated in the $F218W$, $F225W$, and $F275W$ filters by using synthetic photometry and observations of the globular cluster $\omega$ Cen. These are presented and available in CA18. Before applying the offset ($\Delta Mag$) to magnitudes measured on the UVIS2 detector, the photometry must be calibrated by using the provided inverse sensitivities:

\begin{equation}
m(ST) = -21.1 -2.5 log(PHOTFLAM) - \Delta Mag
\end{equation}

The final photometry for both detectors will be in the UVIS1 system. If the observed sources lie in the same detector,
the color term is negligible ($<$ 1\%), and no magnitude offset needs to be applied. 

%Note that if observations are done by using a UVIS2 sub-array, the flux correction can be avoided in WFC3 processing pipeline by setting FLUXCORR = OMIT in the image header and re-running {\it calwf3} on the raw images ({\it \_raw.fits}). In this case, the PHTFLAM2 keyword value in the image header should be used to calibrate the photometry.
%If the observed sources lie in the same detector,
%the color term is negligible ($<$ 1\%), and no magnitude offset needs to be applied. 
%However, due to the significantly different sensitivity of the UVIS1 and UVIS2 detectors in the UV wavelength range, 
%magnitude offsets for the UV filters as a function of the source color need be applied to magnitudes measured on UVIS2 to transform the photometry to the UVIS1 photometric system. These corrections are currently available in \citet{calamida2018}.
For users who require sub-percent photometric calibration accuracy, or if observations are done by using a UVIS2 sub-array only,
we recommend treating each detector separately when observing with 
the UV filters $F218W$, $F225W$, and $F275W$.
UVIS1 magnitudes will be calibrated as usual, by applying PHOTFLAM, while the UVIS2 magnitudes will be calibrated by using PHTFLAM2.
In this case, the MAST downloaded raw images ({\it \_raw.fits}) will have to be re-processed manually through the {\it calwf3} pipeline omitting
the flux correction, i.e. by setting FLUXCORR = OMIT in the image header.

\subsection{An example with observed data}\label{example}
We present here an example of how to perform the photometric calibration of WFC3 data collected at different epochs and in different filters. Observations of the LMC globular cluster NGC~1978 were collected in 2011 in the $F555W$ filter (PID: 12257), in 2016 in $F438W$ (PID: 14069), and in 2019 in $F275W$ and $F814W$ (PID: 15630). We downloaded the {\it \_flc} images processed through {\it calwf3\_v3.5.2} from MAST. PSF photometry was performed with the software ePSF \citep{anderson2006} and corrected to a 10 pixel aperture radius; the new WFC3-UVIS EE corrections were then applied to bring the magnitudes from 10 pixels to {\it infinity}. Magnitudes were also corrected for exposure time and the new ZP for each filter and appropriate epoch of observation, as found in the PHOTFLAM image header keyword, was used to derive the final calibrated magnitudes in the Vega system. Summarizing, we obtained calibrated magnitudes, $M$, for stars in NGC~1978, from the instrumental magnitudes, $m$, as:

\begin{equation}
M = m + AP_{10} + EE_{inf} + 2.5 log(EXPTIME) + ZP_{Vega}
\end{equation}

where $AP_{10}$ is the correction to bring PSF magnitudes to a 10 pixel aperture radius, $EE_{inf}$ in the EE correction from 10 pixel to {\it infinity} and $ZP_{Vega}$ was calculated as explained above.

\begin{figure*}%[ht]
%\centering
\includegraphics[scale=0.36]{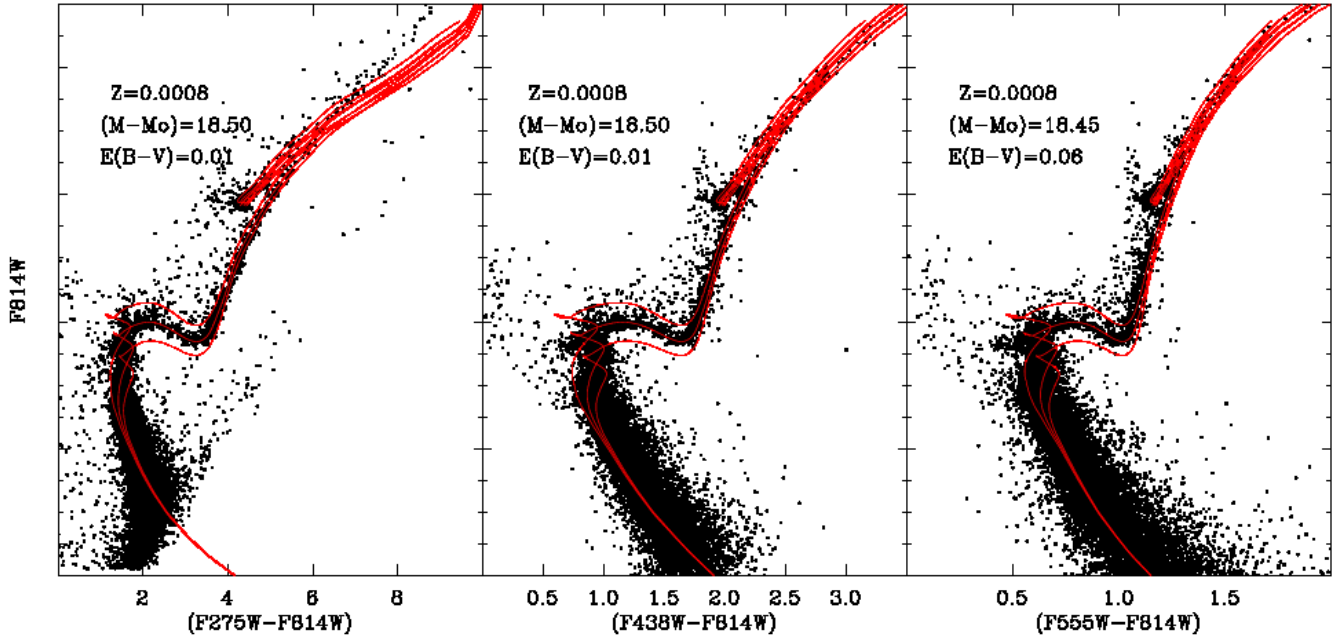}
\caption{$F814W,\ F275W - F814W$ (left panel), $F814W,\ F438W - F814W$ (middle) and $F814W,\ F555W - F814W$ (right) CMD of the LMC globular cluster NGC~1978. A set of BaSTI isochrones with different ages and same metallicity is over-plotted. The assumed distance modulus and reddening are labeled in each panel.  \label{fig:ngc1978} }
\end{figure*}

Fig.~\ref{fig:ngc1978} shows the calibrated $F814W, \ F275W - F814W$ (left panel), $F814W,\ F438W - F814W$ (middle) and $F814W,\ F555W - F814W$ (right) color-magnitude-diagrams (CMDs) of NGC~1978. 
A set of scaled-solar BaSTI\footnote{http://basti-iac.oa-abruzzo.inaf.it/index.html} isochrones for a metallicity of $Z =$ 0.0008 and different ages, namely 2, 2.5 and 3 Gyr, are over-plotted: these models were transformed to the observational plane by using the new WFC3-UVIS filter throughputs\footnote{The updated WFC3 filter throughputs can be downloaded from the WFC3 web page at https://www.stsci.edu/hst/instrumentation/wfc3/performance/throughput . The updated filter curves and other telescope and instrument component files that are needed to run {\it synhpot} simulations can be downloaded from the archive web page at https://archive.stsci.edu/hlsp/reference-atlases or from the {\it STsynphot} web page at https://stsynphot.readthedocs.io/en/latest/stsynphot/data\_hst.html or from the CRDS database web page at https://hst-crds.stsci.edu/}, a distance modulus range of 18.45 $\le \mu_0 \le$ 18.50 mag and reddening of 0.01 $\le E(B-V) \le$ 0.06 mag. The figure shows that theory and observations taken over 8 years of WFC3 lifetime are in very good agreement for all filters from the UV to the optical regime: the isochrones nicely reproduce all the features of the CMD, i.e. the main-sequence, the sub-giant and red-giant branch and the horizontal branch. The comparison between models and observations provides an age range of 2-3 Gyr for NGC~1978, in agreement with previous studies based on Advanced Camera for Surveys (ACS) photometry \citep{martocchia2018}.

\section{Conclusions\label{sec:conclusions}}
In this manuscript we described how new inverse sensitivities were derived
for the WFC3-UVIS and the WFC3-IR detectors.
Time-dependent inverse sensitivities were derived for UVIS1 and UVIS2, 
for the 42 full-frame filters. They provide a photometric internal precision $\lesssim$ 0.5\% for wide-, medium-, and narrow-band filters, with a significant improvement compared to the old values, where the precision was $\lesssim$ 1\% for wide-, $\lesssim$ 2\% for medium-, and $\lesssim$ 5 - 10\% for narrow-band filters. 
The accuracy of the flux calibration is $\approx$ 2-3\% for wide- and medium-band filter, 
and $\approx$ 5\% for narrow-band filters.

In addition to the $\approx$ 1\% error in the absolute flux calibration of the HST standards (BO20), the photometric calibration is also limited by errors in the flat field. These are typically $<$ 1\% in the upper left corner of Amp A (UVIS1) but can be as large as 2\% for a few filters \citep{mack2015,mack2016}.
In this corner of the detector, the PSF focus is highly variable due to telescope breathing effects 
\citep{sabbi2013}, and this affects the accuracy of the aperture correction applied to
the crowded stellar field ($\omega$ Cen) photometry used to derive the in-flight ($L$-flat) correction 
\citep{mack2013}.
Amp A also contains the flare, a wedge-shaped internal reflection in the ground flats 
at a level of 1-2\%, which is strongest in the upper left corner of the UVIS1 detector \citep{mccullough2011, mack2013}.
These uncertainties may also impact the UVIS2 to UVIS1 inverse sensitivity ratio, i.e. PHTRATIO, since it is based on the photometry of standard stars observed in the small 512$\times$512 corner sub-arrays, where the flat field 
is less accurate.
In the future, the WFC3 team plans to improve the flat fields by using more stellar cluster data 
and improved reduction techniques.

Major changes of the new WFC3-UVIS inverse sensitivities compared to the latest values delivered in 2017 can be summarized as follows:\\

1) The new inverse sensitivities are based on new SEDs for the standard stars and 
a new reference flux for Vega (see BO20);\\
2) EE fractions for a few filters were updated by using the time-sensitivity corrections
and a new method for drizzling the standard star images. These were used in the 
computation of the new inverse sensitivities; \\
3) Four extra years (2015 -- 2019) of standard star photometry were used;\\
4) Time-dependent corrections were calculated and standard star photometry was
corrected for the sensitivity changes before deriving the inverse sensitivities. Also, the standard star observed count rates were weighted according to their photometric errors and number of collected measurements.\\

We also provided new inverse sensitivities for the WFC3-UVIS 20 quad filters by using the updated SEDs for the standard stars and the reference Vega flux. These values do not have a time-sensitivity correction, since not enough observations were available to calculate a sensitivity change rate. The accuracy of the calibration for these filters is $\approx$ 10-15\%.

New inverse sensitivities for the 15 WFC3-IR filters were also derived by using the new SEDs, the updated reference Vega flux and 10 years of photometric data for the standard stars. These differ by $\approx$ 1.5\% compared to the latest values delivered in 2011 and provide a photometric internal precision of $\approx$ 1\% for all filters. The accuracy of the flux calibration is $\approx$ 2-3\%.

The new time-dependent WFC3-UVIS photometric calibration was validated by using $F814W$ observations of two CALSPEC standard WDs, namely GD153 and GRW70, collected in staring and scan mode during a 12- and 3-year time frame, respectively. After applying the new time-dependent inverse sensitivities, observed over synthetic count rate ratios are, withing uncertainties, in very good agreement over the entire time intervals.

We also used observations of the globular cluster $\omega$ Cen to validate the WFC3-UVIS time-dependent calibration. Aperture photometry was performed on 162 $F606W$ exposures collected in the 2009 - 2020 time frame and the single epoch inverse sensitivities were used to correct the photometry. The magnitude differences with respect to the first reference image all cluster around zero after the correction, with a dispersion of $\approx$ 0.2\%.

We also showed examples on how to derive ZPs in the AB, ST and Vega photomerric systems and how to apply the time-dependent WFC3-UVIS calibration to real observations of the cluster NGC~1978.

The new in-flight and aperture correction files for WFC3-UVIS, and the new filter curves for both WFC3 detectors were delivered to the 
Calibration Reference Data System (CRDS), and can be downloaded from the CRDS web page\footnote{https://hst-crds.stsci.edu/}, from the STScI archive\footnote{https://archive.stsci.edu/hlsp/reference-atlases} or from the {\it synphot} web page \footnote{https://www.stsci.edu/hst/instrumentation/reference-data-for-calibration-and-tools/synphot-throughput-tables.html} and used
in {\it synphot} simulations. The models of the standard stars used in this analysis and needed for the simulations are available from the CALSPEC data repository\footnote{https://www.stsci.edu/hst/instrumentation/reference-data-for-calibration-and-tools/astronomical-catalogs/calspec} or the aforementioned web pages (see Table~\ref{table:7} for a list of the file names).

New IMPHTTABs for WFC3-UVIS and WFC3-IR were also delivered, {\it 51c1638pi\_imp.fits} and {\it 4af1533ai\_imp.fits}, respectively.
All WFC3 data were re-processed through the new version of the pipeline, {\it calwf3\_v3.5.2} as October 15, 2020, and
%To maintain consistency with the IMPHTTABs, the filter throughputs used in the construction of bandpasses for synthetic photometry with {\it synphot} were updated to reflect the new photometric calibrations. More information regarding the download of these files is available here\footnote{https://www.stsci.edu/hst/instrumentation/reference-data-for-calibration-and-tools/synphot-throughput-tables}. The models of the standard stars used in this analysis are available in the CALSPEC data repository.  Note that these files are not necessary for the re-calibration of images, and are solely for synthetic photometry.
new photometry keyword values (PHOTFLAM, and PHTFLAM1, 
PHTFLAM2, PHTRATIO for WFC3-UVIS) were populated in the image headers.
Therefore, we recommend users to retrieve again data collected before October 2020 so that their headers will be populated with the latest inverse sensitivity values by the WFC3 processing pipeline.

The WFC3 photometric calibration web page\footnote{https://www.stsci.edu/hst/instrumentation/wfc3/data-analysis/photometric-calibration/} 
%uvis-photometric-calibration}
provides the new inverse sensitivity values,
calculated at the reference epoch, i.e. MJD = 55008 (June 26, 2009),
for WFC3-UVIS.
Values of the inverse sensitivities for UVIS1 and UVIS2 at each observing epoch can be found in the image header.
However, at the same web page and on the STScI WFC3 Software Library on Github\footnote{https://github.com/spacetelescope/WFC3Library},
tutorials (Jupyter notebooks) are provided for running {\it synphot} with  the new filter curves in order to derive the inverse sensitivity and ZP values for any detector, observing epoch, 
filter or aperture.  Another notebook describing how to use the new time-dependent solutions to work with WFC3-UVIS data obtained at different observation dates is also available at the same location.

In the future, the WFC3 team plans to improve the EE corrections for more filters, in particular at wavelengths longer than $\lambda \ge$ 8,000\AA~ and for narrow-band and long-pass filters, by using the new method illustrated in this manuscript. Also, new flat fields for the quad filters will be calculated to replace the current set of ground flats, and the inverse sensitivities recomputed. Stare and scan mode observations for different clusters are being collected with WFC3-IR to better characterize the time-sensitivity changes of this detector, if any, and new inverse sensitivities for the WFC3-IR filters will be calculated if needed.

\acknowledgments
The authors would like acknowledge Susana Deustua, Kailash Sahu, Sylvia Bagget and Joel Green for their useful comments and discussions.
This study was supported by NASA through grant P0004.03.06.05 from the Space
Telescope Science Institute, which is operated by AURA, Inc., under NASA contract NAS~5-26555.

\facility{HST (WFC3)}

%%%%%%%%%%%%%%
% References %
%%%%%%%%%%%%%%

\clearpage
\bibliographystyle{aa}

\bibliography{calamida}

%\appendix

\end{document}